\documentclass[twocolumn,amsmath,amssymb,aps,prb]{revtex4-1}

\usepackage{epstopdf}
\usepackage{amsmath}
\usepackage{braket}
\usepackage{amssymb}
\usepackage{color}
\usepackage{amssymb}
\usepackage{appendix}
\usepackage{float}
\usepackage{graphicx}% Include figure files
\usepackage{dcolumn}% Align table columns on decimal point
\usepackage{bm}% bold math
\usepackage{hyperref}% add hypertext capabilities
\usepackage{subfig}
\usepackage{caption}
\usepackage{subcaption}

\hypersetup{
            colorlinks=true,
            linkcolor=blue,
            anchorcolor=blue,
            citecolor=blue}

\begin{document}

\title{Quantum control and the topological properties of the magnon photo-transport in two-dimensional collinear ferromagnet}% Force line breaks with \\

\author{Jun-Cen Li}
\affiliation{College of Sciences, Northeastern University, Shenyang 110004, China}

\author{Corresponding author: An Du}
\email{duan@mail.neu.edu.cn}
\affiliation{College of Sciences, Northeastern University, Shenyang 110004, China}
\affiliation{National Frontiers Science Center for Industrial Intelligence and Systems Optimization, Northeastern University, China.}
\affiliation{Key Laboratory of Data Analytics and Optimization for Smart Industry (Northeastern University), Ministry of Education, China.}

\date{\today}% It is always \today, today,

\begin{abstract}
In this work, we investigate light-induced magnon spin photocurrent (MSPC) and magnon energy photocurrent (MEPC) via the Aharonov-Casher (AC) effect. 
We treat the influence of the electric field through the AC effect as a perturbative term.
Subsequently, we derived explicit expressions for MSPC and MEPC in a two-dimensional, collinear ferromagnetic system.
Our theoretical framework is applied to both Hexagonal and Kagome two-dimensional ferromagnetic lattices.
We demonstrate that MSPC and MEPC can be modulated by tuning the optical frequency and the material's relaxation time.
Furthermore, in the low-frequency and infinite-relaxation-time limit, longitudinal MS(E)PC correlates with the topological properties of the magnon system.
\end{abstract}

\pacs{Valid PACS appear here}% PACS, the Physics and Astronomy
\maketitle

%\tableofcontents

\section{\label{sec:level1}Introduction}
In 1879, E. H. Hall discovered the Hall effect \cite{ref0}.
Then a series of novel electron transport phenomena were discovered \cite{ref0.1,ref0.2,ref0.3,ref0.4}.
Subsequently, 
%in order to develop new electronic devices, 
the quantum Hall effect \cite{ref0.5}, anomalous Hall effect \cite{ref0.6,ref0.7}, electron spin transport \cite{ref0.8,ref0.9,ref0.91,ref0.92,ref0.93}, electron thermoelectric transport \cite{ref0.94,ref0.95}, anomalous Nernst effect \cite{ref0.96}, second-order Hall effect \cite{ref0.97,ref0.98} and the corresponding topological properties\cite{ref0.99,ref0.991,ref0.9911,ref0.9912} of electrons have been extensively studied.
On the other hand, the thermal transport and the topological properties of phonons have also been studied in depth in recent years \cite{ref0.992,ref0.993,ref0.994,ref0.995,ref0.996,ref0.997,ref0.997,ref0.998,ref0.999}.
With the development of spintronics, the transport properties of magnons attracted extensive attention \cite{ref0.9991,ref0.9992,ref0.9993,ref0.9994,ref0.9995}.
Magnon is a quasi-particle equivalent to quantized spin waves. 
Different from electrons, magnons have no charge. Different from phonons, magnons have a magnetic moment. So, magnons do not lose energy through Joule heating, and can be controlled by the magnetic moment\cite{ref1}. 
So magnons have the potential to replace electrons as new information carriers. 

To develop new devices using magnons as information carriers, it is helpful to study the transport of magnons.
But unlike electrons, magnons are neutral particles. So the driving mechanism of magnon transport is different from that of electrons.
Typically, magnon transport is driven by temperature or chemical potential gradients, which create non-equilibrium magnon distributions. Typical transport phenomena, such as the magnon Seebeck effect \cite{ref1.1}, thermal Hall effect  \cite{ref2,ref3,ref4,ref5,ref6,ref8,ref9,ref9.01,ref9.1,ref9.2}, and spin Nernst effect \cite{ref9.3,ref9.4,ref10,ref11,ref11.1}, are all linear-response behaviors under such driving conditions.
In addition, the method of driving magnon transport by strain has also been studied \cite{ref12}.
In contrast, this work focuses on the non-traditional driving method of time-dependent electric fields and explores the topological transport properties of magnons induced thereby.

In order to make it easier to control magnon transport, the control of magnons by an electromagnetic field 
has attracted attention. 
In 2018, I. Proskurin \textit{et al.} theoretically investigated the magnon spin photocurrent (MSPC) generated by the Zeeman coupling of magnons and the magnetic field component of light in antiferromagnetic insulators \cite{ref13}.
Then H. Ishizuka and M. Sato proposed the magnons current induced by the magnetic field component of linearly polarized (LP) light \cite{ref14,ref14.1}.
On the other hand, using the electric field part of light to excite MSPC has also received wide attention.
In 2021, E. V. Boström \textit{et al.} studied the magnon circular photogalvanic effect enabled by two-magnon Raman scattering \cite{ref14.2}.
On the other hand, since the electric field is easier to control compared to temperature gradient, chemical potential gradient, and strain, we hope to achieve the regulation of magnons through the electric field.
In 1984, Y. Aharonov and A. Casher found that neutral particles with magnetic moments acquire a geometric phase when moving in an electric field, an effect known as the Aharonov-Casher (AC) effect \cite{ref15}.
As a kind of neutral quasiparticle with a magnetic moment, magnons can be affected by the AC effect.
In 2017, K. Nakata \textit{et al.} studied the quantum Hall response of magnons to magnetic field gradients in AC effects induced by magnon Landau levels \cite{ref16,ref17}. 
In their work, although the quantum Hall effect of magnons is related to the electric field and AC effect, the magnon current is still a response to the magnetic field gradient, instead of the electric field.
According to the AC effect, the electric field component of light can directly excite magnon spin (energy) current.
Because magnons and electrons have different statistical and electromagnetic properties, the photo-transport of magnons is different from that of electrons\cite{ref18,ref19,ref20,ref21,ref22,ref23}.
In 2024, Y. Wang \textit{et al.} investigated the linear MSPC induced by time-dependent electric field through the AC effect in the SSH model \cite{ref24}.
In the same year, they derived the expression of the magnon photogalvanic effect and studied the properties of the photogalvanic effect of magnons under different polarized light \cite{ref25}.
However, considering the relaxation time of magnons, the linear MSPC and magnon energy photocurrent (MEPC) induced by the time-dependent electric field in two-dimensional collinear ferromagnets have not been studied.

In this work, we investigate the methods for controlling linear MSPC and MEPC (induced by AC effect) in two-dimensional collinear ferromagnetic materials.
Different from the theoretical work of linear magnon transport \cite{ref1.1,ref5,ref6,ref8,ref9,ref9.3,ref9.4}, our study is about MSPC and MEPC.
Compared with the work of Y. Wang \textit{et al.} \cite{ref24,ref25}, the novelty of our work consists of three parts.
Firstly, in Ref. \onlinecite{ref24}, the authors derived the magnon spin photocurrent (MSPC) in one-dimensional materials (the SSH model); In Ref. \onlinecite{ref25}, the authors derived the second-order MSPC in two-dimensional materials. The first innovation of our work lies in our derivation of the first-order MSPC and magnon energy photocurrent (MEPC) in two-dimensional materials.
Secondly, in  Ref. \onlinecite{ref24} and Ref. \onlinecite{ref25}, the authors did not take into account the external effect. However, we considered the external effect through the phenomenological approach of relaxation time.
Thirdly, in the low-frequency limit, the formula of MSPC inherently contains Berry curvature. Thus in this limit MSPC exhibits topological properties. Through the analysis of the topological properties, we can observe that in the low-frequency limit, the direction of MSPC can be controlled by Dzyaloshinskii-Moriya (DM) interaction.

In Sec. \ref{sec:Formalism}, we derive the magnon spin photoconductivity tensor and magnon energy photoconductivity tensor through quantum kinetics theory, and discuss the topological properties of the magnon spin photoconductivity tensors in the case of low optical frequency and infinite relaxation time.
In Sec. \ref{sec:Effective time-reversal symmetry}, we discuss the constraint of effective time-reversal symmetry on the reciprocal space structure of the magnon.
In Sec. \ref{sec:Model calculation}, we take a numerical calculation of the two-dimensional ferromagnetic Hexagonal lattice model and Kagome lattice model to study the method of controlling the MSPC and MEPC. 
We calculate and discuss the variation of magnon spin photoconductivity and magnon energy photoconductivity with optical frequency under different relaxation times.
In addition, under the condition of low optical frequency and infinite relaxation time, the MSPC and MEPC move sinusoidally in the time-dependent electric field. The magnon spin (energy) photoconductivity determines the oscillation amplitude of the MS(E)PC and can hopefully be controlled by the topological properties of the system. Here, MS(E)PC means MSPC or MEPC.

\section{\label{sec:Formalism} Formalism}

The system considered by us is a two-dimensional collinear out-of-plane ferromagnet,
in which the ferromagnet is located in the $x-y$ plane and the spin points in the positive direction of the $z$ axis.

\subsection{\label{sec:level11} The Aharonov-Casher effect and the Hamiltonian of magnon}
Without the time-dependent electric field, the single-magnon Hamiltonian is expressed as $\hat{H}_0$, which satisfies $\hat{H}_0\ket{\psi_n(\boldsymbol{k})}=\varepsilon_n(\boldsymbol{k})\ket{\psi_n(\boldsymbol{k})}$. Here, $\ket{\psi_n(\boldsymbol{k})}=e^{i\boldsymbol{k}\cdot\boldsymbol{r}}\ket{u_n(\boldsymbol{k})}$ is the Bloch state of a magnon, and $\varepsilon_n(\boldsymbol{k})$ is the energy of the magnon band. The perturbation Hamiltonian induced by a time-dependent electric field is expressed as $\hat{H}^{\prime}(t)$.

We consider the effect of a time-dependent electric field on a magnon through the Aharonov-Casher effect (AC) phase.
Magnons acquire a geometric phase when they move in an electric field \cite{ref15,ref24,ref25}.
\begin{equation}
%\begin{split}
\theta_{ij}=\frac{g_J\mu_B}{\hbar c_{lv}^2}\int_{\boldsymbol{r}_i}^{\boldsymbol{r}_j}\left[\boldsymbol{E}(t)\times\hat{\boldsymbol{e}}_{\mu}\right]\cdot d\boldsymbol{r}.
%\end{split}
\label{ACE1}
\end{equation}
Here, $\boldsymbol{E}(t)$ is the electric field of light, $g_J$ is the $Land\acute{e}$ factor, $\mu_B$ is the $Bohr$ magneton, and $c_{lv}$ is light velocity. 
$\hat{\boldsymbol{e}}_{\mu}$ is a unit vector on the magnetic moment  direction of the particle.
We assume that the spatial variation scale of the electric field is much larger than the scale of the lattice constant. So the AC phase can be expressed as \cite{ref25}
\begin{equation}
	%\begin{split}
	\theta_{ij}\approx\frac{g_J\mu_B}{\hbar c_{lv}^2}\left[\boldsymbol{E}(t)\times\hat{\boldsymbol{e}}_{\mu}\right]\cdot \boldsymbol{d}_{ij},
	%\end{split}
	\label{ACE2}
\end{equation}
in which $\boldsymbol{d}_{ij}$ is the displacement from position $i$ to position $j$.
The perturbative Hamiltonian $\hat{H}^{\prime}(t)$ of AC effect can be expressed as \cite{ref24,ref25}
\begin{equation}
	%\begin{split}
	\hat{H}^{\prime}(t)=\frac{g_J\mu_B}{ c_{lv}}\tilde{\boldsymbol{E}}(t)\cdot\boldsymbol{r}.
	\label{Hprimemain}
	%\end{split}
\end{equation}
Here, $\tilde{\boldsymbol{E}}(t)=-\frac{1}{c_{lv}}\partial_t\boldsymbol{E}(t)\times\hat{\boldsymbol{e}}_z$ is the effective electric field, and $\boldsymbol{E}(t)$ is the applied time-dependent electric field which can be expressed as $\boldsymbol{E}(t)=\sum_j\boldsymbol{E}(\omega_j)e^{-i\omega_j t}$, in which $\boldsymbol{E}(\omega_i)$ is the complex amplitude of electric field (detail see Appendix \ref{Hamiltonian}). 
Therefore, the effective electric field can be expressed as $\tilde{\boldsymbol{E}}(t)=\sum_j\tilde{\boldsymbol{E}}(\omega_j)e^{-i\omega_jt}$, in which $\tilde{\boldsymbol{E}}(\omega_j)$ is
\begin{equation}
	%\begin{split}
	\tilde{E}_a(\omega_j)=\sum_b\frac{i\omega_j}{c_{lv}}\epsilon_{abz}E_b(\omega_j).
	%\end{split}
	\label{tildeEE}
\end{equation}
Here, subscripts $a$, $b$ label the direction in Cartesian coordinates, $a,b=x,y$. $E_b(\omega_j)$ is the complex amplitude in the direction of $b$ with the light frequency $\omega_j$. $\epsilon_{abz}$ is Levi-Civita symbol, $c_{lv}$ is light velocity.

In other words, the electric field component of light (the periodically changing electric field) causes a disturbance to magnons through the AC effect, thereby driving the transport of magnons. The AC phase provides the geometric interaction between the electric field and magnons, allowing us to control the magnon transport by manipulating the electric field.

\subsection{The methods for calculating MSPC and MEPC}
In this subsection, we discuss the methods to derive MSPC and MEPC induced by a time-dependent electric field. 
Because the magnon number is not conserved, the chemical potential of the magnon is zero.
Therefore, the energy current of magnons is equal to the heat current of magnons.

Because each magnon excitation carries the spin angular momentum $\hbar$ in systems with conservation
of the total spin along the $z$ direction, the MSPC can be expressed as \cite{ref10,ref23,ref25}
\begin{equation}
	%\begin{split}
	\boldsymbol{j}=\hbar tr\left[\hat{\rho}\hat{\boldsymbol{v}}\right].
	%\end{split}
	\label{jtr}
\end{equation}
Here, $\hat{\rho}$ is the density matrix of one-magnon, which can be expressed by expanding by order $\hat{\rho}=\sum_{\alpha}\hat{\rho}^{(\alpha)}$, in which $\alpha$ labels the order of the correction of density matrix.
$\hat{\boldsymbol{v}}=\frac{1}{i\hbar}\left[\hat{\boldsymbol{r}},\hat{H}_0\right]$ is velocity operator of magnon, in which $\boldsymbol{r}$ and $\hat{H}_0$ are the the position of the magnon and the single magnon Hamiltonian without electric field respectively. The MEPC can be expressed as
\begin{equation}
	%\begin{split}
	\boldsymbol{j}^E=\frac{1}{2}tr\left\{\hat{\rho}\left[\hat{\boldsymbol{v}},\hat{H}_0\right]_+\right\}.
	%\end{split}
	\label{jtrq}
\end{equation}
Here, $\hat{\boldsymbol{v}}^{E}=\frac{1}{2}\left[\hat{\boldsymbol{v}},\hat{H}_0\right]_+$ is   energy velocity operator of magnon, in which $\left[\hat{A},\hat{B}\right]_+=\hat{A}\hat{B}+\hat{B}\hat{A}$ is anticommutation operator \cite{ref9.1}.

As discussed in the Appendix \ref{The expressions of magnon photocurrent and magnon energy photocurrent}, the $\alpha$-th MSPC can be expressed as
\begin{equation}
	%\begin{split}
	\boldsymbol{j}^{(\alpha)}=\hbar\sum_{mn\boldsymbol{k}}\rho_{mn}^{(\alpha)}(\mathbf{k})\boldsymbol{v}_{mn}(\boldsymbol{k}),
	%\end{split}
	\label{j0main}
\end{equation}
and the $\alpha$-th MEPC can be expressed as
\begin{equation}
	%\begin{split}
	\boldsymbol{j}^{E(\alpha)}=\sum_{mn\boldsymbol{k}}\rho_{mn}^{(\alpha)}(\boldsymbol{k})\boldsymbol{v}_{mn}^E(\boldsymbol{k}).
	%\end{split}
	\label{jE0main}
\end{equation}
Here, $\rho_{mn}^{(\alpha)}(\boldsymbol{k},t)$ is the $\alpha$th-order correction of density matrix. $\boldsymbol{v}_{mn}(\boldsymbol{k})$ and $\boldsymbol{v}_{mn}^E(\boldsymbol{k})$ are the matrix elements of velocity and energy velocity in the Bloch representation, respectively.
\subsubsection{The matrix element of velocity and energy velocity in the Bloch representation}
According to Appendix \ref{MVMEV}, the matrix element of magnon velocity is
{\small
	\begin{equation}
		\begin{split}
			\boldsymbol{v}_{mn}(\boldsymbol{k})
			&=\frac{1}{i\hbar}\left\{i\partial_{\boldsymbol{k}}\varepsilon_n(\boldsymbol{k})\delta_{mn}
			+\boldsymbol{\mathcal{A}}_{mn}(\boldsymbol{k})\varepsilon_{nm}(\boldsymbol{k})\right\},\\
		\end{split}
	\end{equation}
}
and the magnon energy velocity is
{\small
\begin{equation}
	\begin{split}
		\boldsymbol{v}^E(\boldsymbol{k})
		&=\frac{1}{2i\hbar}\left[\varepsilon_m(\boldsymbol{k})+\varepsilon_n(\boldsymbol{k})\right]\left\{i\partial_{\boldsymbol{k}}\varepsilon_n(\boldsymbol{k})\delta_{mn}
		+\boldsymbol{\mathcal{A}}_{mn}(\boldsymbol{k})\varepsilon_{nm}(\boldsymbol{k})\right\},\\
	\end{split}
\end{equation}
}
in which $\varepsilon_n(\boldsymbol{k})$ is the energy of band and $\boldsymbol{\mathcal{A}}_{mn}(\boldsymbol{k})=\bra{u_m(\boldsymbol{k})}i\partial_{\boldsymbol{k}}\ket{u_n(\boldsymbol{k})}$ is Berry connection.

\subsubsection{The correction of the density matrix in the Bloch representation}
The correction of the density matrix can be derived by the quantum Liouville equation \cite{ref19,ref22.1,ref22.3,ref23}
\begin{equation}
	%\begin{split}
	\partial_t\hat{\rho}(t)
	=\frac{1}{i\hbar}\left[\hat{H}_0+\hat{H}^{\prime}(t),\hat{\rho}(t)\right]
	+\Gamma\left[\hat{\rho}(t)-\hat{\rho}^{(0)}\right].
	\label{QLE}
	%\end{split}
\end{equation}
The magnon transport can be classified into intrinsic transport and extrinsic transport.
The extrinsic part is derived from the scattering of magnons with phonons and other magnons.
Therefore, we use the relaxation time method to describe the scattering effect of magnons.
Here, $\Gamma=\frac{1}{\tau}$, where $\tau$ is the relaxation time of magnon, which describes the average free movement time of magnons in the material \cite{ref10,ref11,ref23}.
A more detailed derivation and explanation of Eq.\ref{QLE} can be found in Appendix \ref{Magnon scattering effect}.
We assume that the time-dependent electric field  starts at $t=0$, and the effect of scattering on magnons is not considered when $t<0$. 
$\rho^{(0)}=\rho(t=0)$ is the equilibrium density matrix, which satisfies $\partial_t\hat{\rho}^{(0)}=0$. 

By solving the quantum Liouville equation, in the interaction picture, the first-order density matrix is 
\begin{equation}
	%\begin{split}
	\hat{\rho}_I^{(1)}(t)
	=\frac{e^{-\Gamma t}}{i\hbar}\int_0^tdt^{\prime}\left[\hat{H}_{I}^{\prime}(t^{\prime}),e^{\Gamma t^{\prime}}\hat{\rho}_I^{(0)}\right].
	%\end{split}
	\label{rhoI(1)}
\end{equation}
Here, although there is a damping factor $e^{-\Gamma t}$ in Eq. (\ref{rhoI(1)}), at the same time, there is also a divergent factor $e^{\Gamma t^{\prime}}$. Therefore, the damping factor can not make the density matrix turn to zero in the long-term limit. (for details, see Appendix \ref{Calculation details of the density matrix}.)

In Bloch representation, for the zero-order magnon density matrix,   $\rho_{mn}^{(0)}(\boldsymbol{k})=f_n^B(\boldsymbol{k})\delta_{mn}$, in which $f_n^B(\boldsymbol{k})=1/\left[e^{\varepsilon_n(\boldsymbol{k})/k_BT}-1\right]$ is Bose distribution. 
The first-order magnon density matrix can be expressed as  $\rho_{mn}^{(1)}(\boldsymbol{k},t)=\rho_{mn}^{(1)i}(\boldsymbol{k},t)+\rho_{mn}^{(1)e}(\boldsymbol{k},t)$ in Bloch representation, in which $\rho_{mn}^{(1)i}(\boldsymbol{k},t)$ is the intraband part of the first-order density matrix, and $\rho_{mn}^{(1)e}(\boldsymbol{k},t)$ is the interband part of the first-order density matrix. 
The expression of $\rho_{mn}^{(1)i}(\boldsymbol{k},t)$ and $\rho_{mn}^{(1)e}(\boldsymbol{k},t)$ is Eq. \ref{rho1ie} in Appendix \ref{Calculation details of the density matrix}.

\subsection{The zero-order MS(E)PC}
We substitute the zero-order density matrix into Eq. \ref{j0main} and Eq. \ref{jE0main} to obtain the zero-order magnon spin current (for details, see Appendix \ref{j(0)} and Appendix \ref{Zero-order magnon energy photocurrent})
\begin{equation}
	%\begin{split}
		\boldsymbol{j}^{(0)}
		=\sum_{n}\int[dk]f_n^B(\boldsymbol{k})\partial_{\boldsymbol{k}}\varepsilon_n(\boldsymbol{k}),
	%\end{split}
\end{equation}
and the zero-order magnon energy current
\begin{equation}
	%\begin{split}
		\boldsymbol{j}^{E(0)}
		=\frac{1}{\hbar}\sum_{n}\int[dk]f_n^B(\boldsymbol{k})
		\varepsilon_n(\boldsymbol{k})\partial_{\boldsymbol{k}}\varepsilon_n(\boldsymbol{k}),
	%\end{split}
\end{equation}
in which $f_n^B(\boldsymbol{k})=1/\left[e^{\varepsilon(\boldsymbol{k})/k_BT}-1\right]$ is Bose distribution, and $\int[dk]=\frac{1}{(2\pi)^2}\int d^2k$.
When the band energy satisfies $\varepsilon_n(\boldsymbol{k})=\varepsilon_n(-\boldsymbol{k})$, $\boldsymbol{j}^{(0)}=0$ and $\boldsymbol{j}^{E(0)}=0$.

\subsection{\label{The first-order magnon transport induced by time-dependent electric field}The first-order MS(E)PC induced by time-dependent electric field}

\subsubsection{\label{MSPC} MSPC}

The first-order MSPC is $\boldsymbol{j}^{(1)}=\hbar\sum_{mn}\rho_{nm}^{(1)}(\boldsymbol{k},t)\boldsymbol{v}_{mn}(\boldsymbol{k})$, which can be divided in to oscillating term $\boldsymbol{j}_O^{(1)}$ and damping term $\boldsymbol{j}_D^{(1)}$. 
These two parts can be expressed as 
\begin{equation}
	\begin{split}
		j_{O,a}^{(1)}=\sum_{ib}\left[\chi_{O,ab}^i(\omega_i)+\chi_{O,ab}^e(\omega_i)\right]E_b(\omega_i)e^{-i\omega_it},
	\end{split}
\end{equation}
and
\begin{equation}
	%\begin{split}
		j_{D,a}^{(1)}=\sum_{ib}\left[\chi_{D,ab}^i(\omega_i)+\chi_{D,ab}^e(\omega_i)\right]E_b(\omega_i)e^{-\Gamma t}.
	%\end{split}
	\label{jDa(1)main}
\end{equation}
Here, the corresponding magnon spin photoconductivity is 
{\small
	\begin{equation}
		\begin{cases}
			\chi_{O,ab}^i(\omega)=
			\nu\sum_{n,c}\int[dk]\epsilon_{bcz}\frac{\omega}{\omega+i\Gamma}\frac{\partial f_n(\boldsymbol{k})}{\partial k_c}\frac{\partial \omega_n(\boldsymbol{k})}{\partial k_a}\\
			\chi_{O,ab}^e(\omega)=\nu
			\sum_{n\ne m,c}\int[dk]\epsilon_{bcz}\omega
			\frac{\mathcal{A}_{a,mn}(\boldsymbol{k})\mathcal{A}_{c,nm}(\boldsymbol{k})}{\omega-\omega_{nm}(\boldsymbol{k})+i\Gamma}f_{nm}(\boldsymbol{k})\omega_{nm}(\boldsymbol{k}) \\
			\chi_{D,ab}^i(\omega)
			=-\chi_{O,ab}^i(\omega)\\
			\chi_{D,ab}^e(\omega,t)
			=-\chi_{O,ab}^e(\omega)e^{-i\omega_{nm}(\boldsymbol{k})t},\\
		\end{cases}
		\label{chi(1)main}
	\end{equation}
}
and $E_b$ is the complex amplitude in the direction of $b$.
Here, $\omega$ is optical frequency, $\nu=\frac{g_J\mu_B}{c_{lv}^2}$, $\mathcal{A}_{a,mn}(\boldsymbol{k})$ is the interband Berry connection of band $m$ and $n$ in the direction of $a$, $\omega_n(\boldsymbol{k})=\varepsilon_n(\boldsymbol{k})/\hbar$ and $\omega_{nm}(\boldsymbol{k})=\omega_n(\boldsymbol{k})-\omega_m(\boldsymbol{k})$.
$a$, $b$, $c$ mean the direction of Cartesian coordinates.
According to Eq. \ref{jDa(1)main}, $\boldsymbol{j}_D^{(1)}$ decreases exponentially with time. When the relaxation time $\tau$ is short enough ($\Gamma>>\omega_{gap}$) or the time evolution is long enough, we can ignore the damping part $\boldsymbol{j}_D^{(1)}$ \cite{ref23}. 
In this work,  we only consider the MSPC after the system is stabilized, so $\boldsymbol{j}_D^{(1)}$ is not considered.
Therefore, we omit the $O$ and $D$ marks in the subscript of MSPC and magnon spin photoconductivity.
From here, we can see that the magnon spin photoconductivity is related to the optical frequency, the relaxation time, and the reciprocal space structure of the material.

When $\Gamma\to 0$ and $\omega<<\omega_{gap}$, $\omega\omega_{nm}/(\omega-\omega_{nm}+i\Gamma)\approx-\omega$, so
\begin{equation}
	\begin{split}
		\chi_{ab}^e(\omega)
		&\approx-\nu\sum_{n\ne m,c}\int[dk]\epsilon_{bcz}\omega
		\mathcal{A}_{a,mn}(\boldsymbol{k})\mathcal{A}_{c,nm}(\boldsymbol{k})f_{nm}(\boldsymbol{k})\\
		&=-i\nu\omega\sum_{c}\epsilon_{bcz}
		\sum_{n}\int[dk]\Omega_n^{ac}(\boldsymbol{k})f_{n}^B(\boldsymbol{k}),\\
	\end{split}
	\label{chiOeab}
\end{equation}
in which $\Omega_n^{ac}(\boldsymbol{k})$ is the Berry curvature (for details, see Appendix \ref{The expressions of magnon spin photocurrent}).
Here, Berry curvature represents the geometric properties of the magnon reciprocal space and serves as the source of magnon intrinsic topological transport. 
Now $\chi_{xy}^e(\omega)=\chi_{yx}^e(\omega)=0$, and $\chi_{xx}^e(\omega)=-i\nu\omega
\sum_{n}\int[dk]\Omega_n^{xy}(\boldsymbol{k})f_{n}^B(\boldsymbol{k})$.
Formally, Eq. \ref{chiOeab} is very similar to the Hall conductivity of electrons
\begin{equation}
	\sigma_{ab}=-\frac{e}{\hbar}\sum_n\int [dk]f_n^F(\boldsymbol{k})\Omega_n^{ab}(\boldsymbol{k}).
\end{equation}
There are two main differences between the magnon spin photoconductivity $\chi_{ab}^e(\omega)$ and the Hall conductivity of electrons $\sigma_{ab}$. The first difference is that the Hall conductivity is a transverse transport. However, because the effective electric field $\tilde{\boldsymbol{E}}(t)=-\frac{1}{c_{lv}}\partial_t\boldsymbol{E}(t)\times\boldsymbol{e}_z$ is perpendicular to the actual electric field, the topological MSPC is a longitudinal transport.
The second difference is that electrons obey Fermi-Dirac statistics. At low temperatures, electrons can be regarded as filling all the energy levels below the Fermi level, thus having perfect topological properties. That is to say, the electron Hall conductivity can be directly expressed in terms of the Chern number at low temperatures \cite{ref0.5}
\begin{equation}
	\begin{split}
	\sigma_{ab}
	&=-\frac{e}{\hbar}\sum_n\int [dk]f_n^F(\boldsymbol{k})\Omega_n^{ab}(\boldsymbol{k})\\
	&\approx -\frac{e}{\hbar}\sum_n\int [dk]\Omega_n^{ab}(\boldsymbol{k})\\
	&=-\frac{e}{2\pi \hbar}\sum_nC_n.
	\end{split}
\end{equation}
However, the magnon is a boson and obey Bose-Einstein statistics. The weighting effect of the Bose distribution function $f_n^B(\boldsymbol{k})$ on the magnon spin photoconductivity cannot be completely eliminated. So different with the conventional topological conductivity, the magnon spin photoconductivity cannot be directly expressed in terms of the Chern number. We can only control the direction of the magnon spin photocurrent through topological properties.
We will discuss this problem in Sec. \ref{sec:Model calculation}. 
The result of the magnon thermal Hall effect by R. Matsumoto and S. Murakami \cite{ref5} is structurally similar to the result of the magnon spin photocurrent we obtained at the low-frequency limit. The difference is mainly that in the results of R. Matsumoto and S. Murakami, the weight function before the Berry curvature is $c_2(\rho_n^B)$; while in our results (MSPC), the weight function before the Berry curvature is the Bose-Einstein distribution.

In addition, different from the Hall effect of electrons, according to Eq. \ref{Hprimemain}, the $\omega_i$ can not be zero, otherwise the perturbation term is zero. 
So the effect of the topological properties on MSPC is approximate.

\subsubsection{\label{MEPC}MEPC}

Similar to the discussion of MSPC, the MEPC induced by time-dependent electric field is $\boldsymbol{j}^{E(1)}=\sum_{mn}\rho_{nm}^{(1)}(\boldsymbol{k},t)\boldsymbol{v}_{mn}^E(\boldsymbol{k})$. 
MEPC can be expressed as $\boldsymbol{j}^{E(1)}=\boldsymbol{j}_O^{E(1)}+\boldsymbol{j}_D^{E(1)}$, in which
\begin{equation}
	\begin{split}
		j_{O,a}^{E(1)}=\sum_{ib}\left[\chi_{O,ab}^{E,i}(\omega_i)+\chi_{O,ab}^{E,e}(\omega_i)\right]E_b(\omega_i)e^{-i\omega_it}.
	\end{split}
\end{equation}
and
\begin{equation}
	\begin{split}
		j_{D,a}^{E(1)}=\sum_{ib}\left[\chi_{D,ab}^{E,i}(\omega_i)+\chi_{D,ab}^{E,e}(\omega_i)\right]E_b(\omega_i)e^{-\Gamma t}.
	\end{split}
\end{equation}
The corresponding magnon energy photoconductivity can be expressed as
\begin{widetext}
\begin{equation}
	\begin{cases}
		\chi_{O,ab}^{E,i}(\omega_i)
		=\nu\sum_{nc}\int[dk] \frac{\omega\epsilon_{bcz}}{\omega+i\Gamma} \omega_n(\boldsymbol{k})\frac{\partial f_n(\boldsymbol{k})}{\partial k_c}\frac{\partial\omega_n(\boldsymbol{k})}{\partial k_a}\\
		\chi_{O,ab}^{E,e}(\omega)
		=\frac{\nu}{2}\sum_{n\ne m,c}\int[dk] \left[\frac{\omega\epsilon_{bcz}}{\omega-\omega_{nm}(\boldsymbol{k})+i\Gamma}\mathcal{A}_{a,mn}(\boldsymbol{k})\mathcal{A}_{c,nm}(\boldsymbol{k})f_{nm}(\boldsymbol{k})\left[\omega_n^2(\boldsymbol{k})-\omega_m^2(\boldsymbol{k})\right]\right]\\
		\chi_{D,ab}^{E,i}(\omega)
		=-\chi_{O,ab}^{E,i}(\omega)\\
		\chi_{D,ab}^{E,e}(\omega,t)
		=-\chi_{O,ab}^{E,e}(\omega)e^{-i\omega_{nm}(\boldsymbol{k})t}.\\
	\end{cases}
	\label{chi(1)Emain}
\end{equation}
\end{widetext}
Therefore, $\boldsymbol{j}_D^{E(1)}$ also decreases exponentially with time. When the relaxation time $\tau$ is short enough ($\Gamma>>\omega_{gap}$) or the time evolution is long enough, we can ignore the damping part $\boldsymbol{j}_D^{E(1)}$ \cite{ref23}.
So we also omit the subscripts $O$ and $D$.
When $\Gamma\to 0$ and $\omega<<\omega_{gap}$, 
the magnon energy photoconductivity can be expressed as (for details, see Appendix \ref{The expression of magnon energy photocurrent})
\begin{equation}
	\begin{split}
		\chi_{ab}^{E,e}(\omega)
		&\approx\frac{\nu}{2}\omega\sum_{n\ne m,c}\int[dk] \left\{\epsilon_{cbz}\mathcal{A}_{a,mn}(\boldsymbol{k})\mathcal{A}_{c,nm}(\boldsymbol{k})\right.\\
		&\left.\times f_{nm}(\boldsymbol{k})\left[\omega_n(\boldsymbol{k})+\omega_m(\boldsymbol{k})\right]\right\}.\\
		\label{chiEOeab}
	\end{split}
\end{equation}
Different from magnon spin photoconductivity, since the energy velocity involves the anticommutation of velocity and the Hamiltonian $\hat{\boldsymbol{v}}^E=\frac{1}{2}\left[\hat{v},\hat{H}_0\right]_+$, the expression of magnon energy photoconductivity is related to the inter-band relationship (the average value of the two bands $\frac{1}{2}\left[\omega_n(\boldsymbol{k})+\omega_m(\boldsymbol{k})\right]$), so it can not be written in the form related to Berry curvature directly.
Therefore, in most cases, it is difficult to discuss the control of MEPC through the topological properties directly.
However, when the average value of the bands is a constant, the expression of magnon energy photoconductivity in the low-frequency limit is very similar to that of magnon spin photoconductivity, 
{\small
\begin{equation}
	\begin{split}
		\chi_{ab}^{E,e}(\omega)
		&\approx\mu\omega_0\omega\sum_{n\ne m,c}\int[dk] \left\{\epsilon_{cbz}\mathcal{A}_{a,mn}(\boldsymbol{k})\mathcal{A}_{c,nm}(\boldsymbol{k})f_{nm}(\boldsymbol{k})\right\}\\
		&=-i\mu\omega\omega_0\sum_{c}\epsilon_{bcz}\sum_n\int[dk]\Omega_n^{ac}(\boldsymbol{k})f_n^B(\boldsymbol{k})\\
		&=\omega_0\chi_{ab}^e(\omega)
		\label{chiEOeabomega0}
	\end{split}
\end{equation}
}
differing only by an energy constant $\omega_0=\frac{1}{2}\left[\omega_n(\boldsymbol{k})+\omega_m(\boldsymbol{k})\right]$. 
At this point, we can use the topological properties of the system to study MEPC.

In conclusion, the AC phase provides the geometric interaction between the electric field and magnons, thereby driving the magnon transport.
The magnon transport induced by light can be classified into intrinsic part and extrinsic part. 
Here, The intrinsic part of magnon transport originates from the geometric properties of the magnon reciprocal space.
The external part of magnon transport originates from the scattering effect of magnons. The scattering effect can be described by the relaxation time method.

%Different from the magnon spin photoconductivity, the magnon energy photoconductivity can not be written in the form related to Berry curvature. 
%
%So it is difficult to discuss the control of MEPC through the topological properties directly.

\subsubsection{Phenomenological representation of MS(E)PC}\label{Phenomenological representation of magnon photo-transport}
Then we consider the monochromatic light.
For the magnon spin photoconductivity and magnon energy photoconductivity, we can define $\chi_{ab}^{(E)}(\omega)=\chi_{ab}^{(E)i}(\omega)+\chi_{ab}^{(E)e}(\omega)$, so
%{\footnotesize
\begin{equation}
	\begin{split}
		j_{a}^{(E)(1)}(\omega)
		&=\sum_{b}\left\{Re\left[\chi_{ab}^{(E)}(\omega)\right]\cos(-\omega t+\phi_b)\right.\\
		&\left.-Im\left[\chi_{ab}^{(E)}(\omega)\right]\sin(-\omega t+\phi_b)\right\}E_{0,b}.\\
	\end{split}
\end{equation}
Here, $E_{0,b}$ is the amplitude of the electric field, $\omega$ is the optical frequency, and $\phi_b$ is the initial phase of the direction of $b$, in which $b$ is the direction of Cartesian coordinates (for details, see Appendix \ref{appendixPhenomenological}). 
What we're talking about here is a phenomenological representation of MS(E)PC, where the $(E)$ in superscript means that we're thinking about either MSPC or MEPC.
Here, we assume that $\phi_x=0$ and $\phi_y-\phi_x=\phi_y=\delta$, so the MS(E)PC can be expressed as
%{\small
	\begin{equation}
		\begin{split}
			j_{a}^{(E)(1)}(\omega)
			&=\left\{Re\left[\chi_{ax}^{(E)}(\omega)\right]\cos(-\omega t)\right.\\
			&\left.-Im\left(\chi_{ax}^{(E)}(\omega)\right)\sin(-\omega t)\right\}E_{0,x}\\
			&+\left\{Re\left[\chi_{ay}^{(E)}(\omega)\right]\cos(-\omega t+\delta)\right.\\
			&\left.-Im\left(\chi_{ay}^{(E)}(\omega)\right)\sin(-\omega t+\delta)\right\}E_{0,y}.\\
			\label{j(E)ellipsemain}
		\end{split}
	\end{equation}
%}

For monochromatic linearly polarized (LP) light, we can take $\phi_x=\phi_y=\delta = 0$. 
Then the MS(E)PC can be expressed as
%{\small
\begin{equation}
	\begin{split}
		j_{a}^{LP(E)(1)}(\omega)
		&=\sum_{b}\left\{Re\left[\chi_{O,ab}^{(E)}(\omega)\right]\cos(\omega t)\right.\\
		&\left.+Im\left[\chi_{O,ab}^{(E)}(\omega)\right]\sin(\omega t)\right\}E_{0,b}.\\
		\label{jaLP}
	\end{split}
\end{equation}
%}
%
%
For monochromatic circularly polarized (CP) light, we can take $E_{0,x}=E_{0,y}=E_0$, $\phi_x=0$ and  $\delta=\pm\frac{\pi}{2}$. 
When the light is left-handed circularly polarized, $\delta=\frac{\pi}{2}$; when the light right-handed circularly polarized, $\delta=-\frac{\pi}{2}$.
The MS(E)PC of circularly polarized light $j_a^{CP(E)(1)}(\omega)$ is the difference between MS(E)PC of left-handed polarized light and the MS(E)PC of right-handed polarized light,  $j_a^{CP(E)(1)}(\omega)=j_a^{LCP(E)(1)}(\omega)-j_a^{RCP(E)(1)}(\omega)$ \cite{ref23}
%{\footnotesize
	\begin{equation}
		\begin{split}
			j_{a}^{CP(E)(1)}(\omega)
			&=2\left\{ Re\left[\chi_{O,ay}^{(E)}(\omega)\right]\sin(\omega t)\right.\\ &\left.-Im\left[\chi_{O,ay}^{(E)}(\omega)\right]\cos(\omega t)\right\}E_{0}.\\
		\end{split}
	\end{equation}
%}

\section{\label{sec:Effective time-reversal symmetry} Effective time-reversal symmetry $\hat{\mathcal{T}}^{\prime}$}\label{Tprime}
The effective time-reversal symmetry (ETRS) $\hat{\mathcal{T}}^{\prime}$ 
is the operator of reversing time and the direction of spin \cite{ref9.3,ref25,ref26}.  
A general Hamiltonian is transformed with the ETRS operator as \cite{ref25}
\begin{eqnarray}
	%\begin{split}
	\hat{\mathcal{T}}^{\prime}\mathcal{H}(\boldsymbol{k})\left(\hat{\mathcal{T}}^{\prime}\right)^{-1}
	=\mathcal{H}(-\boldsymbol{k})
	=\mathcal{H}^{\ast}(\boldsymbol{k}).
	\notag
\end{eqnarray}
So
\begin{eqnarray}
	%\begin{split}
	U^{\dag}(\boldsymbol{k})\hat{\mathcal{H}}(\boldsymbol{k})U(\boldsymbol{k})
	=U^{\dag}(\boldsymbol{k})\hat{\mathcal{H}}^{\ast}(-\boldsymbol{k})U(\boldsymbol{k})
	=\varepsilon(\boldsymbol{k}),
	\notag
\end{eqnarray}
and
{\small
\begin{eqnarray}
	%\begin{split}
	U^{T}(-\boldsymbol{k})\hat{\mathcal{H}}^{\ast}(-\boldsymbol{k})U^{\ast}(-\boldsymbol{k})
	=\left[U^{\ast}(-\boldsymbol{k})\right]^{\dag}\hat{\mathcal{H}}(\boldsymbol{k})U^{\ast}(-\boldsymbol{k})
	=\varepsilon(-\boldsymbol{k}).
	\notag
\end{eqnarray}
}
Under the condition of  $\varepsilon(\boldsymbol{k})=\varepsilon(-\boldsymbol{k})$, we can see that $U^{\dag}(\boldsymbol{k})\hat{\mathcal{H}}(\boldsymbol{k})U(\boldsymbol{k})=\left[U^{\ast}(-\boldsymbol{k})\right]^{\dag}\hat{\mathcal{H}}(\boldsymbol{k})U^{\ast}(-\boldsymbol{k})=\varepsilon(\boldsymbol{k})$.
So we can replace $U(\boldsymbol{k})$ by $U^{\ast}(-\boldsymbol{k})$ (only one phase factor apart).
Therefore, under the ETRS, the Berry connection satisfies 
\begin{eqnarray}
	\mathcal{A}_{a,mn}(\boldsymbol{k})
	&=& \sum_pU_{mp}^{\dag}(\boldsymbol{k})i\partial_{k_a}U_{pn}(\boldsymbol{k})\\
	\notag 
	&=& \sum_pU_{pm}(-\boldsymbol{k})i\partial_{k_a}U_{pn}^{\ast}(-\boldsymbol{k})\\
	\notag 
	&=& \sum_pU_{np}^{\dag}(-\boldsymbol{k})i\partial_{-k_a}U_{pm}(-\boldsymbol{k})\\
	\notag
	&=& \mathcal{A}_{a,nm}(-\boldsymbol{k}).
	\label{ATprime}
\end{eqnarray} 
In particular, when $m=n$, intraband Berry connection satisfies $\mathcal{A}_{a,n}(\boldsymbol{k})=\mathcal{A}_{a,n}(-\boldsymbol{k})$. 
So Berry curvature $\boldsymbol{\Omega}_n(\boldsymbol{k})=\boldsymbol{\nabla}_{\boldsymbol{k}}\times\boldsymbol{\mathcal{A}}_n(\boldsymbol{k})$ satisfies $\boldsymbol{\Omega}_n(\boldsymbol{k})=-\boldsymbol{\Omega}_n(-\boldsymbol{k})$. 
Therefore, under the ETRS, the longitudinal magnon spin photoconductivity in Eq. \ref{chiOeab} is equal to zero.
And under the ETRS, 
\begin{widetext}
\begin{equation}
	\begin{split}
	\chi_{ab}^{E,e}(\omega)
	&=\frac{\nu}{2}\omega\sum_{n\ne m,c}\int [dk]\epsilon_{cbz}\mathcal{A}_{a,mn}(\boldsymbol{k})\mathcal{A}_{c,nm}(\boldsymbol{k})f_{nm}(\boldsymbol{k})\left[\omega_n(\boldsymbol{k})+\omega_m(\boldsymbol{k})\right]\\
	&=\frac{\nu}{2}\omega\sum_{n\ne m,c}\int [dk]\epsilon_{cbz}\mathcal{A}_{a,nm}(-\boldsymbol{k})\mathcal{A}_{c,mn}(-\boldsymbol{k})f_{nm}(-\boldsymbol{k})\left[\omega_n(-\boldsymbol{k})+\omega_m(-\boldsymbol{k})\right]\\
	&=-\frac{\nu}{2}\omega\sum_{n\ne m,c}\int [dk]\epsilon_{cbz}\mathcal{A}_{a,nm}(\boldsymbol{k})\mathcal{A}_{c,mn}(\boldsymbol{k})f_{mn}(\boldsymbol{k})\left[\omega_n(\boldsymbol{k})+\omega_m(\boldsymbol{k})\right]\\
	&=-\frac{\nu}{2}\omega\sum_{n\ne m,c}\int [dk]\epsilon_{cbz}\mathcal{A}_{a,mn}(\boldsymbol{k})\mathcal{A}_{c,nm}(\boldsymbol{k})f_{nm}(\boldsymbol{k})\left[\omega_m(\boldsymbol{k})+\omega_n(\boldsymbol{k})\right]\\
	&=-\chi_{ab}^{E,e}(\omega).
	\end{split}
\end{equation} 
So Eq. \ref{chiEOeabomega0} is also equal to zero.
\end{widetext}
In order to study the control of the longitudinal MS(E)PC, we need to break the ETRS of the system.

\section{\label{sec:Model calculation} Model calculation}
\begin{figure}[tb]
	\centering
	\centering
	% 适当减小图片scale值，比如从0.27改为0.24
	\subfloat[Ferromagnetic Kagome lattice]{
		\includegraphics[scale=0.4]{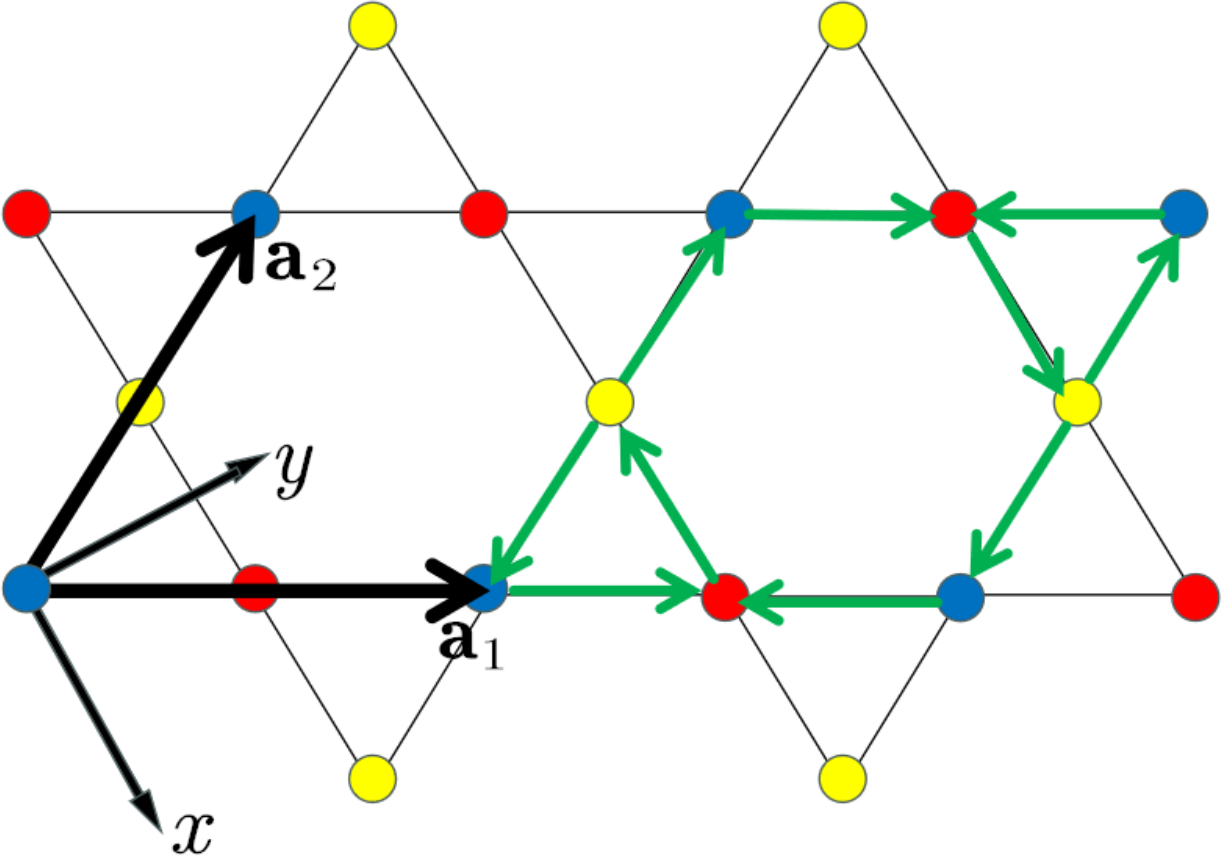}}
	\hfill % 使用\hfill让图片水平排列，自动调整间距
	\subfloat[Ferromagnetic Hexagonal lattice]{
		\includegraphics[scale=0.5]{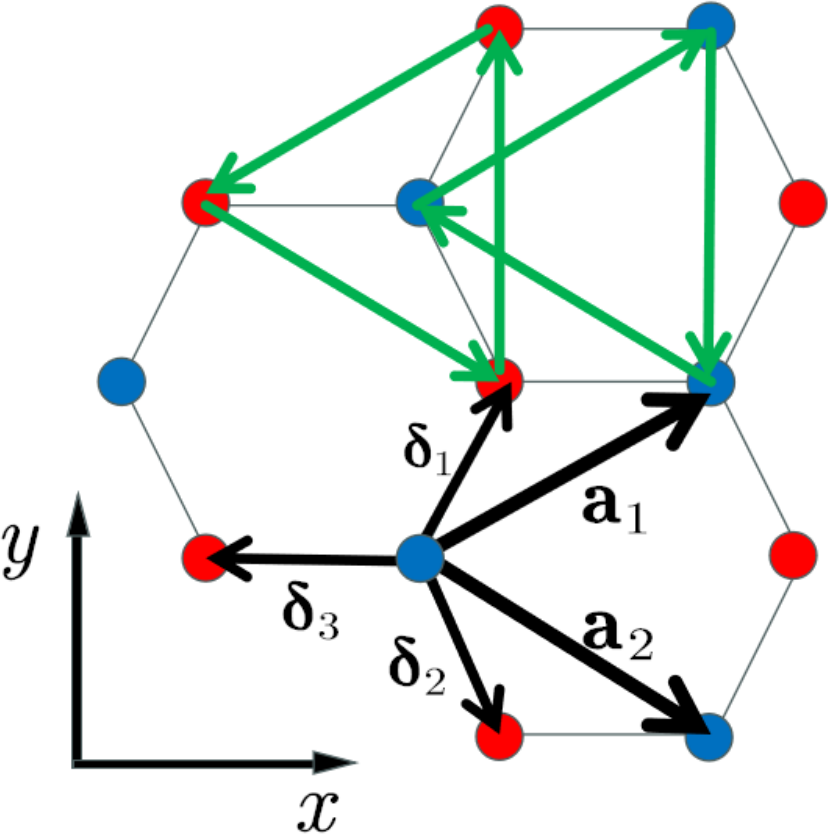}}
	\hfill
	\caption{Ferromagnetic Kagome (a) and Hexagonal (b) lattice in real space.
		$\mathbf{a}_1$ and $\mathbf{a}_2$ are basis vectors in real space. The DM vectors $\mathbf{D}_{ij}=D\nu_{ij}\mathbf{e}_z$ are parallel to the z axis, in which $\mathbf{e}_z$ represents the unit vector pointing in the positive direction of the z axis. As shown in the figure, $\nu_{ij}=1$ along the orange arrows.
		Here, $i$ and $j$ represent the nearest lattice points in (a) and represent the next-nearest neighbor lattice point in (b).}
	\label{fig:lattice}
\end{figure}

In this section, we take a model calculation of the MS(E)PC in  two-dimensional collinear ferromagnets.
As shown in Fig. \ref{fig:lattice}, the models we calculate are a two-dimensional Hexagonal lattice\cite{ref9.1,ref9.2} and a Kagome lattice \cite{ref3,ref9,ref25}.
Here, $\mathbf{a}_1$ and $\mathbf{a}_2$ are the basis vectors for the real space lattice. 
The Hamiltonian of models can be expressed as
\begin{eqnarray}
	%\begin{split}
	\hat{H} &=& -J\sum_{\langle ij\rangle}\hat{\boldsymbol{S}}_i\cdot\hat{\boldsymbol{S}}_j+\sum_{\ll ij\gg }\boldsymbol{D}_{ij}\cdot\left(\hat{\boldsymbol{S}}_i\times\hat{\boldsymbol{S}}_j\right) \notag\\
	& + & g_{J}\mu_B\sum_i\hat{\boldsymbol{S}}_i\cdot\boldsymbol{B}.
	\label{model-study}
	%\end{split}
\end{eqnarray}
The first term is the Heisenberg interaction, the second term is the Dzyaloshinskii-Moriya (DM) interaction, and the third term is the Zeeman interaction. 
Because the perturbation part and the scattering part are considered in the first order density matrix (Eq.(\ref{rho1ie})), the Hamiltonian do not include the perturbation part and the scattering part in Eq. (\ref{model-study}). See details in Appendix \ref{Quantum Liouville equation} and Appendix \ref{First-order correction of the density matrix}.

\subsection{Model calculation of MS(E)PC with different optical frequencies and different relaxation times}

Firstly, we take a model calculation in the Kagome lattice with different optical frequencies and different relaxation times. 
%
%%%%%%%%%%%%%%%%%%%%%%%%%%%%%%%%%%%%%%%%%%%%%%%%%%%%%%%%%%%%%%%%%%%%%%%%%%%%%%%%%%%%%%%%%%%%%%%
%%%%%%%%%%%%%%%%%%%%%%%%%%%%%%%%%%%%%%%%%%%%%%%%%%%%%%%%%%%%%%%%%%%%%%%%%%%%%%%%%%%%%%%%%%%%%%%
After performing the Holstein-Primakoff (H-P) transformation and the  Fourier transformation on Eq. \ref{model-study}, the Hamiltonian can be expressed as
\begin{equation}
	%\begin{split}
	\hat{H}
	=\sum_{\mathbf{k}}\hat{\Phi}^{\dag}(\mathbf{k})\mathcal{H}(\mathbf{k})\hat{\Phi}(\mathbf{k}). 
	%\end{split}
\end{equation}
Here, $\hat{\Phi}(\mathbf{k})=(\hat{a}_{\mathbf{k}},\hat{b}_{\mathbf{k}}, \hat{c}_{\mathbf{k}})^{T}$, in which $\hat{a}_{\boldsymbol{k}}$, $\hat{b}_{\boldsymbol{k}}$, and $\hat{c}_{\boldsymbol{k}}$ are the magnon annihilation operators of sublattice sites in reciprocal space, 
%\begin{equation}
%	 \hat{\Phi}(\mathbf{k})=(\hat{a}_{\mathbf{k}},\hat{b}_{\mathbf{k}},\hat{c}_{\mathbf{k}})^{T}\notag \\
%\end{equation}
and
\begin{equation}
	\begin{split}
		\mathcal{H}(\boldsymbol{k})
		&=
		\begin{pmatrix}
			H_{11}(\boldsymbol{k}) & H_{12}(\boldsymbol{k}) & H_{13}(\boldsymbol{k})\\
			H_{12}^{\ast}(\boldsymbol{k}) & H_{22}(\boldsymbol{k}) & H_{23}(\boldsymbol{k})\\
			H_{13}^{\ast}(\boldsymbol{k}) & H_{23}^{\ast}(\boldsymbol{k}) & H_{33}(\boldsymbol{k})\\
		\end{pmatrix},
		\label{Hkagome}
	\end{split}
\end{equation}
in which $H_{11}(\boldsymbol{k})=H_{22}(\boldsymbol{k})=H_{33}(\boldsymbol{k})=4JS-g_J\mu_BB^z$, $H_{12}(\boldsymbol{k})=-2(J+iD)S\cos(\boldsymbol{k}\cdot\boldsymbol{\delta}_1)$, $H_{13}(\boldsymbol{k})=-2(J-iD)S\cos(\boldsymbol{k}\cdot\boldsymbol{\delta}_2)$ and $H_{23}(\boldsymbol{k})=-2(J+iD)S\cos(\boldsymbol{k}\cdot(\boldsymbol{\delta}_2-\boldsymbol{\delta}_1))$.
Here, $\boldsymbol{\delta}_1=\frac{1}{2}\boldsymbol{a}_1$, $\boldsymbol{\delta}_2=\frac{1}{2}\boldsymbol{a}_2$.

\begin{figure}[h]
	\centering
	\includegraphics[width=1\columnwidth]{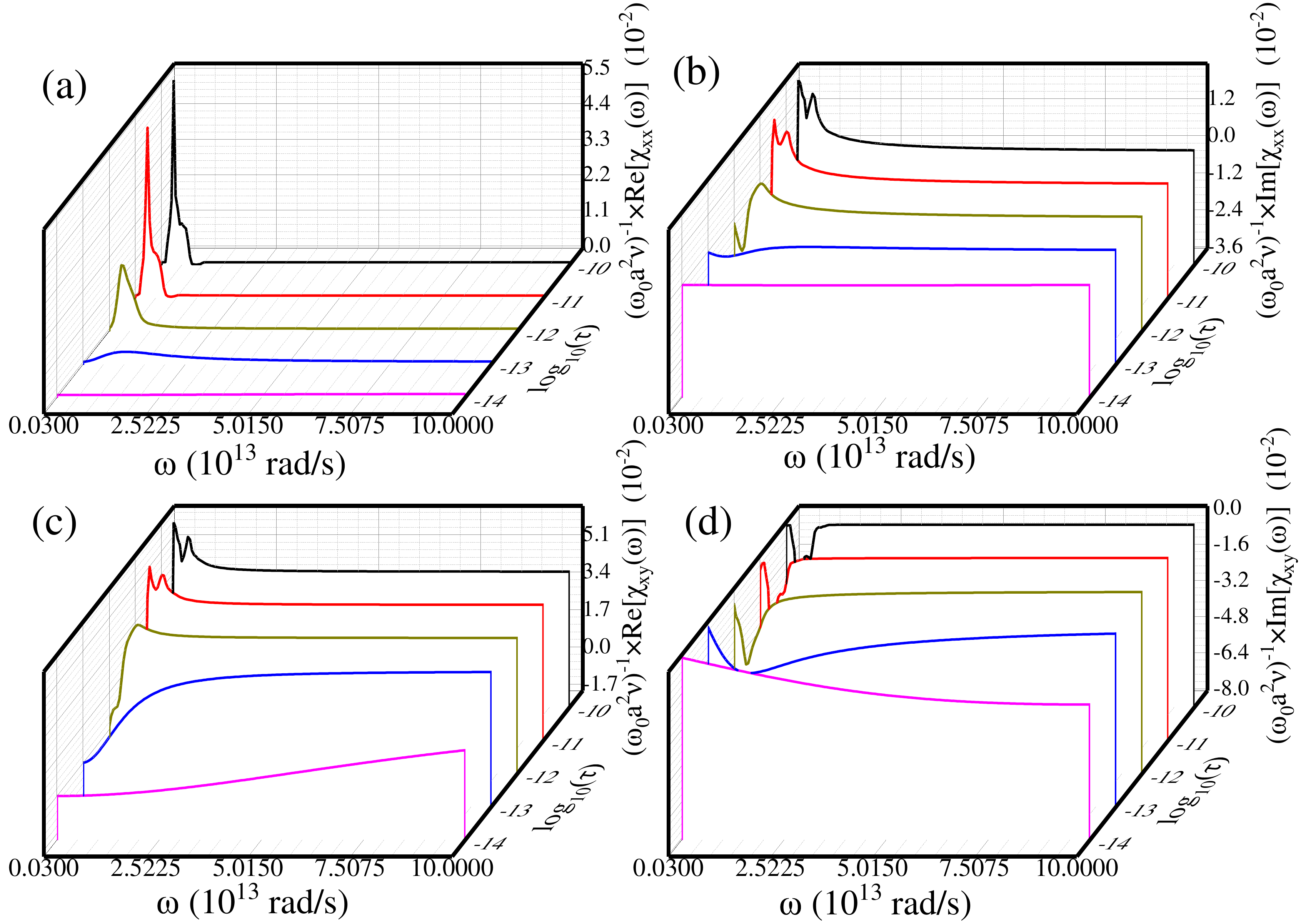}
	\caption{Waterfall plot of the change of the magnon spin photoconductivity with optical frequency under different relaxation times, where the optical frequency is in about infrared range.
	Here, we take $J=1$ meV, $D=0.32$ meV, $S=1$, $g_J\mu_BB^z=-3$ meV and $T = 20$ K.
    The waterfall plot describe the real part of longitudinal magnon spin photoconductivity (a), the imaginary part of longitudinal magnon spin photoconductivity (b), the real part of transverse magnon spin photoconductivity (c) and the imaginary part of transverse magnon spin photoconductivity (d).}
	\label{fig:Kagomechi}
\end{figure}

\begin{figure}[h]
	\centering
	\includegraphics[width=1\columnwidth]{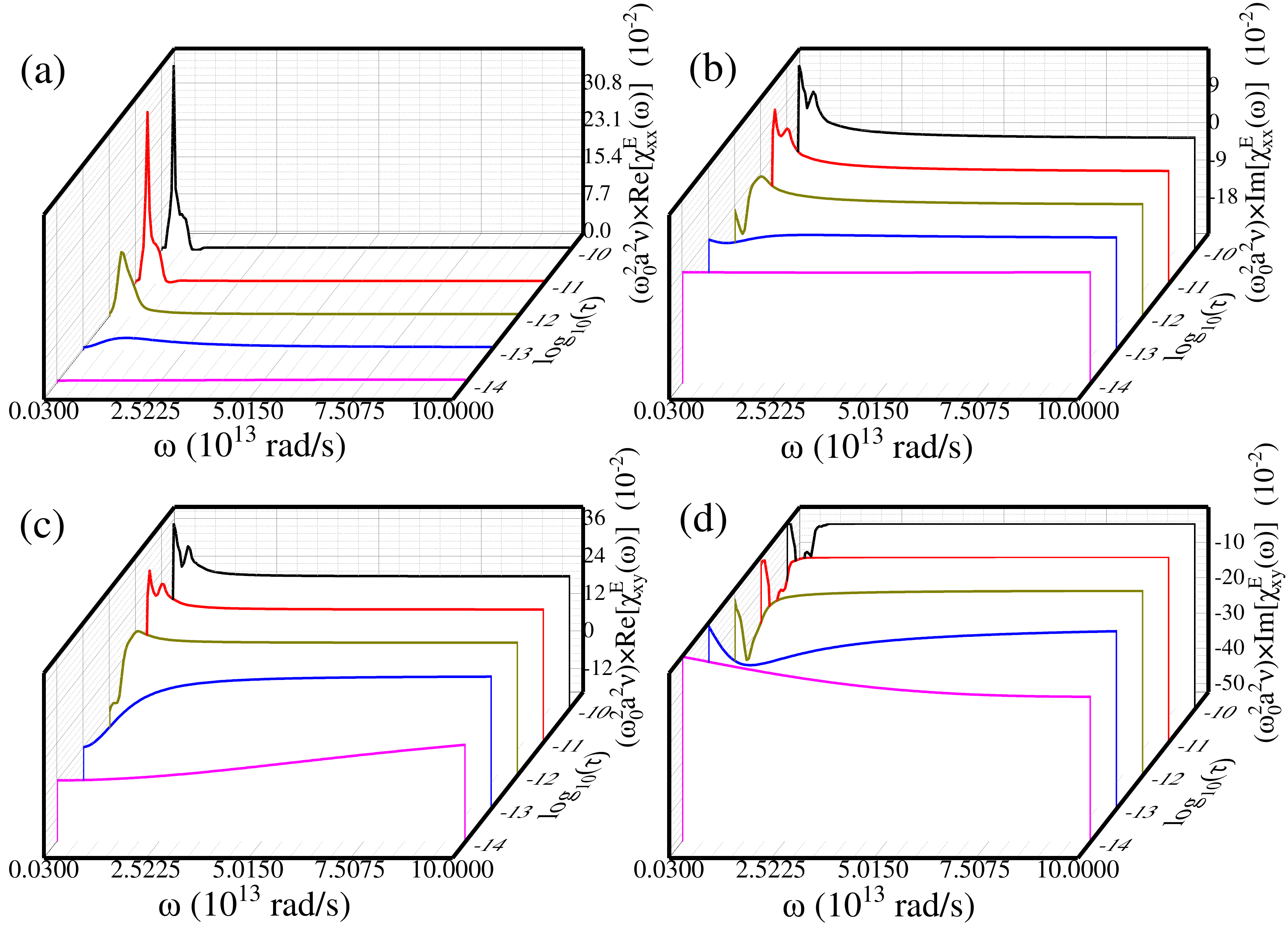}
	\caption{Waterfall plot of the change of the magnon energy photoconductivity with optical frequency under different relaxation times, where the optical frequency is in about infrared range. Here, we take $J=1$ meV, $D=0.32$ meV, $S=1$, $g_J\mu_BB^z=-3$ meV and $T = 20$ K. The waterfall plot describe the real part of longitudinal magnon energy photoconductivity (a), the imaginary part of longitudinal magnon energy photoconductivity (b), the real part of transverse magnon energy photoconductivity (c) and the imaginary part of transverse magnon energy photoconductivity (d).}
	\label{fig:KagomechiE}
	%\label{lattice}
\end{figure}

In Fig. \ref{fig:Kagomechi} and Fig. \ref{fig:KagomechiE}, we plot the change of magnon spin photoconductivity and magnon energy photoconductivity with optical frequency at different relaxation times, where the optical frequency is approximately in the infrared range. 
Here, the relaxation time $\tau$ exists as a parameter. The relaxation times are taken as $10^{-14}$s, $10^{-13}$s, $10^{-12}$s, $10^{-11}$s and $10^{-10}$s. The range roughly refers to the work of H. Varshney et al. \cite{ref9.1}, where the relaxation time is 1 picosecond.

As shown in Fig. \ref{fig:Kagomechi} (a), (b) and Fig.  \ref{fig:KagomechiE} (a), (b), when the relaxation time is $10^{-10}$ s, $10^{-11}$ s, $10^{-12}$ s and $10^{-13}$ s the longitudinal magnon spin (energy) photoconductivity $\chi_{xx}^{(E)}(\omega)$ (including the real part and the imaginary part), as the increase of optical frequency, increases rapidly at first, and decreases after it reaches a peak, and then changes slowly.
Here, for the longitudinal magnon spin (energy) photoconductivity, only interband part is nonzero, so the peaks of longitudinal magnon spin (energy) photoconductivity come from the minimal value of the denominator of $\chi_{O,xx}^{(E)e}(\omega)$ in Eq. (\ref{chi(1)main}) and Eq. (\ref{chi(1)Emain}). 
In Fig. \ref{figbandkagome} (b), we can observe that the interband energy difference is in the range of about 1.1 meV - 4.9 meV. In other words, the interband angular frequency difference $\omega_{nm}(\boldsymbol{k})=\varepsilon_{nm}(\boldsymbol{k})/\hbar$ is in the range about $0.17\times 10^{13} $ - $0.74\times 10^{13}$ rad/s. 
According to Eq. (\ref{chi(1)main}) and Eq. (\ref{chi(1)Emain}), when the optical frequency $\omega\approx\omega_{nm}(\boldsymbol{k})$, magnon spin (energy) photoconductivity reach the peak.
In addition, as $\tau$ increases (scattering effect decreases), the peak values of  $\chi_{O,xx}^{(E)e}(\omega)$ increase. 
When the relaxation time is infinite (without scattering effect), the magnon spin photoconductivity will diverge at the peak. It is consistent with physics because when the relaxation time is infinite, the excited magnons will not return to equilibrium.
By comparison, as shown in Fig. \ref{fig:Kagomechi} (c), (d) and Fig.  \ref{fig:KagomechiE} (c), (d), the peak is not obvious.
Because the transverse magnon spin (energy) photoconductivity $\chi_{O,xx}^{(E)}(\omega)$ depends not only on the contribution of the interband part, but also the intraband part.

%%%%%%%%%%%%%%%%%%%%%%%%%%%%%%%%%%%%%%%%%%%%%%%%%%%%%%%%%%%%%%%%%%%%%%%%%%%%%%%%%%%%%%%%%%%%%%%
%%%%%%%%%%%%%%%%%%%%%%%%%%%%%%%%%%%%%%%%%%%%%%%%%%%%%%%%%%%%%%%%%%%%%%%%%%%%%%%%%%%%%%%%%%%%%%%

So, in the frequency range that we calculate, we can get the maximum value of magnon spin (energy) photoconductivity in the range of roughly $0.17\times 10^{13} $ to $0.74\times 10^{13}$ rad/s.
According to Eq. \ref{j(E)ellipsemain}, in this range, we can expect to find the strongest MS(E)PC.

%%%%%%%%%%%%%%%%%%%%%%%%%%%%%%%%%%%%%%%%%%%%%%%%%%%%%%%%%%%%%%%%%%%%%%%%%%%%%%%%%%%%%%%%%%%%%%%
%%%%%%%%%%%%%%%%%%%%%%%%%%%%%%%%%%%%%%%%%%%%%%%%%%%%%%%%%%%%%%%%%%%%%%%%%%%%%%%%%%%%%%%%%%%%%%%

\subsection{\label{subsec:pure}The topological control of MS(E)PC}

When we can ignore the scattering of magnons in the material, 
%and the temperature is relatively low, 
we can assume that the relaxation time of magnons tends to infinity, that is, $\Gamma\to 0$.
As discussed in Subsec. \ref{The first-order magnon transport induced by time-dependent electric field},  
under the condition of low optical frequency and zero $\Gamma$, 
the magnon spin photoconductivity satisfies Eq. \ref{chiOeab}, 
and the magnon energy photoconductivity satisfies Eq. \ref{chiEOeab}.
Here, Eq. \ref{chiOeab} makes it possible to control the magnon spin photoconductivity through the topological properties of the magnon system, but Eq. \ref{chiEOeab} does not.
Next, to discuss the topological control of the MS(E)PC under the condition of low optical frequency and $\Gamma=0$. We take a model calculation of Eq. \ref{chiOeab} on the Hexagonal lattice and the Kagome lattice in this subsection. 

We take LP light, and we orient the direction of the LP light along the $x$-axis.
According to Eq. \ref{chi(1)main}, for the transverse MS(E)PC ($j_{y}^{(1)}$ and $j_{y}^{E(1)}$), the parts of intraband and interband are both not zero. 
However, for the longitudinal MS(E)PC ($j_{x}^{(1)}$ and $j_{x}^{E(1)}$), under the relation $\varepsilon_n(\boldsymbol{k})=\varepsilon_n(-\boldsymbol{k})$, the intraband part of MS(E)PC is zero, leaving only the interband part. 
So we only consider MS(E)PC in the $x$ direction induced by LP light in the $x$ direction (longitudinal part of MS(E)PC) in this subsection.
According to Eq. \ref{chiOeab} and Eq. \ref{chiEOeab}, in the case discussed in this subsection,  the longitudinal magnon spin photoconductivity and the longitudinal magnon energy photoconductivity are purely imaginary.
So, according to Eq. \ref{jaLP}, under the LP light along $x$-axis, the MS(E)PC along $x$-axis can be expressed as
\begin{equation}
	\begin{split}
		j_{x}^{LP(E)(1)}(\omega)
		&=Im\left[\chi_{xx}^{(E)}(\omega)\right]\sin(\omega t)E_{0,x}\\
		&=-iE_{0,x}\chi_{xx}^{(E)e}(\omega)\sin(\omega t).\\
	\end{split}
	\label{jxLPLFPM}
\end{equation}
Here, $j_{x}^{LP(E)(1)}(\omega)$ has the form of a sinusoidal function. 
The amplitude of the electric field $E_{0,x}$ and $\chi_{xx}^{(E)e}(\omega)$ determine the magnitude of $j_{x}^{LP(E)(1)}(\omega)$, and the optical frequency determines the oscillation frequency of $j_{x}^{LP(E)(1)}(\omega)$.
\subsubsection{Hexagonal lattice}
Firstly, as shown in Fig. \ref{fig:lattice} (a), we consider a model with the Hexagonal lattice. 

After the linear Holstein–Primakoff transform and Fourier transform, the Hamiltonian of the Hexagonal lattice can be written as \cite{ref9.1,ref9.2,ref28}
\begin{equation}
%\begin{split}
\hat{H}
=\sum_{\mathbf{k}}\hat{\Phi}^{\dag}(\mathbf{k})\mathcal{H}(\mathbf{k})\hat{\Phi}(\mathbf{k})
%\end{split}
\end{equation}
Here, $\hat{\Phi}(\mathbf{k})=(\hat{a}_{\mathbf{k}},\hat{b}_{\mathbf{k}})^{T}$, in which $\hat{a}_{\boldsymbol{k}}$ and $\hat{b}_{\boldsymbol{k}}$ are the magnon annihilation operators of sublattice sites in reciprocal space, and
\begin{equation}
%\begin{split}
\mathcal{H}(\boldsymbol{k})=\begin{pmatrix}
\Delta+\mathcal{D}(\boldsymbol{k}) & -3JS\gamma_{\boldsymbol{k}}\\
-3JS\gamma_{-\boldsymbol{k}} & \Delta-\mathcal{D}(\boldsymbol{k})\\
\end{pmatrix}=d_0+\boldsymbol{d}(\boldsymbol{k})\cdot\boldsymbol{\sigma}.
\label{Hamiltonian}
%\end{split}
\end{equation}
Here, $\Delta=3JS-g_J\mu_BB^z$, 
$\gamma_{\mathbf{k}}=\frac{1}{3}\sum_ie^{i\mathbf{k}\cdot\mathbf{\boldsymbol{\delta}_i}}$ is the structure factor, and $\mathcal{D}(\mathbf{k})=2SD\left[\sin{\frac{1}{2}(k_{y}-k_{x}\sqrt{3})}-\sin{k_{y}}+\sin{\frac{1}{2}(k_{y}+k_{x}\sqrt{3})}\right]$.  $\boldsymbol{\sigma}=\left(\sigma_x,\sigma_y,\sigma_z\right)$.
So $d_0=\Delta$, $d_x=-3JSRe\left[\gamma_{\boldsymbol{k}}\right]$, $d_y=3JSIm\left[\gamma_{\boldsymbol{k}}\right]$, and $d_z=\mathcal{D}(\boldsymbol{k})$.
So the energy of mgnon bands is 
\begin{equation}
	%\begin{split}
	\varepsilon_{\pm}(\boldsymbol{k})=d_0\pm|\boldsymbol{d}(\boldsymbol{k})|,
	\label{band}
	%\end{split}
\end{equation}
in which $|\boldsymbol{d}(\boldsymbol{k})|=\sqrt{d_x^2+d_y^2+d_z^2}$. The corresponding Bloch state can be expressed as
\begin{equation}
	%\begin{split}
	\ket{u_{\pm}(\boldsymbol{k})}=\frac{1}{\sqrt{2|\boldsymbol{d}|^2\mp2|\boldsymbol{d}|d_z}}
	\begin{pmatrix}
		-d_x+id_y\\
		d_z\mp|\boldsymbol{d}|
	\end{pmatrix}.
	\label{BlochD}
	%\end{split}
\end{equation}
Because of the inversion symmetry of the Hexagonal model, and the magnon bands satisfy $\varepsilon_n(\boldsymbol{k})=\varepsilon_n(-\boldsymbol{k})$, the Bose distribution satisfies $f_n^B(\boldsymbol{k})=f_n^B(-\boldsymbol{k})$.
So the intraband part of longitudinal magnon spin (energy) photoconductivity is zero.

In Subsec. \ref{MEPC}, we have discussed that, except when the average value of the energy band is constant,  magnon energy photoconductivity under the condition of low optical frequency and $\Gamma=0$  (Eq. \ref{chiEOeab}) can not be expressed as the weighted integral of Berry curvature with respect to Bose distribution.
According to Eq. \ref{band}, the average value of two bands of the Hexagonal lattice model is a constant.
So the magnon energy photoconductivity can be expressed as
%In Subsec. \ref{MEPC}, we have discussed that magnon energy photoconductivity under the condition of low optical frequency and $\Gamma=0$  (Eq. \ref{chiEOeab}) can not be expressed as the weighted integral of Berry curvature with respect to Bose distribution.
%
%However, according to Eq. \ref{band}, the sum of two bands is a constant.
%
%So Eq. \ref{chiEOeab} can be expressed as
{\small
\begin{equation}
	\begin{split}
		\chi_{ab}^{E,e}(\omega)
		&\approx\frac{\nu}{2}\omega\sum_{n\ne m,c}\int[dk] \left\{\epsilon_{cbz}\mathcal{A}_{a,mn}(\boldsymbol{k})\mathcal{A}_{c,nm}(\boldsymbol{k})\right.\\
		&\left.\times f_{nm}(\boldsymbol{k})\left[\omega_n(\boldsymbol{k})+\omega_m(\boldsymbol{k})\right]\right\}\\
		&=\nu\omega \frac{d_0}{\hbar}\sum_{n\ne m,c}\int[dk] \left\{\epsilon_{cbz}\mathcal{A}_{a,mn}(\boldsymbol{k})\mathcal{A}_{c,nm}(\boldsymbol{k})\right.\\
		&\left.\times f_{nm}(\boldsymbol{k})\right\}\\
		&=i\nu\omega \frac{d_0}{\hbar}\sum_{c}\epsilon_{cbz}\sum_n\int[dk] \Omega_n^{ac}(\boldsymbol{k})f_n^B(\boldsymbol{k})\\
		&=\frac{d_0}{\hbar}\chi_{ab}^{e}(\omega).\\
		\label{d0chi}
	\end{split}
\end{equation}
}
In this situation, magnon energy photoconductivity and magnon spin photoconductivity differ by only a constant $\omega_0=\frac{d_0}{\hbar}$ with the dimension of frequency.
Here, $d_0=\Delta=3JS+g_J\mu_BB^z$ is independent on $D$, but is dependent on $J$.
So the relationship between $\chi_{ab}^{E,e}(\omega_i)$ and $D$ is similar to that between $\chi_{ab}^{e}(\omega_i)$ and $D$.

According Eq. \ref{band} and Eq. \ref{BlochD}, 
when $D>0$ meV, the Chern number set of the ferromagnetic Hexagonal lattice satisfies $(C_-,C_+)=(1,-1)$, in which $C_-$ is the Chern number of the lower band and $C_+$ is the Chern nember of the higher band (for details, see Appendix \ref{appchern}). 
When $D<0$ meV, the Chern number satisfies $(C_-,C_+)=(-1,1)$.
Therefore, the exchange interaction $J$ and the magnetic field $B^z$ have no effect on the Chern number of the magnon system.

Because Bose distribution $f_n^B(\varepsilon)=\frac{1}{e^{\varepsilon/k_BT}-1}$ decreases with increasing energy $\varepsilon$, magnons are more widely distributed in the lower band $\varepsilon_-(\boldsymbol{k})$.
Therefore, according to Eq. \ref{chiOeab}, we can control the direction of the MSPC by changing the direction of the DM interaction roughly.
In order to discuss this problem more concretely, we introduce the Chern number isoenergy surfaces \cite{ref9}
\begin{equation}
	\begin{split}
		C_n(\varepsilon)=\frac{1}{2\pi}\int_{BZ}d^2k\delta\left(\varepsilon_n(\boldsymbol{k})-\varepsilon\right)\Omega_n^z(\boldsymbol{k}),
	\end{split}
\end{equation}
the magnon spin photoconductivity isoenergy surfaces \cite{ref9}
{\small
	\begin{equation}
		%\begin{split}
		\chi_{xx}^e(\omega,\varepsilon)
		=-i\nu\omega
		\sum_{n}\int[dk]\delta\left(\varepsilon_n(\boldsymbol{k})-\varepsilon\right)\Omega_n^{xy}(\boldsymbol{k})f_{n}^B(\boldsymbol{k}),
		%\end{split}
		\label{chixxmain}
	\end{equation}
}
and the magnon energy photoconductivity isoenergy surfaces 
{\small
	\begin{equation}
		%\begin{split}
		\chi_{xx}^{E,e}(\omega,\varepsilon)
		=-i\frac{d_0}{\hbar}\nu\omega
		\sum_{n}\int[dk]\delta\left(\varepsilon_n(\boldsymbol{k})-\varepsilon\right)\Omega_n^{xy}(\boldsymbol{k})f_{n}^B(\boldsymbol{k}).
		%\end{split}
		\label{chiExxmain}
	\end{equation}
}
\begin{widetext}
	
	\begin{figure}[htbp]
		\centering
		\subfloat[$\varepsilon_n(\boldsymbol{k})$ with $D>0$]{
			\includegraphics[scale=0.26]{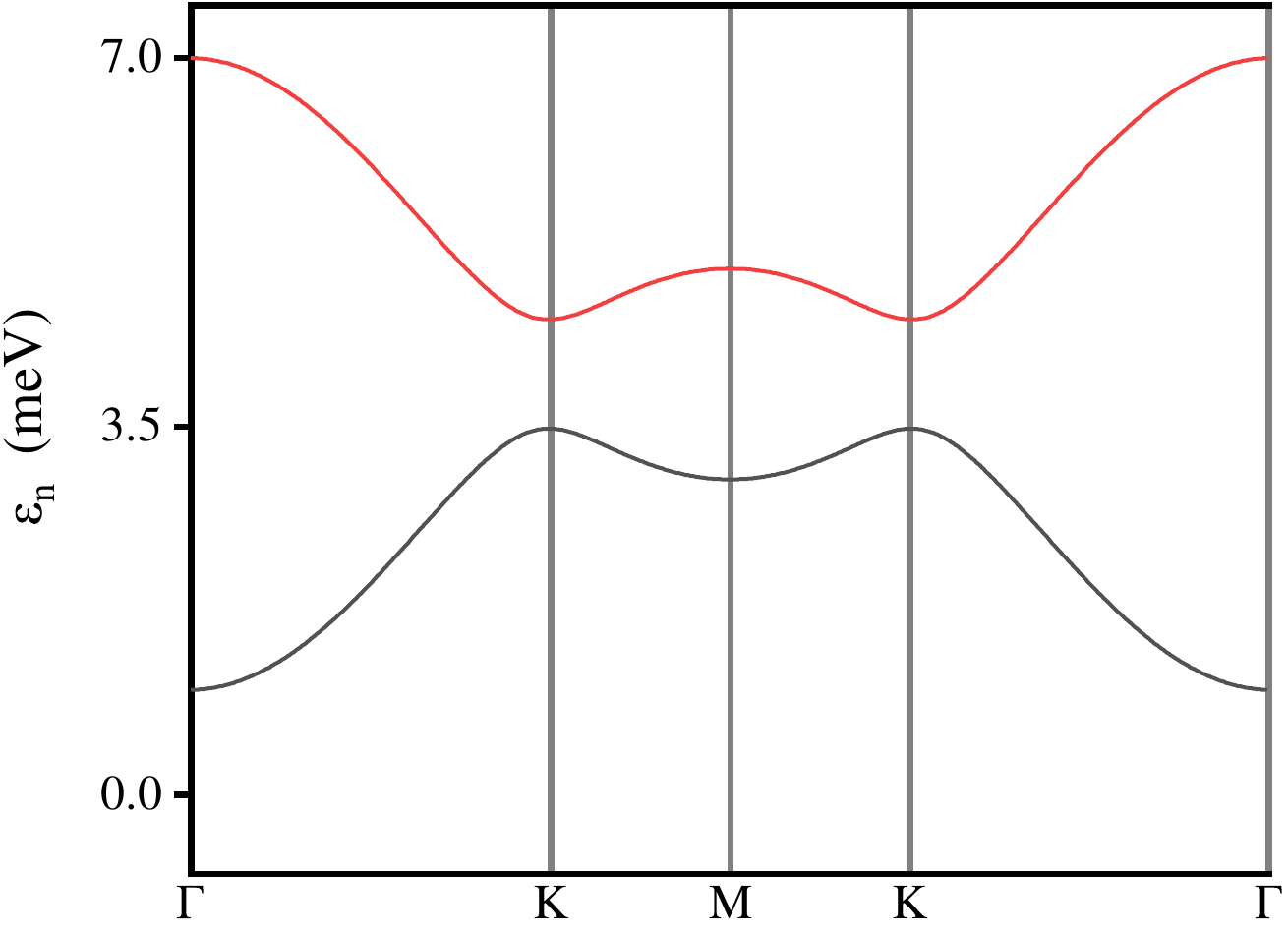}}
		\hfill 
		\subfloat[$C_n(\varepsilon)$ with $D=0.1$ meV]{
			\includegraphics[scale=0.26]{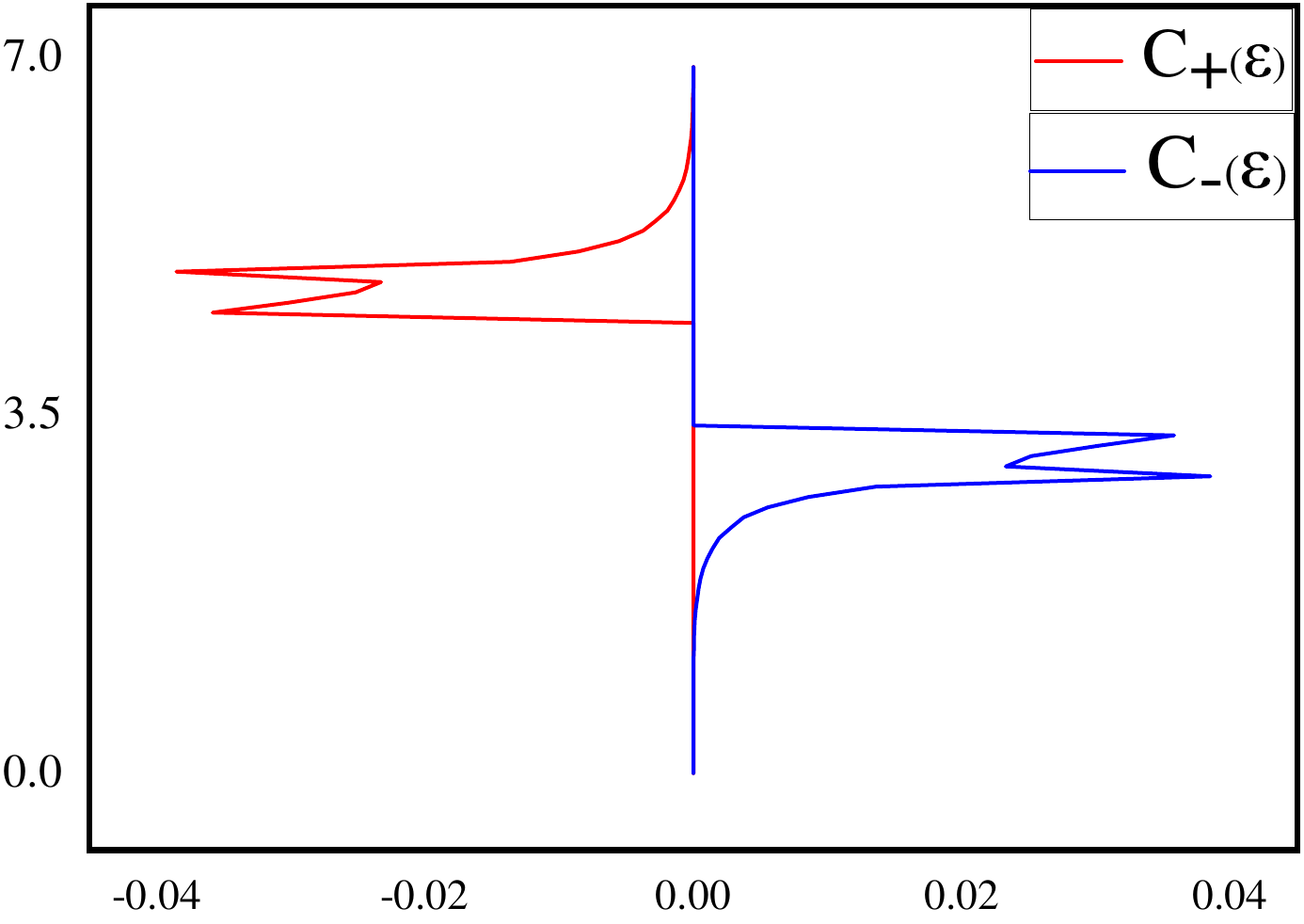}}
		\hfill
		\subfloat[$\left(i/\nu\omega a^2\right)\times\chi_{xx}^e(\omega,\varepsilon)$ or $\left(i\hbar/d_0\nu\omega a^2\right)\times\chi_{xx}^{E,e}(\omega,\varepsilon)$ with $D=0.1$ meV]{
			\includegraphics[scale=0.26]{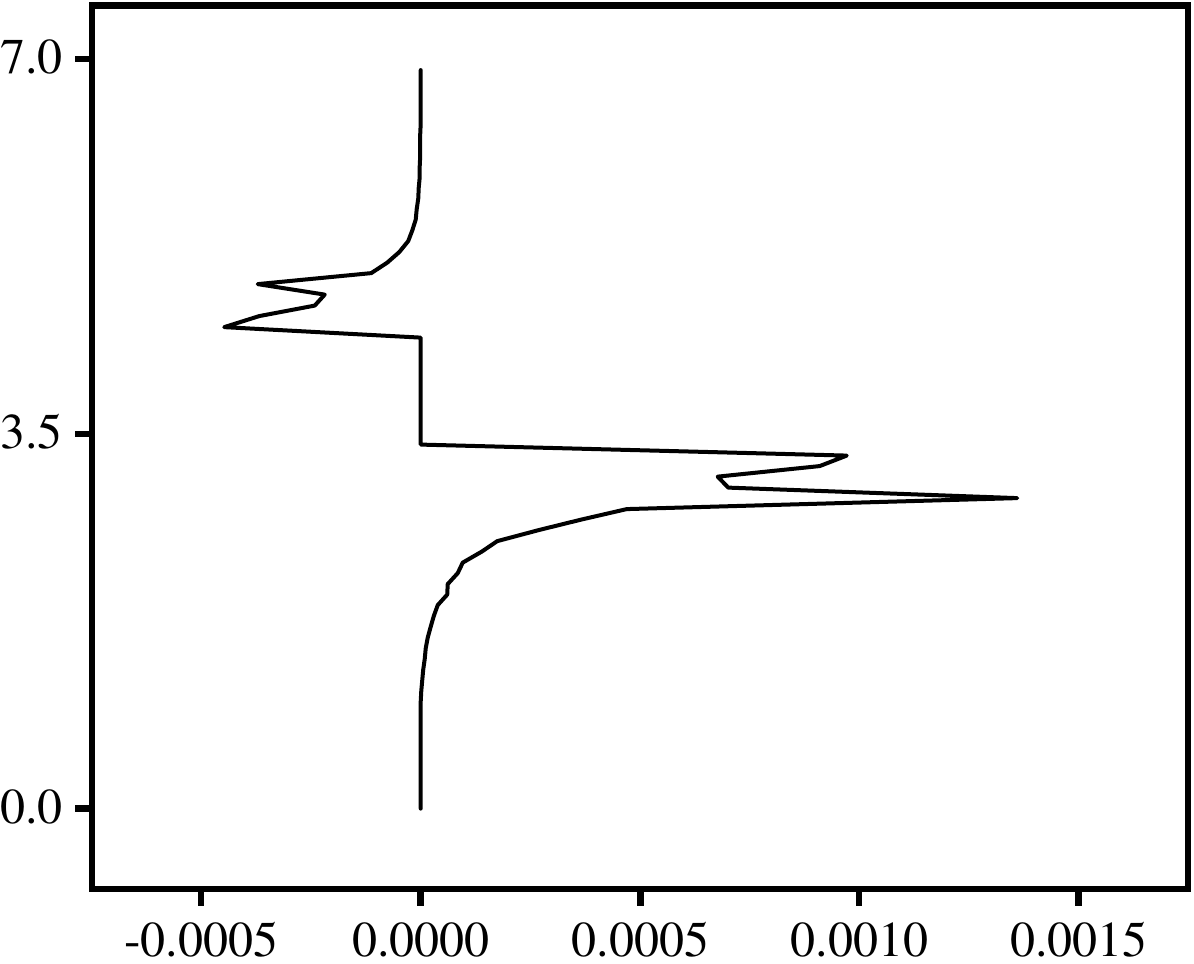}}
		\hfill
		\subfloat[$\varepsilon_n(\boldsymbol{k})$ with $D=-0.1$ meV]{
			\includegraphics[scale=0.26]{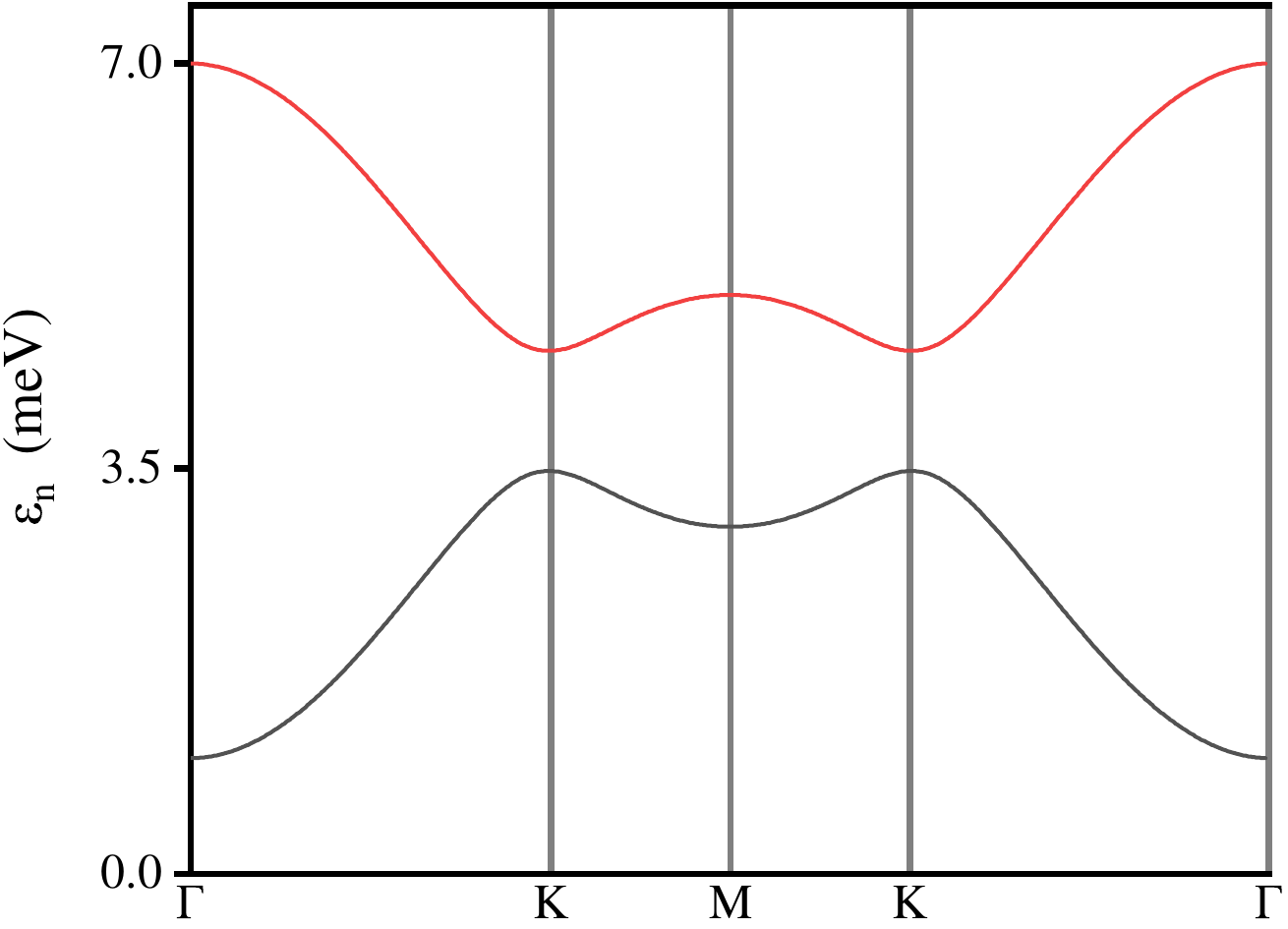}}
		\hfill
		\subfloat[$C_n(\varepsilon)$ with $D=-0.1$ meV]{
			\includegraphics[scale=0.26]{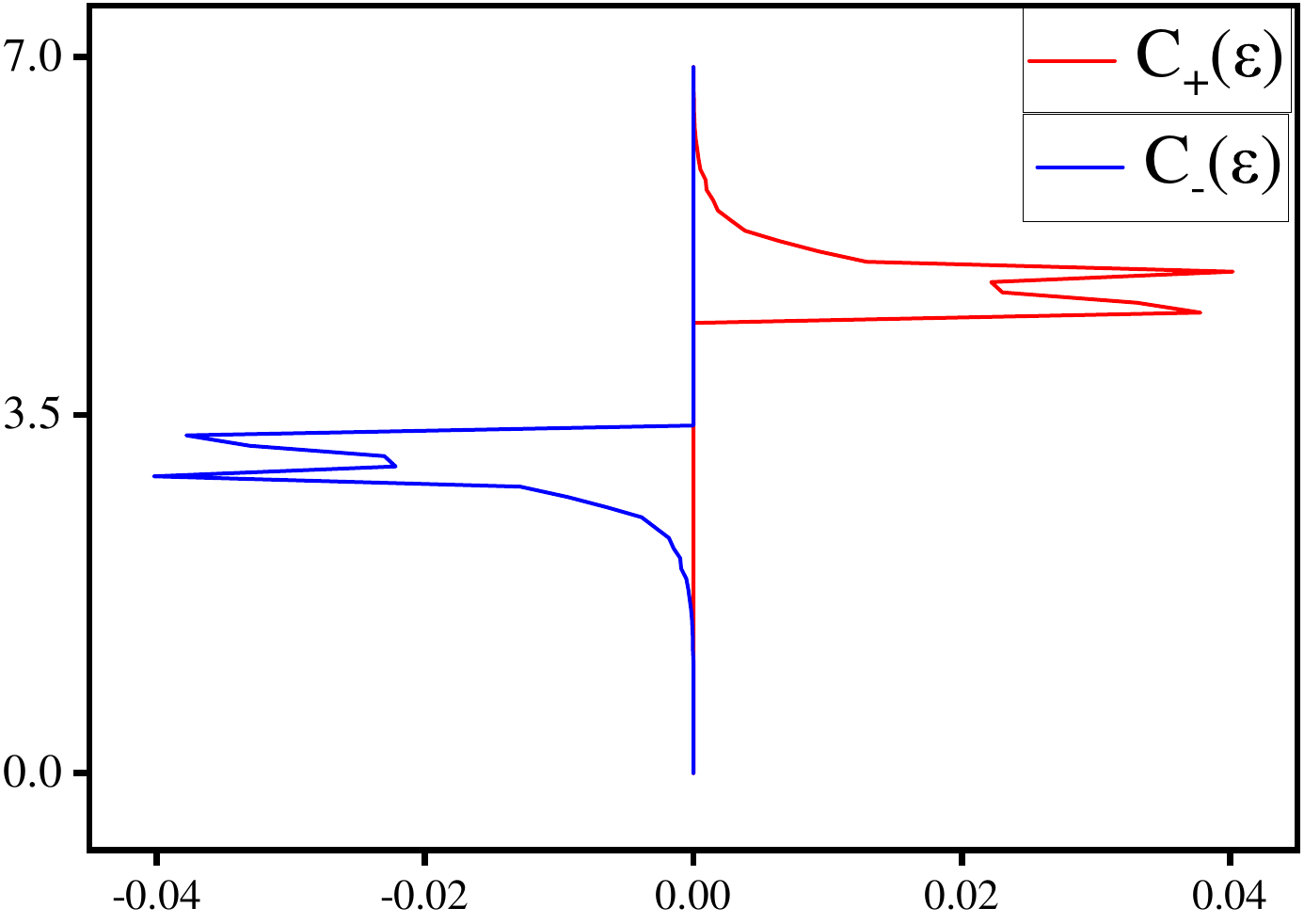}}
		\hfill
		\subfloat[$\left(i/\nu\omega a^2\right)\times\chi_{xx}^e(\omega,\varepsilon)$ or $\left(i\hbar/d_0\nu\omega a^2\right)\times\chi_{xx}^{E,e}(\omega,\varepsilon)$ with $D=-0.1$ meV]{
			\includegraphics[scale=0.26]{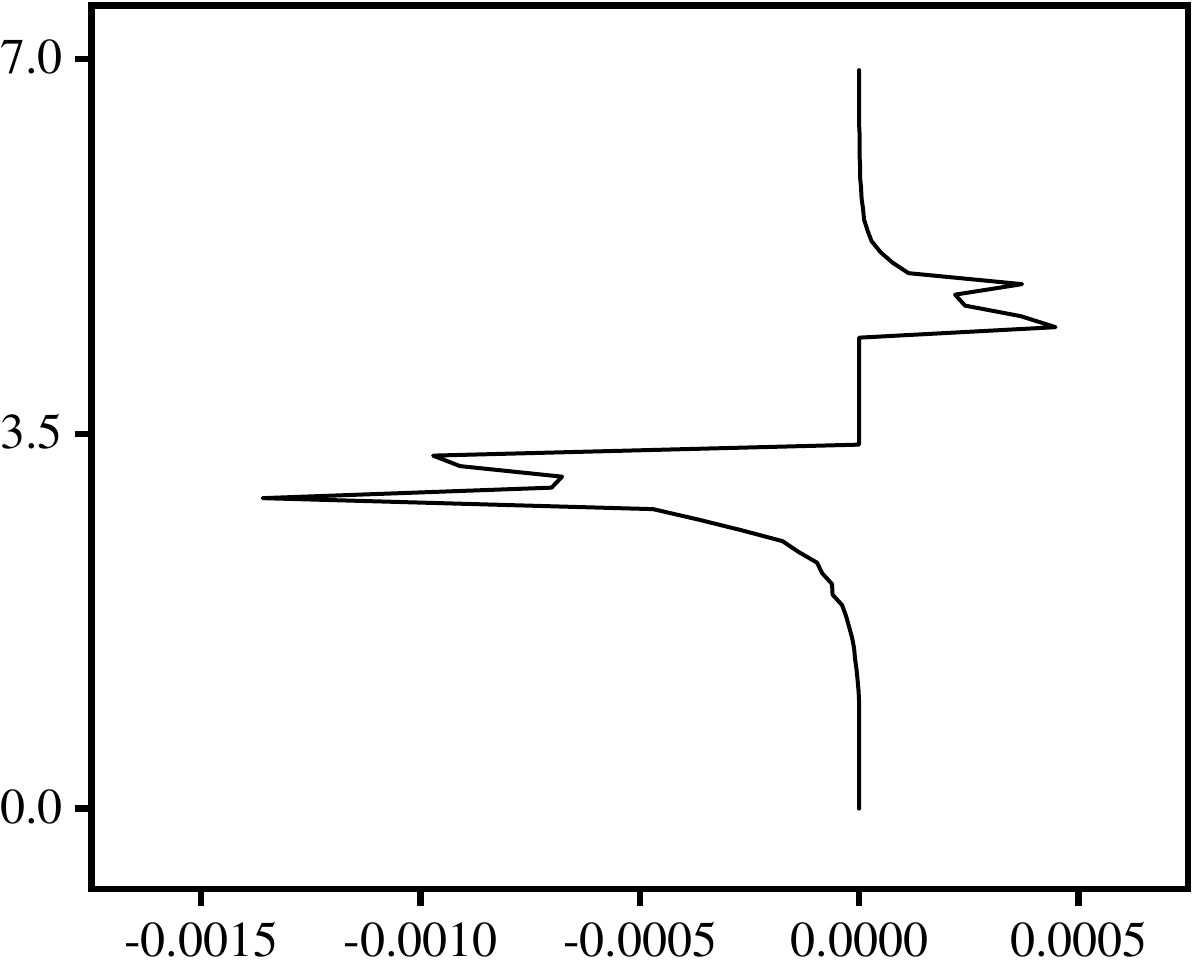}}
		\vspace{0.1cm} 
		\captionsetup{width=\textwidth}
		\caption{Relation between magnon band structure (a) and (d), Chern number isoenergy surfaces (b) and (e), and magnon spin (energy) photoconductivity isoenergy surfaces under the condition of low optical frequency and $\Gamma=0$ (c) and (f) in ferromagnetic Hexagonal lattice. Here, we take $J=1$ meV, $g_J\mu_BB_z=-1$ meV, $T=20$ K,  and $D=\pm 0.1$ meV. For the condition of $D=0.1$ meV, magnon bands (a), Chern number isoenergy surfaces (b), and magnon spin (energy) photoconductivity isoenergy surfaces are shown in the first line. For the condition of $D=-0.1$ meV, magnon bands (d), Chern number isoenergy surfaces (e), and magnon spin (energy) photoconductivity isoenergy surfaces (f) are shown in the second line.  Here,  the vertical coordinates of all figures label energy. And the horizontal coordinates in (b) and (e) label Chern number isoenergy surfaces $C_n(\varepsilon)$; the horizontal coordinates in (c) and (f) label magnon spin (energy) photoconductivity isoenergy surfaces.}
		\label{figbandCchi}
	\end{figure}
\end{widetext}
For $D>0$ meV, as shown in Fig. \ref{figbandCchi} (a) and (b), $C_-(\varepsilon)>0$ meV and $C_+(\varepsilon)<0$. 
Because of the expression of Berry curvature (Eq. \ref{Berrycurvature}), the main contributions to the Chern numbers appear at the band edges with the smallest energy gap.
According to Eq. \ref{chiOeab}, $\chi_{xx}^e(\omega)$ is roughly given by the Chern numbers weighted by the Bose distribution. 
Additionally, because the Bose distribution decreases with the increase of energy, the Chern number of the lower band has a greater contribution to the magnon spin photoconductivity (Fig. \ref{figbandCchi} (c)). 
Therefore, when $D>0$ meV, $(i/\nu\omega a^2)\chi_{xx}^{e}(\omega)>0$ and $(i\hbar/d_0\nu\omega a^2)\chi_{xx}^{E,e}(\omega)>0$. 
In a similar way, for $D<0$ meV, because the change of Chern numbers, $(i/\nu\omega a^2)\chi_{xx}^{e}(\omega)<0$ and $(i\hbar/d_0\nu\omega a^2)\chi_{xx}^{E,e}(\omega)<0$.
To sum up, $D$ can be used to control the symbol of $\chi_{xx}^e(\omega)$ and $\chi_{xx}^{E,e}(\omega)$.
According to Eq. \ref{jxLPLFPM}, the change of the sign of $D$ will change the direction of MSPC and MEPC.
%%%%%%%%%%%%%%%%%%%%%%%%%%%%%%%%%%%%%%%%%%%%%%%%%%%%%%%%%%%%%%%%%%%%%%%%%%%%%%%%%%%%%%%%%%%%%%%
\begin{figure}[h]
	\centering
	\includegraphics[width=1\columnwidth]{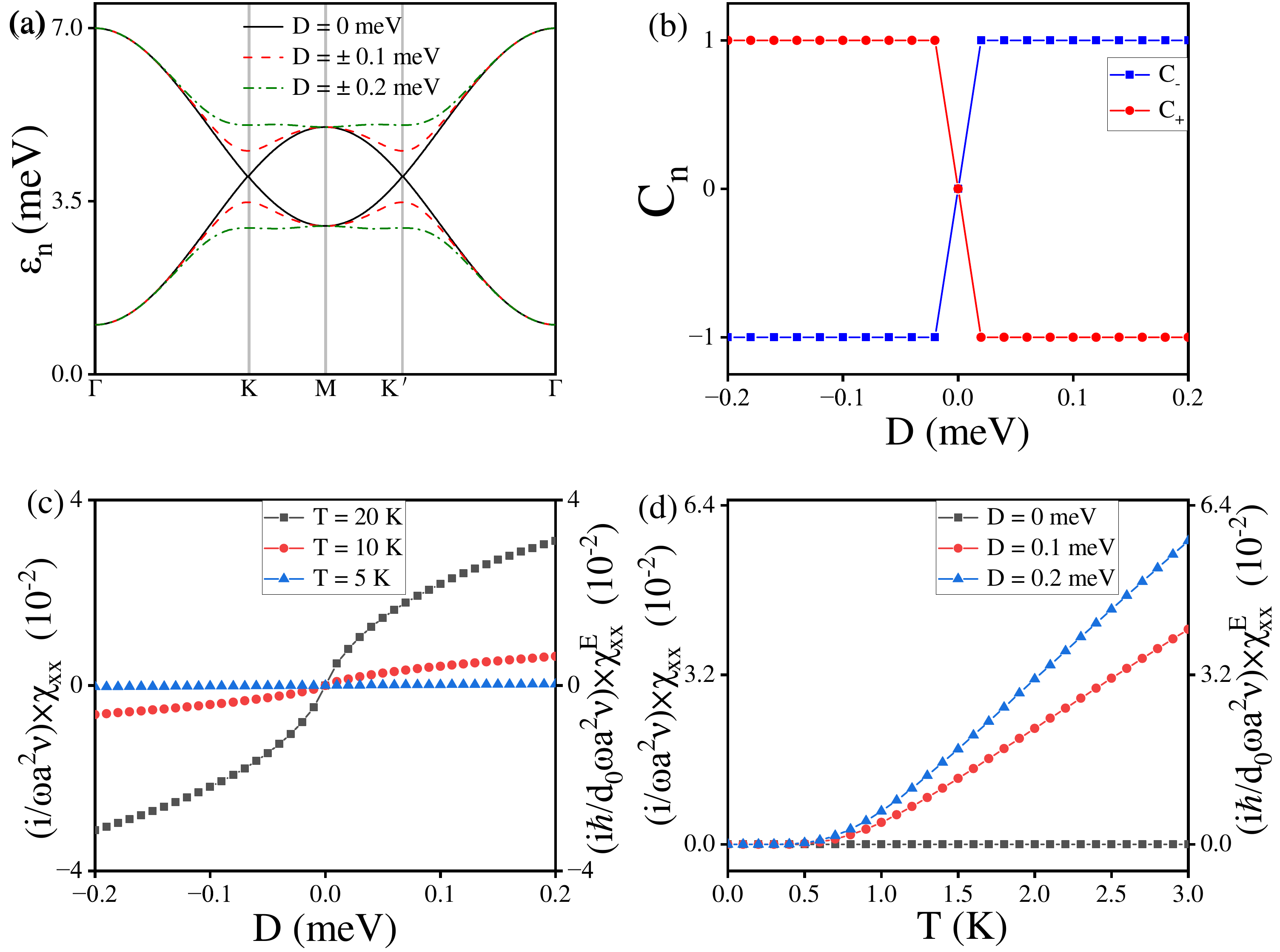}
	\caption{The calculation results of the Hexagonal lattice. (a) Magnon bands with different $D$. (b) Chern number $C_n$ of magnon bands.  
		(c) Magnon spin (energy) photoconductivity as a function of $D$ under different temperatures. (d) Magnon spin (energy) photoconductivity as a function of temperature with different $D$. }
	\label{fig:bccc}
\end{figure}
%%%%%%%%%%%%%%%%%%%%%%%%%%%%%%%%%%%%%%%%%%%%%%%%%%%%%%%%%%%%%%%%%%%%%%%%%%%%%%%%%%%%%%%%%%%%%%%

From another point of view, we can control the magnitude of $\chi_{xx}^e(\omega)$ and $\chi_{xx}^{E,e}(\omega)$ by changing DM strength $D$. 
As shown in Fig. \ref{figbandCchi} (c) and (f), Chern numbers of different bands have opposite sign, leading to a partial cancellation in the total magnon spin (energy) photoconductivity \cite{ref9}. 
Roughly speaking, 
because the Bose distribution decreases with the increase of energy, the effect of the partial cancellation can be reduced by increasing the interband energy difference.  %energy difference of bands. 
According to Eq. \ref{band} and the expression of $\boldsymbol{d}(\boldsymbol{k})$, we can increase the magnitude of $D$ ($|D|$) to increase the difference of bands.
Therefore, we can hopefully increase the magnitude of $\chi_{xx}^e(\omega)$ and $\chi_{xx}^{E,e}(\omega)$ by increasing $|D|$. 
In other words, under the condition of low optical frequency and $\Gamma=0$, the magnitude of MSPC and MEPC (Eq. \ref{jxLPLFPM}) can be increased by increasing the $|D|$.

The calculation results of magnon spin (energy) photoconductivity in the Hexagonal lattice are shown in Fig. \ref{fig:bccc}.
In Fig. \ref{fig:bccc} (b), as mentioned earlier, when the sign of $D$ is reversed, the sign of the Chern number is also reversed.
So in Fig. \ref{fig:bccc} (c), the reversing of $D$ reverses the sign of magnon spin (energy) photoconductivity.
When $D=0$ meV, the magnon spin (energy) photoconductivity is zero because of the ETRS $\hat{\mathcal{T}}$ (for details, see Sec. \ref{Tprime}).
Additionally, in Fig. \ref{fig:bccc} (a) 
, the %band difference
interband energy difference around points $K$ and $K^{\prime}$ increases with the increase of $|D|$.
Therefore, in Fig. \ref{fig:bccc} (c), the magnitude of magnon spin (energy) photoconductivity increases with $|D|$.
In addition, as shown in Fig. \ref{fig:bccc} (d), the magnitude of magnon spin (energy) photoconductivity increases with the increasing of temperature, because the increase temperature can excite more spin vibrations.

L. Chotorlishvili \textit{et al.} \cite{ref1.1} investigated the spin Seebeck effect in one-dimensional antiferromagnets. In their research, the research platform is a one-dimensional magnetic system, the source of excitation for the magnon current is the uneven temperature, and the mechanism of the magnon current is from the influence of the temperature gradient on statistical distribution. But in our research, the research platform is a two-dimensional magnetic system, the source of excitation for the magnon spin (energy) current is the periodic electric field, and the mechanism is  the Aharonov-Casher (AC) effect. 
In the result of L. Chotorlishvili \textit{et al.} \cite{ref1.1}, the contributions to spin Seebeck currents contributed from the two magnon modes are opposite. So the spin Seebeck current vanishes when there is no DM interaction and external magnetic field in the system. When they apply DM interaction and an external magnetic field to break the degeneracy and the left-right symmetry, the spin Seebeck current is nonzero. Additionally, they found that the enhancement of DMI and the magnetic field can increase the density of magnons. In our work, under low-frequency conditions, the system requires a non-zero DM interaction to break the effective time reversal symmetry; otherwise, the MS(E)PC is zero. The magnitude of MS(E)PC increases as $|D|$ increases in the Hexagonal lattice. In addition, we also studied the topological properties of MS(E)PC as the DM interaction changes.

\subsubsection{Kagome lattice}
\begin{widetext}
	\begin{figure}[htbp]
		\centering
		{
			\includegraphics[scale=1.2]{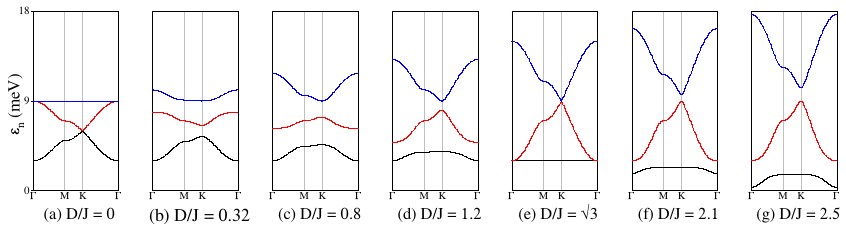}}
		\hfill 
		\vspace{0.1cm} 
		\captionsetup{width=\textwidth}
		\caption{Magnon bands of Kagome lattice with different $D/J$. Here, we take $S=1$, $J = 1$ meV, $g_J\mu_BB^z = -3$ meV.}
		\label{figbandkagome}
	\end{figure}
\end{widetext}
In particular, when the lowest band is a flat band $\varepsilon_0(\boldsymbol{k})=\varepsilon$ 
, Bose distribution $f_0^B(\boldsymbol{k})=1/\left(e^{\varepsilon/k_BT}-1\right)$ will be independent on $\boldsymbol{k}$.
And at low temperature ($k_BT$ $\ll$ band gap), the influence of higher bands on the magnon spin photoconductivity is relatively low.
So under these circumstances, the magnon spin photoconductivity can be approximated as quantized. 
\begin{equation}
	\begin{split}
		\chi_{xx}^e(\omega_i)
		&\approx-i\nu\omega_i
		f_0^B\int[dk]\Omega_0^{ac}(\boldsymbol{k})
		=-\frac{i\nu\omega_i}{2\pi}
		f_0^B\mathcal{C}_0.
	\end{split}
	\label{qchixxmain}
\end{equation}
To better discuss this point,
next, we discuss the magnon spin photoconductivity in the kagome lattice.

The Hamiltonian terms of the Kagome lattice is shown as Eq. \ref{Hkagome}.
So on the point $K$, 
the eigenenergy can be expressed as
\begin{equation}
	\begin{cases}
		\varepsilon_0(\boldsymbol{k}_K)=-2JS-g_J\mu_BB^z\\
		\varepsilon_1(\boldsymbol{k}_K)=JS-\sqrt{3}|DS|-g_J\mu_BB^z\\
		\varepsilon_2(\boldsymbol{k}_K)=JS+\sqrt{3}|DS|-g_J\mu_BB^z.\\
	\end{cases}
\end{equation}
Therefore, the topological phase boundary satisfied $D=0$ meV or $J/D=\pm1/\sqrt{3}$.
In our work on the Kagome lattice, we only consider the condition of $D \geq 0$. 
As shown in Fig. \ref{figbandkagome} (a), when $D/J=0$, 
the band gaps are both zero.
Then, as the system turns to $D/J=\sqrt{3}$, the highest band gradually increases, and the lowest band tends approximately to the flat band. (Fig. \ref{figbandkagome} (b) (c) and (d)). 
In Fig. \ref{figbandkagome} (e), when $D/J=\sqrt{3}$,  
the band gaps are both zero again, and the lowest band is a flat band. 
Then, as $D/J$ increases, band gaps increase, and the lowest band decreases.
%
%%%%%%%%%%%%%%%%%%%%%%%%%%%%%%%%%%%%%%%%%%%%%%%%%%%%%%%%%%%%%%%%%%%%%%%%%%%%%%%%%%%%%%%%%%%%%%%
\begin{figure}[h]
	\centering
	\includegraphics[width=1.05\columnwidth]{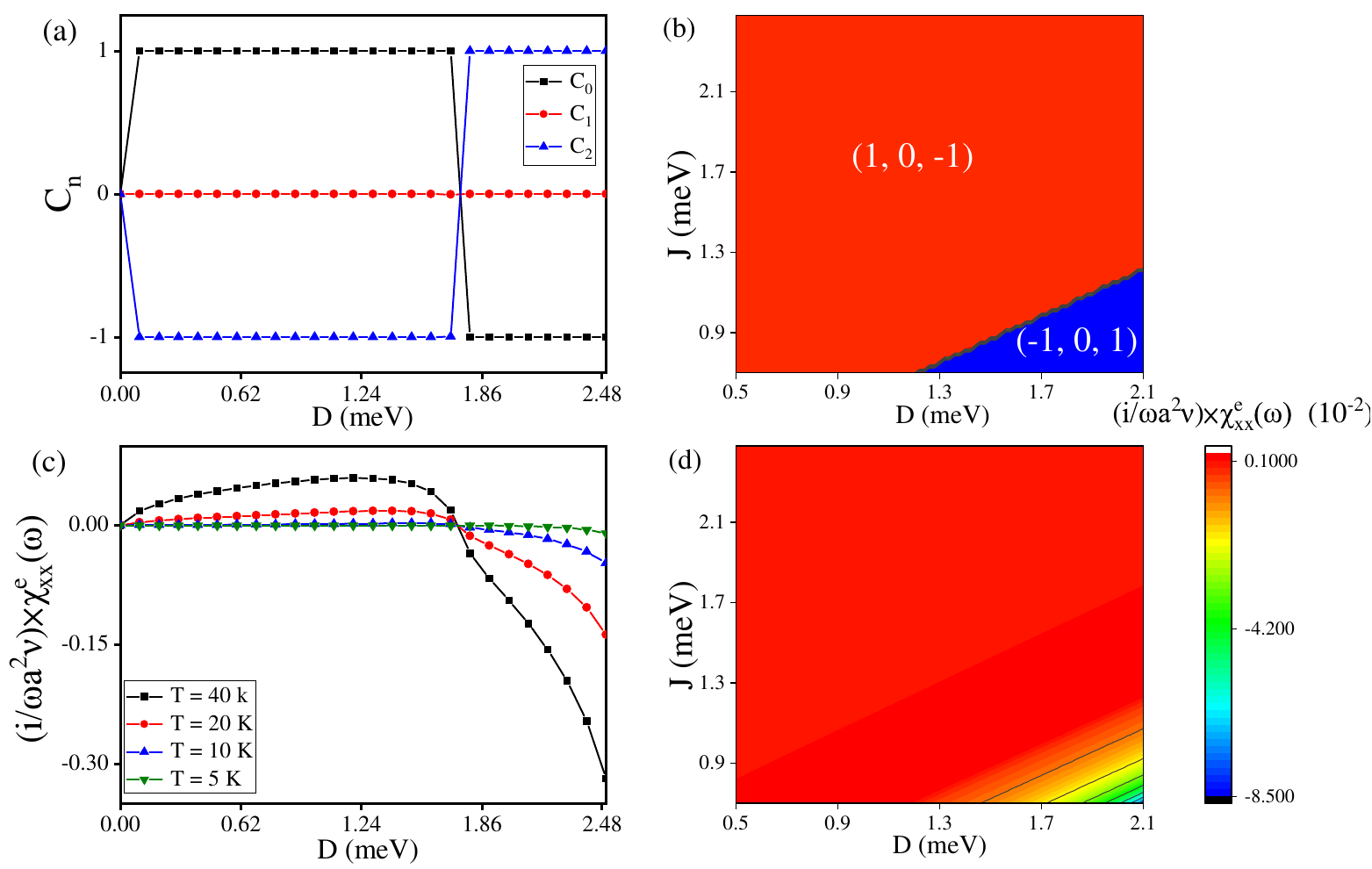}
	\caption{The effect of topological properties on magnon spin photoconductivity in ferromagnetic Kagome lattice. Here, we take $S=1$, $J = 1$ meV, and $g_J\mu_BB^z=-3$ meV. (a) Chern number curve with $D$ of different bands. 
		(b) Topological phase diagram with regions characterized by Chern number set $(C_0,C_1,C_2)$, where the horizontal coordinate represents $D$ and the vertical coordinate represents exchange interaction $J$. Here, the set of the red area is $(C_0,C_1,C_2)=(1,0,-1)$, and the set of the blue area is $(C_0,C_1,C_2)=(-1,0,1)$. 
		(c) Magnon spin photoconductivity curve with $D$ under different temperatures. (d) Magnon spin photoconductivity with $D$ and exchange interaction $J$, in which the temperature $T = 10$ K.}
	\label{fig:kagomecurrent}
\end{figure}
%%%%%%%%%%%%%%%%%%%%%%%%%%%%%%%%%%%%%%%%%%%%%%%%%%%%%%%%%%%%%%%%%%%%%%%%%%%%%%%%%%%%%%%%%%%%%%%

Next, we study the variation of the magnon spin photoconductivity with $D$ through model calculation, and discuss the relationship between the magnon spin photoconductivity and the topological properties of the magnon system.
In Fig. \ref{fig:kagomecurrent} (a), we take $J = 1$ meV, and as $D$ increases, when $D < \sqrt{3}$ meV ($D/J < \sqrt{3}$, corresponding to Fig. \ref{figbandkagome} (a)-(d)), the Chern number set is $(1,0, -1)$; when $D > \sqrt{3}$ meV ($D/J > \sqrt{3}$, corresponding to Fig. \ref{figbandkagome} (f) and (g)), the Chern number set is $(-1,0,1)$. 
Similar to the case of the Hexagonal lattice, because magnons conform to the Bose distribution, relatively more magnons cluster on the lowest band, and the Chern number of the lowest band has a greater contribution to the magnon spin photoconductivity. 
Therefore, in Fig. \ref{fig:kagomecurrent} (c),  $\frac{i}{\omega a^2\nu}\chi_{xx}^e(\omega)$ is greater than zero when the Chern number set is $(1,0,-1)$ ($D/J < \sqrt{3}$) and is less than zero when the Chern number set is $(-1,0,1)$. 
When $D=0$ meV, the magnon spin conductivity is zero because of the ETRS (for details, see Sec. \ref{Tprime}).
In Fig. \ref{figbandkagome} (b)-(d), as $D/J$ increases, the lowest band remains relatively flat, and the lowest band basically does not rise or fall as a whole.
So when $D/J < \sqrt{3}$ (Chern number set is $(1,0,-1)$), the magnon spin photoconductivity varies little with $D$, but it does change with temperature according to Eq. \ref{qchixxmain}.
However, when $D/J > \sqrt{3}$, the lowest band as a whole falls with the increase of $D$.
Because the Bose distribution $f^B(\varepsilon)=1/(e^{\varepsilon/k_BT}-1)$ increases with the decrease of energy $\varepsilon$, the magnitude of magnon spin photoconductivity increases as the decrease of the lowest band (the band that makes a major contribution to the magnon spin photoconductivity) according to Eq. \ref{qchixxmain}. 
Therefore, when $D/J > \sqrt{3}$, different from the case of $D/J < \sqrt{3}$, the magnitude of magnon spin  photoconductivity increases dramatically with the increase of $D$.
This is shown more intuitively in Fig. \ref{fig:kagomecurrent} (b) and (c). 
In Fig. \ref{fig:kagomecurrent} (b), the Chern number set of the red area is $(1,0,-1)$, the Chern number set of the blue area is $(-1,0,1)$. The topological phase transition occurs on the line $D/J = \sqrt{3}$.
In Fig. \ref{fig:kagomecurrent} (c), in the region where the Chern number set is $(1,0,-1)$, the magnon spin photoconductivity changes little with $J$ and $D$; when the Chern number set is $(1,0,-1)$, the magnitude of magnon spin photoconductivity increase with the increasing of $D/J$.

In the work of thermal Hall effect by A. Mook \textit{et al.}, authors investigated the influence of topological properties on the thermal Hall effect of magnon in Kagome lattice \cite{ref9}. While in our work, we investigate the effect of topological properties on the MS(E)PC. In the calculation result of us and A. Mook \textit{et al.} \cite{ref9}, the topological properties of the system has a significant impact on the direction of magnon transport. That is, under the condition of unchanged excitation, the change in the topological properties of the system will largely affect the direction of the current of the magnon. 
However, when calculating the topological properties of the system, we focus on the variation of the topological properties with DM strength and nearest neighbor exchange interaction, while A. Mook \textit{et al.} \cite{ref9} focus on the variation of topological properties with the ratio of DM strength to nearest neighbor exchange interaction and the ratio of second nearest neighbor exchange interaction to nearest neighbor exchange interaction.

For the Kagome lattice, magnon energy photoconductivity can not be expressed by Berry curvature.
Therefore, it is difficult to discuss the magnon energy photoconductivity of the Kagome lattice through the topological properties of the system.
So we do not discuss the magnon energy photoconductivity of Kagome under the condition of low optical frequency and $\Gamma=0$ in this work.

In the work of YuanDong Wang \textit{et al.} \cite{ref24}, the authors investigated the magnon spin current induced by the Aharonov-Casher effect and the magnon Landau-Zener tunneling effect. Their research platform is a one-dimensional magnetic system, but our research platform is a two-dimensional magnetic system. Moreover, there is no DM interaction system in the work of YuanDong Wang \textit{et al.} \cite{ref24}. But in our result, under the low-frequency condition, both in the Hexagonal lattice and Kagome lattice, we need a non-zero DM interaction to break the effective time-reversal symmetry; otherwise, the MS(E)PC would be zero. In addition, we derive not only the spin magnon current induced by the oscillating electric field, but also the energy current. Then we also investigated the topological properties of the magnon spin current and energy current induced by the oscillating electric field under low-frequency conditions.

\subsection{Materials realization}
\begin{figure}[h]
	\centering
	\includegraphics[width=1.05\columnwidth]{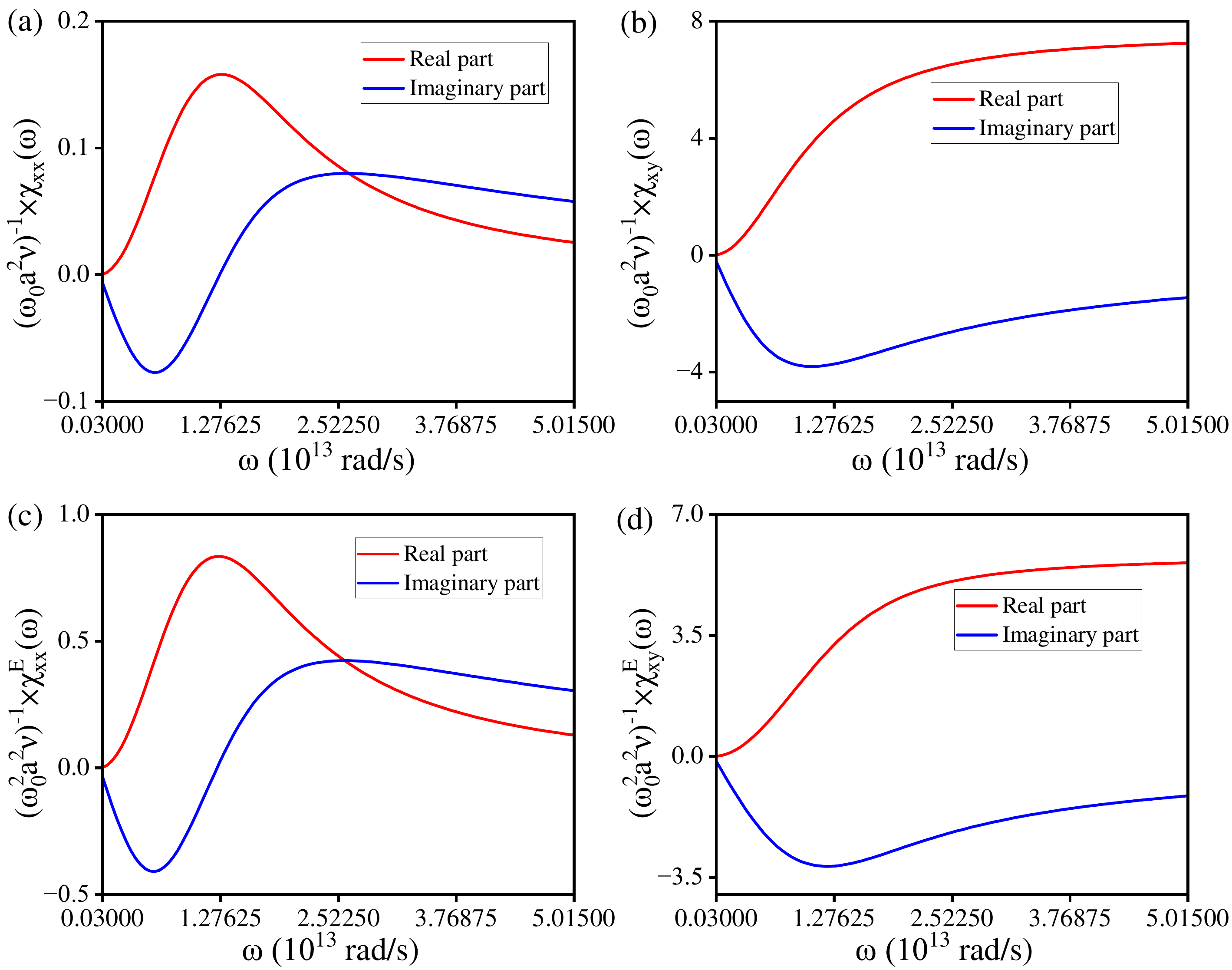}
	\caption{The longitudinal magnon spin photoconductivity (a), transverse magnon spin photoconductivity (b), the longitudinal magnon energy photoconductivity (c), and the transverse magnon energy photoconductivity (d) of $Lu_2V_2O_7$.
		Here, $\omega$ is optical frequency and $\omega_0$ satisfies $\hbar\omega_0=1$ meV.}
	\label{fig:Lu2V2O7}
	%\label{lattice}
\end{figure}

Next, we discuss the magnon spin (energy) photoconductivity in $Lu_2V_2O_7$.
Here, we do not consider the three-dimensional structure of $Lu_2V_2O_7$, we treat the system as a stack of non-interacting layers \cite{ref3,ref4,ref9,ref25}.
The Curie temperature of $Lu_2V_2O_7$ is 70 K \cite{ref3,ref25}. 
In addition, $Lu_2V_2O_7$ satisfies $a=0.7024$ nm, $S=\frac{1}{2}$, $J=3.405$ meV and $D/J=0.32$ \cite{ref3,ref9,ref25}.
Here, we take the temperature as 50 K and $g_J\mu_BB^z=-10^{-4}$ meV.
We assume the magnon lifetime to be of the order of $\tau\approx 10^{-13}$ s.
As shown in Fig. \ref{fig:Lu2V2O7}, in the optical frequency range that we calculate, the transverse magnon spin (energy) conductivity of $Lu_2V_2O_7$ is stronger than the longitudinal magnon spin (energy) conductivity. 
The magnon spin (energy) conductivity reaches its peak in the frequency range of approximately $0.03\times 10^{13}$ rad/s to $2.5335\times 10^{13}$ rad/s, and then turn to stabilize.
In this optical frequency range, 
the magnitude of transverse magnon spin photoconductivity can reach the scale of $10^{-9} e(nm)^2$, when we apply a light with the electric field $E_{0,y}\sim 1 V/nm$, the MSPC can reach the scale of $j_x\sim 10^{-6} (meV)(nm)$.
The magnitude of transverse magnon energy photoconductivity can reach $10^{3} e(nm^2)/s$, when we apply a light with the electric field $E_{0,y}\sim 10^{-6} V/nm$, the MEPC can reach the scale of $j_x^E\sim 1 (meV)(nm)/s$.
%$j_x^E\sim 10^2 eVnm/s=10^5 meVnm/s=10^{-4} meVm/s$.

MSPC can be detected via the inverse spin Hall effect (ISHE) \cite{ref25,ref25.01}. Magnons are difficult to measure because they are electrically neutral. However, the ISHE method achieves the detection of spin magnon current by converting the magnon transport into a electrical signal. The researchers brought the ferromagnetic insulator into contact with a heavy metal material (such as Pt), and then exposed the ferromagnetic insulator to light (time-depended electric field) to generate MSPC. When magnons transported to the interface between Pt and the ferromagnetic insulator, the spin angular momentum of magnons would be transferred to the conduction electrons in the heavy metal, thereby generating a spin current in the heavy metal. The spin current would produce a measurable electrical signal through the ISHE. When the intensity of MSPC reached its peak, the intensity of the electrical signal would also reach its peak. If a low-frequency electric field is applied, the direction of MSPC can be reversed by reversing the DM vector, which can also be measured using the ISHE method. MEPC can be detected through the spin Seebeck effect and ISHE. 
%%%%%%%%%%%%%%%%%%%%%%%%%%%%%%%%%%%%%%%%%%%%%%%%%%%%%%%%%%%%%%%%%%%%%%%%%%%%%%%%%%%%%%%%%%%%%%%
%%%%%%%%%%%%%%%%%%%%%%%%%%%%%%%%%%%%%%%%%%%%%%%%%%%%%%%%%%%%%%%%%%%%%%%%%%%%%%%%%%%%%%%%%%%%%%%
\section{\label{sec:level1} conclusion}
In conclusion, we have derived the linear magnon spin (energy) photoconductivity in two-dimensional collinear ferromagnetic systems.
To investigate the method of controlling the linear magnon spin (energy) photoconductivity, we take a model calculation in the two-dimensional collinear ferromagnetic Kagome lattice and the Hexagonal lattice.
After the model calculation, we find that the linear magnon spin (energy) photoconductivity can be controlled by changing the optical frequency and the relaxation time.
Then we also find that under the condition of low optical frequency and infinite relaxation time, the magnon spin (energy) photoconductivity can be controlled by DM interaction through the topological properties of the magnon system.
Under this condition, when we apply a LP
light, the direction of longitudinal MSPC and MEPC can be controlled by the sign of $D$, and the magnitude of longitudinal MSPC and MEPC can be controlled by the magnitude of $D$ in the two-dimensional collinear ferromagnetic Hexagonal lattice.
In the two-dimensional collinear ferromagnetic Kagome lattice, the magnon spin photoconductivity changes with the Chern number set.

To sum up, our work mainly focused on the methods of controlling the linear MS(E)PC in two-dimensional ferromagnetic materials, providing a new idea for the development of new devices.

\section{Acknowledgements}
This paper is supported by the 111 Project (B16009).
National Frontiers Science Center for Industrial Intelligence and Systems Optimization, Northeastern University, China. 
Key Laboratory of Data Analytics and Optimization for Smart Industry (Northeastern University), Ministry of Education, China.

\section{Author contributions}
J-C. L  carried out the work of theoretical derivation and model calculation, wrote the main manuscript text, and prepared all the figures.

A. D provided guidance for J-C. L's work and reviewed the manuscript.
\section{Competing interests}
All authors declare no financial or non-financial competing interests.

\appendix
\begin{widetext}

\section{The perturbative magnon Hamiltonian induced by the electromagnetic field
} \label{Hamiltonian}

When we take a time-dependent electric field on the magnon, the AC effect on the magnon is \cite{ref24,ref25}
\begin{equation}
	%\begin{split}
	\theta_{in,jm}=\frac{g_J\mu_B}{\hbar c_{lv}^2}\left[\boldsymbol{E}(t)\times\boldsymbol{e}_z\right]\cdot \boldsymbol{l}_{ij}
	=\frac{g_J\mu_B}{\hbar c_{lv}}\boldsymbol{A}_E(t)\cdot\boldsymbol{l}_{in,jm}.
	%\end{split}
	\label{thetaA}
\end{equation}
Here,  $\boldsymbol{l}_{in,jm}=\boldsymbol{R}_j+\boldsymbol{\tau}_m-\boldsymbol{R}_i-\boldsymbol{\tau}_n$, in which $\boldsymbol{R}_i$ represents the $i$-th lattice position, and $\boldsymbol{\tau}_n$ denotes the $n$-th spin.  $\boldsymbol{A}_E(t)=\frac{1}{c_{lv}}\boldsymbol{E}(t)\times\boldsymbol{e}_z$ is the effective vector potential, and $\boldsymbol{e}_z$ is the unit vector along $z$ axis. \cite{ref24}
In general, the two body spin interaction Hamiltonian can be expressed as \cite{ref24}
\begin{equation}
	\begin{split}
	\hat{H}
	&=\frac{1}{2}\sum_{i,j}\sum_{n,m}\sum_{\alpha,\beta}\hat{S}_{i,n}^{\alpha}H_{nm}^{\alpha\beta}(j-i)\hat{S}_{j,m}^{\beta}.\\
	\end{split}
\end{equation}
For a two-dimensional collinear ferromagnet, after H-P transformation, the Hamiltonian can be expressed as
\begin{equation}
	\begin{split}
	\hat{H}
	&=\frac{1}{2}\sum_{i,j}\sum_{n,m}A^{mn}(\boldsymbol{R}_j-\boldsymbol{R}_i)\hat{a}_{i,n}^{\dag}\hat{a}_{j,m}
	%&=\frac{1}{2}\sum_{i}\sum_{\Delta}\sum_{n,m}A^{mn}(\Delta)\hat{a}_{i,n}^{\dag}\hat{a}_{i+\Delta,m}\\
	%&=\frac{1}{2N}\sum_{i}\sum_{\boldsymbol{k}\boldsymbol{k}^{\prime}}\sum_{\Delta}\sum_{n,m}A^{mn}(\Delta)\hat{a}_{n}^{\dag}(\boldsymbol{k})e^{-i\boldsymbol{k}\cdot(\boldsymbol{R}_i+\boldsymbol{\tau}_n)}\hat{a}_{m}(\boldsymbol{k}^{\prime})e^{i\boldsymbol{k}^{\prime}\cdot(\boldsymbol{R}_i+\boldsymbol{\Delta}+\boldsymbol{\tau}_m)}\\
	%&=\frac{1}{2N}\sum_{i}\sum_{\boldsymbol{k}\boldsymbol{k}^{\prime}}\sum_{\Delta}\sum_{n,m}A^{mn}(\Delta)\hat{a}_{n}^{\dag}(\boldsymbol{k})\hat{a}_{m}(\boldsymbol{k}^{\prime})e^{-i(\boldsymbol{k}-\boldsymbol{k}^{\prime})\cdot\boldsymbol{R}_i}e^{i\boldsymbol{k}^{\prime}\cdot(\boldsymbol{\Delta}+\boldsymbol{\tau}_m)}e^{-i\boldsymbol{k}\cdot\boldsymbol{\tau}_n}\\
	%&=\frac{1}{2}\sum_{\boldsymbol{k}}\sum_{n,m}\sum_{\Delta}A^{mn}(\Delta)e^{i\boldsymbol{k}\cdot(\boldsymbol{\Delta}-\boldsymbol{\tau}_n+\boldsymbol{\tau}_m)}\hat{a}_{n}^{\dag}(\boldsymbol{k})\hat{a}_{m}(\boldsymbol{k})\\
	=\frac{1}{2}\sum_{\boldsymbol{k}}\sum_{n,m}B^{mn}(\boldsymbol{k})\hat{a}_{n}^{\dag}(\boldsymbol{k})\hat{a}_{m}(\boldsymbol{k})\\
	\end{split}
\end{equation}
in which $B^{mn}(\boldsymbol{k})=\sum_{\boldsymbol{\Delta}}A^{mn}(\boldsymbol{\Delta})e^{i\boldsymbol{k}\cdot(\boldsymbol{\Delta}-\boldsymbol{\tau}_n+\boldsymbol{\tau}_m)}$, and $\boldsymbol{\Delta}$ is the lattice vector difference.
In the presence electric field, the magnon acquires a phase $\theta_{ij}$ (Eq. \ref{thetaA}) while travelling between the mth and the nth spin on the ith and jth site \cite{ref24}
\begin{equation}
	\begin{split}
	\hat{H}
	&\to\sum_{i,j}\sum_{n,m}A^{mn}(\boldsymbol{R}_j-\boldsymbol{R}_i)\hat{a}_{i,n}^{\dag}\hat{a}_{j,m}e^{i\frac{g_J\mu_B}{\hbar c_{lv}}\boldsymbol{A}_E(t)\cdot\boldsymbol{l}_{ij}^{nm}}
	%&=\frac{1}{2}\sum_{i}\sum_{\Delta}\sum_{n,m}A^{mn}(\Delta)\hat{a}_{i,n}^{\dag}\hat{a}_{i+\Delta,m}e^{i\frac{g_J\mu_B}{\hbar c_{lv}}\boldsymbol{A}_E(t)\cdot\left(\boldsymbol{\Delta}-\boldsymbol{\tau}_n+\boldsymbol{\tau}_m\right)}\\
	%&=\frac{1}{2N}\sum_{i}\sum_{\boldsymbol{k}\boldsymbol{k}^{\prime}}\sum_{\Delta}\sum_{n,m}A^{mn}(\Delta)\hat{a}_{n}^{\dag}(\boldsymbol{k})e^{-i\boldsymbol{k}\cdot(\boldsymbol{R}_i+\boldsymbol{\tau}_n)}\hat{a}_{m}(\boldsymbol{k}^{\prime})e^{i\boldsymbol{k}^{\prime}\cdot(\boldsymbol{R}_i+\boldsymbol{\Delta}+\boldsymbol{\tau}_m)}e^{i\frac{g_J\mu_B}{\hbar c_{lv}}\boldsymbol{A}_E(t)\cdot\left(\boldsymbol{\Delta}-\boldsymbol{\tau}_n+\boldsymbol{\tau}_m\right)}\\
	%&=\frac{1}{2N}\sum_{i}\sum_{\boldsymbol{k}\boldsymbol{k}^{\prime}}\sum_{\Delta}\sum_{n,m}A^{mn}(\Delta)\hat{a}_{n}^{\dag}(\boldsymbol{k})\hat{a}_{m}(\boldsymbol{k}^{\prime})e^{-i(\boldsymbol{k}-\boldsymbol{k}^{\prime})\cdot\boldsymbol{R}_i}e^{i\boldsymbol{k}^{\prime}\cdot(\boldsymbol{\Delta}+\boldsymbol{\tau}_m)}e^{-i\boldsymbol{k}\cdot\boldsymbol{\tau}_n}e^{i\frac{g_J\mu_B}{\hbar c_{lv}}\boldsymbol{A}_E(t)\cdot\left(\boldsymbol{\Delta}-\boldsymbol{\tau}_n+\boldsymbol{\tau}_m\right)}\\
	%&=\frac{1}{2}\sum_{\boldsymbol{k}}\sum_{n,m}\sum_{\Delta}A^{mn}(\Delta)e^{i\boldsymbol{k}\cdot(\boldsymbol{\Delta}-\boldsymbol{\tau}_n+\boldsymbol{\tau}_m)}e^{i\frac{g_J\mu_B}{\hbar c_{lv}}\boldsymbol{A}_E(t)\cdot\left(\boldsymbol{\Delta}-\boldsymbol{\tau}_n+\boldsymbol{\tau}_m\right)}\hat{a}_{n}^{\dag}(\boldsymbol{k})\hat{a}_{m}(\boldsymbol{k})\\
	=\frac{1}{2}\sum_{\boldsymbol{k}}\sum_{n,m}B^{mn}\left[\boldsymbol{k}+\frac{g_J\mu_B}{\hbar c_{lv}}\boldsymbol{A}_E(t)\right]\hat{a}_{n}^{\dag}(\boldsymbol{k})\hat{a}_{m}(\boldsymbol{k})\\
	\end{split}
\end{equation}
Therefore, Under the action of a time-dependent electric field, the single-magnon Hamiltonian can be expressed as \cite{ref24,ref25}
\begin{equation}
	%\begin{split}
	\hat{H}=\hat{H}_0\left[-i\boldsymbol{\nabla}+\frac{g_J\mu_B}{\hbar c_{lv}}\boldsymbol{A}_E(t),\boldsymbol{r}\right],
	%\end{split}
\end{equation}
%Therefore, Under the action of a time-dependent electric field, the single-magnon Hamiltonian can be expressed as \cite{ref24,ref25}
%\begin{equation}
	%\begin{split}
%	\hat{H}(\boldsymbol{k})=\hat{H}_0\left[\boldsymbol{k}+\frac{g_J\mu_B}{c_{lv}}\boldsymbol{A}_E(t)\right],
	%\end{split}
%\end{equation}
in which $\hat{H}_0$ is the single-magnon Hamiltonian without an electric field. 

The time-dependent Schr\"odinger equation is
\begin{equation}
	%\begin{split}
	i\hbar\partial_t\ket{\psi(t)}=\hat{H}\ket{\psi(t)}.
	%\end{split}
\end{equation}
We take a gauge transformation $\hat{U}=e^{i\hat{\mathcal{K}}(t)}$ on $\ket{\psi(t)}$
\begin{equation}
	%\begin{split}
	\ket{\psi_g(t)}=\hat{U}\ket{\psi(t)}
	%\end{split}
\end{equation}
We require that the time-dependent Schr\"odinger equation after the transformation still holds, that is, 
\begin{equation}
	%\begin{split}
	i\hbar\partial_t\ket{\psi_g(t)}=\hat{H}_g\ket{\psi_g(t)}.
	%\end{split}
\end{equation}
The time-dependent Schr\"odinger equation transforms as
\begin{equation}
	\begin{split}
	i\hbar\partial_t\ket{\psi_g(t)}
	&=i\hbar\partial_t\hat{U}\ket{\psi(t)}
	=i\hbar\left(\partial_t\hat{U}\right)\ket{\psi(t)}+i\hbar \hat{U}\partial_t\ket{\psi(t)}\\
	&=i\hbar\left(\partial_t\hat{U}\right)\ket{\psi(t)}+ \hat{U}\hat{H}\ket{\psi(t)}\\
	&=i\hbar\left(\partial_t\hat{U}\right)\hat{U}^{\dag}\ket{\psi_g(t)}+ \hat{U}\hat{H}\hat{U}^{\dag}\ket{\psi_g(t)}\\
	&=\hat{H}_g\ket{\psi_g(t)}.
	\end{split}
\end{equation}
So $\hat{H}_g=i\hbar\left(\partial_t\hat{U}\right)\hat{U}^{\dag}+ \hat{U}\hat{H}\hat{U}^{\dag}$.
According to the Baker-Campbell-Hausdorff identity 
\begin{equation}
	\begin{split}
	\hat{U}\hat{H}\hat{U}^{\dag}
	&=e^{i\hat{\mathcal{K}}}\hat{H}e^{-i\hat{\mathcal{K}}}\\
	&=\hat{H}
	+i\left[\hat{\mathcal{K}},\hat{H}\right]
	-\frac{1}{2!}\left[\hat{\mathcal{K}},\left[\hat{\mathcal{K}},\hat{H}\right]\right]
	+...+
	\frac{i^n}{n!}\left[\hat{\mathcal{K}},...,\left[\hat{\mathcal{K}},\hat{H}\right]\right]+...
	\end{split}
\end{equation}
We take $\hat{\mathcal{K}}=\frac{g_J\mu_B}{\hbar c_{lv}}\boldsymbol{A}_{E}(t)\cdot\hat{\boldsymbol{r}}$. Here, $\hat{H}$ can expressed as $\hat{H}\left(-i\boldsymbol{\nabla}+\frac{g_J\mu_B}{\hbar c_{lv}}\boldsymbol{A}_E(t),\boldsymbol{r}\right)$, in which
\begin{equation}
	\begin{split}
		e^{i\hat{\mathcal{K}}}\boldsymbol{r}e^{-i\hat{\mathcal{K}}}
		=\boldsymbol{r},
	\end{split}
\end{equation}
and 
%\begin{equation}
%	\begin{split}
%		\left[\hat{\mathcal{K}},-i\partial_b-\frac{g_J\mu_B}{\hbar c_{lv}}A_{E,b}(t)\right]
%		=-\frac{g_J\mu_B}{\hbar c_{lv}}A_{E,a}(t)\left[r_a,-i\partial_b-\frac{g_J\mu_B}{\hbar c_{lv}}A_{E,b}(t)\right]
%		=-\frac{g_J\mu_B}{\hbar c_{lv}}A_{E,a}(t)i\delta_{ab}
%		=-\frac{ig_J\mu_B}{\hbar c_{lv}}A_{E,b}(t)
%		\\
%	\end{split}
%\end{equation}
\begin{equation}
	\begin{split}
		e^{i\hat{\mathcal{K}}}\left[-i\boldsymbol{\nabla}+\frac{g_J\mu_B}{\hbar c_{lv}}\boldsymbol{A}_E(t)\right]e^{-i\hat{\mathcal{K}}} 
		=-i\boldsymbol{\nabla}+\frac{g_J\mu_B}{\hbar c_{lv}}\boldsymbol{A}_E(t)
		-\frac{g_J\mu_B}{\hbar c_{lv}}\boldsymbol{A}_{E}(t)
		=-i\boldsymbol{\nabla}.
		\\
	\end{split}
\end{equation}
So 
\begin{equation}
	\begin{split}
		\hat{U}\hat{H}\hat{U}^{\dag}
		&=e^{i\hat{\mathcal{K}}}\hat{H}\left[-i\boldsymbol{\nabla}+\frac{g_J\mu_B}{\hbar c_{lv}}\boldsymbol{A}_E(t),\boldsymbol{r}\right]e^{-i\hat{\mathcal{K}}}\\
		&=\hat{H}\left(-i\boldsymbol{\nabla},\boldsymbol{r}\right)\\
		&=\hat{H}_0.
	\end{split}
\end{equation}
Therefore, under the gauge transformation, the Hamiltonian becomes
\begin{equation}
	\begin{split}
	\hat{H}_g
	&=i\hbar\left(\partial_t\hat{U}\right)\hat{U}^{\dag}+\hat{U}\hat{H}\hat{U}\\
	&=i\hbar\left(i\frac{g_J\mu_B}{\hbar c_{lv}}\partial_t\boldsymbol{A}_E(\boldsymbol{r})\cdot\boldsymbol{r}\right)+\hat{H}_0\\
	&=\hat{H}_0-\frac{g_J\mu_B}{c_{lv}}\partial_t\boldsymbol{A}_E(t)\cdot\boldsymbol{r}\\
	&=\hat{H}_0+\frac{g_J\mu_B}{c_{lv}}\tilde{\boldsymbol{E}}(t)\cdot\boldsymbol{r}.\\
	\end{split}
\end{equation}
in which $\tilde{\boldsymbol{E}}(t)=-\partial_t\boldsymbol{A}_E(t)=-\frac{1}{c}\partial_t\boldsymbol{E}(t)\times\boldsymbol{e}_z$ is the effective electric field.
We apply a plane wave perpendicular to a two-dimensional collinear ferromagnet, so the direction of the electric field vector of light is parallel to the two-dimensional ferromagnet. (We take the direction of light propagation as the negative direction of the $z$ axis.) On the two-dimensional ferromagnet, the electric field vector of a monochromatic plane wave is \cite{ref23}
\begin{equation}
	%\begin{split}
		E_{0,x}\cos(-\omega t+\phi_x)\boldsymbol{e}_x+E_{0,y}\cos(-\omega t+\phi_y)\boldsymbol{e}_y\\
		=\boldsymbol{E}(\omega)e^{-i\omega t}+\boldsymbol{E}(-\omega)e^{i\omega t}.
	%\end{split}
\end{equation}
Here, $\boldsymbol{E}(\omega)=\frac{1}{2}|\boldsymbol{E}_0|\boldsymbol{\epsilon}$ is the complex amplitude of the electric field, in which $|\boldsymbol{E}_0|=\sqrt{E_{0,x}^2+E_{0,y}^2}$, $\boldsymbol{\epsilon}=\frac{E_{0,x}e^{i\phi_{x}}}{|\boldsymbol{E}_0|}\boldsymbol{e}_x+\frac{E_{0,y}e^{i\phi_{y}}}{|\boldsymbol{E}_0|}\boldsymbol{e}_y$ is complex unit polarization vector and $\boldsymbol{E}(-\omega)=\boldsymbol{E}^{\ast}(\omega)$\cite{ref23,ref28.1}. 
Because we consider a plane wave that is superimposed by different monochromatic plane waves, the plane wave can be expressed as
\begin{equation}
	%\begin{split}
		\boldsymbol{E}(t)
		=\sum_i\boldsymbol{E}(\omega_i)e^{-i\omega_i t},
	%\end{split}
\end{equation}
in which $\omega_i$ contains positive frequency and negative frequency. Therefore, the effective electric field of the plane wave is
\begin{equation}
	%\begin{split}
	\tilde{\boldsymbol{E}}(t)
	=-\frac{1}{c_{lv}}\partial_t\boldsymbol{E}(t)\times\boldsymbol{e}_z
	=\frac{1}{c_{lv}}\sum_ii\omega_i\boldsymbol{E}(\omega_i)e^{-i\omega_i t}\times\boldsymbol{e}_z.
	%\end{split}
\end{equation}
To sum up, in our work, $\boldsymbol{E}_0$ is the amplitude of the electric field components of light, $\boldsymbol{E}$ is the electric field complex amplitude, and $\tilde{\boldsymbol{E}}$ is the effective electric field. 
\section{Phenomenological representation of MS(E)PC}\label{appendixPhenomenological}
As we discussed in Subsec. \ref{Phenomenological representation of magnon photo-transport}, the magnon spin photoconductivity and magnon energy photoconductivity can be expressed as $\chi_{ab}^{(E)}(\omega)=\chi_{ab}^{(E)i}(\omega)+\chi_{ab}^{(E)e}(\omega)$.
So MSPC (MEPC) under a monochromatic light can be expressed as
\begin{equation}
	\begin{split}
		j_{a}^{(E)(1)}(\omega)
		&=\sum_{b}\left[\chi_{ab}^{(E)}(\omega)E_b(\omega)e^{-i\omega t}+\chi_{ab}^{(E)}(-\omega)E_b(-\omega)e^{i\omega t}\right]\\
		&=\sum_{b}\left[\chi_{ab}^{(E)}(\omega)E_b(\omega)e^{-i\omega t}+\chi_{ab}^{(E)\ast}(\omega)E_b^{\ast}(\omega)e^{i\omega t}\right]\\
		&=\sum_{b}\left\{\left[Re\left(\chi_{ab}^{(E)}(\omega)\right)+iIm\left(\chi_{ab}^{(E)}(\omega)\right)\right]\frac{1}{2}E_{0,b}e^{i\phi_b}e^{-i\omega t}
		+\left[Re\left(\chi_{ab}^{(E)}(\omega)\right)-iIm\left(\chi_{ab}^{(E)}(\omega)\right)\right]\frac{1}{2}E_{0,b}e^{-i\phi_b}e^{i\omega t}\right\}\\
		&=\frac{1}{2}\sum_{b}\left\{\left[Re\left(\chi_{ab}^{(E)}(\omega)\right)+iIm\left(\chi_{ab}^{(E)}(\omega)\right)\right]e^{i(-\omega t+\phi_b)}
		+\left[Re\left(\chi_{ab}^{(E)}(\omega)\right)-iIm\left(\chi_{ab}^{(E)}(\omega)\right)\right]e^{-i(-\omega t+\phi_b)}\right\}E_{0,b}\\
		&=\sum_{b}\left[Re\left(\chi_{ab}^{(E)}(\omega)\right)\cos(-\omega t+\phi_b)-Im\left(\chi_{ab}^{(E)}(\omega)\right)\sin(-\omega t+\phi_b)\right]E_{0,b}.\\
	\end{split}
\end{equation}
Here we use the relation $\chi_{ab}^{(E)}(-\omega)=\chi_{ab}^{(E)\ast}(\omega)$, which can be obtained from Eq. \ref{chi(1)main} and \ref{chi(1)Emain}.

\section{Calculation details of the density matrix} \label{Calculation details of the density matrix}
\subsection{Magnon scattering effect}\label{Magnon scattering effect}
We denote the second-quantized magnon Hamiltonian by $\hat{h}=\hat{h}_0+\hat{v}$. Here, $\hat{h}_0=\sum_{n\boldsymbol{k}}\varepsilon_n(\boldsymbol{k})\hat{a}_n^{\dag}(\boldsymbol{k})\hat{a}_n(\boldsymbol{k})$ is the one-body part, and $\hat{v}$ is the many-body part.
The many-body part describes interactions between quasi-particles, such as magnon-phonon interactions and magnon-magnon interactions. 
It is the source of scattering effects and relaxation times.
Because the relaxation time method is a phenomenological approach, we did not clearly classify the scattering effects of magnons in a specific manner.
The retarded Green’s function is
\begin{equation}
	\begin{split}
		\langle\langle \hat{a}_n(\boldsymbol{k}); \hat{a}_n^{\dag}(\boldsymbol{k}) \rangle\rangle^{R}(t)
		&=-i\theta(t)\braket{\left[\hat{a}_n(\boldsymbol{k},t),\hat{a}_n^{\dag}(\boldsymbol{k},0)\right]}.\\
	\end{split}
\end{equation}
In this subsection, we use the Heisenberg picture. For brevity, we abbreviate $\hat{a}_n(\boldsymbol{k},0)$ as $\hat{a}_n(\boldsymbol{k})$.
The dynamical equation of retarded Green function is
\begin{equation}
	\begin{split}
		i\hbar\partial_t\langle\langle \hat{a}_n(\boldsymbol{k}); \hat{a}_n^{\dag}(\boldsymbol{k}) \rangle\rangle^R(t)
		&=\hbar\delta(t)\braket{\left[\hat{a}_n(\boldsymbol{k},t), \hat{a}_n^{\dag}(\boldsymbol{k})\right]}
		+\langle\langle \left[\hat{a}_n(\boldsymbol{k}),\hat{h}\right]; \hat{a}_n^{\dag}(\boldsymbol{k}) \rangle\rangle^R\\
		&=\hbar\delta(t)
		+\langle\langle \left[\hat{a}_n(\boldsymbol{k}),\hat{h}_0\right]; \hat{a}_n^{\dag}(\boldsymbol{k}) \rangle\rangle^R
		+\langle\langle \left[\hat{a}_n(\boldsymbol{k}),\hat{v}\right]; \hat{a}_n^{\dag}(\boldsymbol{k}) \rangle\rangle^R.\\
	\end{split}
\end{equation}
After the Fourier transformation, the dynamic equation of the frequency-domain retarded Green's function is
\begin{equation}
	\begin{split}
		\hbar \omega \langle\langle \hat{a}_n(\boldsymbol{k}); \hat{a}_n^{\dag}(\boldsymbol{k}) \rangle\rangle^R(\omega)
		&=\hbar
		+\varepsilon_n(\boldsymbol{k})\langle\langle \hat{a}_n(\boldsymbol{k}); \hat{a}_n^{\dag}(\boldsymbol{k}) \rangle\rangle^R(\omega)
		+\langle\langle \left[\hat{a}_n(\boldsymbol{k}),\hat{v}\right]; \hat{a}_n^{\dag}(\boldsymbol{k}) \rangle\rangle^R(\omega)\\
		&=\hbar
		+\varepsilon_n(\boldsymbol{k})\langle\langle \hat{a}_n(\boldsymbol{k}); \hat{a}_n^{\dag}(\boldsymbol{k}) \rangle\rangle^R(\omega)
		+\Sigma_n^R(\boldsymbol{k},\omega)\langle\langle \hat{a}_n(\boldsymbol{k}); \hat{a}_n^{\dag}(\boldsymbol{k}) \rangle\rangle^R(\omega).\\
	\end{split}
	\label{fdde}
\end{equation}
Here, $\langle\langle \left[\hat{a}_n(\boldsymbol{k}),\hat{v}\right]; \hat{a}_n^{\dag}(\boldsymbol{k}) \rangle\rangle^R(\omega)=\Sigma_n^R(\boldsymbol{k},\omega)\langle\langle \hat{a}_n(\boldsymbol{k}); \hat{a}_n^{\dag}(\boldsymbol{k}) \rangle\rangle^R(\omega)$, in which $\Sigma_n^R(\boldsymbol{k},\omega)$ is self-energy.
The self-energy describes the scattering effect of magnons.
So, when the scattering effect is not considered, the frequency-domain retarded Green's function is
\begin{equation}
	\begin{split}
		\langle\langle \hat{a}_n(\boldsymbol{k}); \hat{a}_n^{\dag}(\boldsymbol{k}) \rangle\rangle^{R(0)}(\omega)
		&=\frac{\hbar}{\hbar \omega-\varepsilon_n(\boldsymbol{k})+i0^+}.
	\end{split}
\end{equation}
So, the spectral function is 
\begin{equation}
	\begin{split}
		A_n(\boldsymbol{k},\omega)
		&=-\frac{1}{\pi}Im\left[\langle\langle \hat{a}_n(\boldsymbol{k}); \hat{a}_n^{\dag}(\boldsymbol{k}) \rangle\rangle^{R(0)}(\omega)\right]\\
		&=-\frac{1}{\pi}Im\left[\frac{\hbar}{\hbar \omega-\varepsilon_n(\boldsymbol{k})+i0^+}\right]\\
		&=\hbar\delta(\hbar \omega-\varepsilon_n(\boldsymbol{k})).\\
	\end{split}
	\label{rGf(0)}
\end{equation}
It means that magnons have a well-defined energy, and according to the time-energy uncertainty relation, their lifetime is infinite in this case.
Considering the interaction effect, the frequency-domain retarded Green's function is
 \begin{equation}
 	\begin{split}
 		\langle\langle \hat{a}_n(\boldsymbol{k}); \hat{a}_n^{\dag}(\boldsymbol{k}) \rangle\rangle^R(\omega)
 		&=\frac{\hbar}{\hbar \omega-\varepsilon_n(\boldsymbol{k})-\Sigma_{n}^R(\boldsymbol{k},\omega)}.\\
 	\end{split}
 	\label{rGf}
 \end{equation}
Here, $\Sigma_{n}^R(\boldsymbol{k},\omega)$ is the self-energy in Eq. (\ref{fdde}) which describes the interaction effect.
We simplify the self energy to $\braket{\Sigma_n^R(\boldsymbol{k},\omega)}\approx-\frac{i\hbar}{2}\Gamma_{nn}$ \cite{ref28.2}.
%Since the self-energy describes scattering processes, it can be simplified to $\braket{\Sigma_n^R(\boldsymbol{k},\omega)}\approx-\frac{i\hbar}{2\tau_n}=-\frac{i\hbar}{2}\Gamma_{nn}$ .
%Here, $\tau_n$ is magnon relaxation time in band $n$, and $\Gamma_{nn}=\frac{1}{\tau_n}$.
%
The spectral function changes to   
\begin{equation}
	\begin{split}
		A_n(\boldsymbol{k},\omega)
		&=-\frac{\hbar Im\left[\Sigma_{n}^R(\boldsymbol{k},\omega)\right]}{\pi\left(\hbar \omega-\varepsilon_m(\boldsymbol{k})\right)^2+\pi Im\left[\Sigma_{n}^R(\boldsymbol{k},\omega)\right]^2}\\
		&=\frac{\Gamma_{n}/2}{\pi\left( \omega-\varepsilon_m(\boldsymbol{k})/\hbar\right)^2+\pi(\Gamma_{n}/2)^2}.\\
	\end{split}
\end{equation}
It is no longer a $\delta$ function, but now takes the form of a resonance peak function with a peak width of $\Gamma_{n}$. 
As a result, the relaxation time $\tau_n$ can be defined as $\tau_n=1/\Gamma_{n}$ based on the energy-time uncertainty relation.
Therefore, the scattering effect of magnons can be expressed by the relaxation time $\tau_n$, that is, $\braket{\Sigma_n^R(\boldsymbol{k},\omega)}\approx-\frac{i\hbar}{2\tau_n}=-\frac{i\hbar}{2}\Gamma_{n}$.
Furthermore, we can make an additional approximation by assuming that the magnon relaxation time is independent of the band, $\Gamma_n=\Gamma=\frac{1}{\tau}$.

Compared with Eq. (\ref{rGf(0)}), Eq. (\ref{rGf}) indicate that self-energy makes corrections to the magnon band.
So, we introduce a scattering-induced perturbation to the single-magnon Hamiltonian
\begin{equation}
	\begin{split}
		\hat{H}=\hat{H}_0+\hat{U}.
	\end{split}
\end{equation}
Here, $\hat{H}_0$ describes the intrinsic properties of the magnon, and $\hat{U}$ is the scattering-induced perturbation (Non-Hermitian part).
The Hamiltonian in the crystal momentum $\boldsymbol{k}$ representation is $\hat{H}(\boldsymbol{k})=e^{i\boldsymbol{k}\cdot\boldsymbol{r}}\hat{H}e^{-i\boldsymbol{k}\cdot\boldsymbol{r}}=\hat{H}_0(\boldsymbol{k})+\hat{U}(\boldsymbol{k})$.
So, in Bloch representation, the quantum Liouville equation can be expressed as \cite{ref23}
\begin{equation}
	\begin{split}
		\partial_t\hat{\rho}(t)
		=\frac{1}{i\hbar}\left[\hat{H}_0+\hat{U},\hat{\rho}(t)\right]
		=\frac{1}{i\hbar}\left[\hat{H}_0,\hat{\rho}(t)\right]
		+\frac{1}{i\hbar}\left[\hat{U},\hat{\rho}(t)\right].
	\end{split}
\end{equation}
Here, $\frac{1}{i\hbar}\left[\hat{H}_0,\hat{\rho}(t)\right]$ implies a reversible process without scattering, and $\frac{1}{i\hbar}\left[\hat{U},\hat{\rho}(t)\right]$ implies the irreversible process caused by scattering.
The irreversible processes can be represented through relaxation time phenomenologically
$\frac{1}{i\hbar}\left[\hat{H}_0,\hat{\rho}(t)\right]=-\frac{\hat{\rho}(t)-\hat{\rho}^{(0)}}{\tau}$, in which $\hat{\rho}^{(0)}$ is magnon density in the equilibrium state.

\subsection{Quantum Liouville equation}\label{Quantum Liouville equation}

The density matrix of single-magnon is $\hat{\rho}(t)$, we can calculate the correction of the density matrix by the quantum Liouville equation (Eq. (\ref{QLE})) \cite{ref19,ref22.1,ref22.3,ref23}
\begin{equation}
	%\begin{split}
	\partial_t\hat{\rho}(t)
	=\frac{1}{i\hbar}\left[\hat{H}_0+\hat{H}_E^{\prime}(t)+\hat{U},\hat{\rho}(t)\right]
	=\frac{1}{i\hbar}\left[\hat{H}_0+\hat{H}_E^{\prime}(t),\hat{\rho}(t)\right]-\frac{\left[\hat{\rho}(t)-\hat{\rho}^{(0)}\right]}{\tau}.
	%\end{split}
\end{equation}
Here, $\hat{H}_0$ describes the intrinsic properties of the system (magnetic system), $\hat{H}_E^{\prime}(t)$ is the perturbation term of the Hamiltonian, and $\hat{U}$ is the scattering term. In phenomenological, $\frac{1}{i\hbar}\left[\hat{U},\hat{\rho}(t)\right]=-\frac{\left[\hat{\rho}(t)-\hat{\rho}^{(0)}\right]}{\tau}$\cite{ref23}.
Then we take $\Gamma=\frac{1}{\tau}$, in which $\tau$ is the relaxation time of magnons \cite{ref9.1,ref10}. 
We assume that light (time-dependent electric field) is applied when $t=0$, and we ignore the effect of scattering when $t<0$ \cite{ref23}.
$\hat{\rho}^{(0)}$ is the equilibrium density matrix.
For ease of derivation, we take damping density matrix $\hat{\tilde{\rho}}(t)=e^{\Gamma t}\hat{\rho}(t)$ \cite{ref19,ref23}.
In the interaction picture, the operator can be expressed as $\hat{O}_I(t)=e^{i\hat{H}_0t/\hbar}\hat{O}e^{-i\hat{H}_0t/\hbar}$. We consider the quantum Liouville equation in the picture of interaction  \cite{ref23}
{\small
\begin{equation}
	\begin{split}
	i\hbar\partial_t\hat{\tilde{\rho}}_I(t)
	&=i\hbar\frac{de^{\Gamma t}e^{i\hat{H}_0t/\hbar}\hat{\rho}(t)e^{-i\hat{H}_0t/\hbar}}{dt}\\
	&=i\hbar\left[\Gamma e^{\Gamma t}e^{i\hat{H}_0t/\hbar}\hat{\rho}(t)e^{-i\hat{H}_0t/\hbar}+e^{\Gamma t}e^{i\hat{H}_0t/\hbar}\frac{i}{\hbar}\hat{H}_0\hat{\rho}(t)e^{-i\hat{H}_0t/\hbar}+e^{\Gamma t}e^{i\hat{H}_0t/\hbar}\left(\partial_t\hat{\rho}(t)\right)e^{-i\hat{H}_0t/\hbar}-e^{\Gamma t}e^{i\hat{H}_0t/\hbar}\hat{\rho}(t)\frac{i}{\hbar}\hat{H}_0e^{-i\hat{H}_0t/\hbar}\right].\\
	\end{split}
	\label{solve1}
\end{equation}}
We substitute the quantum Liouville equation into Eq. (\ref{solve1})
{\small
\begin{equation}
	\begin{split}
		&i\hbar\partial_t\hat{\tilde{\rho}}_I(t)\\
		&=i\hbar\left\{\Gamma e^{\Gamma t}e^{i\hat{H}_0t/\hbar}\hat{\rho}(t)e^{-i\hat{H}_0t/\hbar}
		+\frac{1}{i\hbar}e^{\Gamma t}e^{i\hat{H}_0t/\hbar}\left[\hat{\rho}(t),\hat{H}_0\right]e^{-i\hat{H}_0t/\hbar}+e^{\Gamma t}e^{i\hat{H}_0t/\hbar}\left[\frac{1}{i\hbar}\left[\hat{H}_0+\hat{H}_E(t),\hat{\rho}(t)\right]-\Gamma\left(\hat{\rho}(t)-\hat{\rho}^{(0)}\right)\right]e^{-i\hat{H}_0t/\hbar}\right\}\\
		&=\left[\hat{H}_{E,I}(t),\hat{\tilde{\rho}}_I(t)\right]+i\hbar\Gamma\hat{\tilde{\rho}}^{(0)}.
	\end{split}
	\label{solve2}
\end{equation}}
So the density matrix can be expressed as
\begin{equation}
	\begin{split}
		\hat{\tilde{\rho}}_I(t)
		&=\frac{1}{i\hbar}\int_0^tdt^{\prime}\left[\hat{H}_{E,I}(t^{\prime}),\hat{\tilde{\rho}}_I(t^{\prime})\right]+\Gamma\hat{\tilde{\rho}}^{(0)}.
	\end{split}
	\label{solve3}
\end{equation}
To calculate the MS(E)PC for each order of perturbation, we need to take the density matrix at each order. Because $\hat{H}_{E,I}^{\prime}(t)$ is the perturbation, the first order density matrix is
\begin{equation}
	\begin{split}
		\hat{\tilde{\rho}}_I^{(1)}(t)
		&=\frac{1}{i\hbar}\int_0^tdt^{\prime}\left[\hat{H}_{E,I}(t^{\prime}),\hat{\tilde{\rho}}_I^{(0)}(t^{\prime})\right].
	\end{split}
	\label{solve4}
\end{equation}
Because $\hat{\tilde{\rho}}(t)=e^{\Gamma t}\hat{\rho}(t)$, the first order of density matrix $\hat{\rho}(t)$ can be expressed as Eq. (\ref{rhoI(1)})
\begin{equation}
	\begin{split}
		\hat{\rho}_I^{(1)}(t)
		&=\frac{e^{-\Gamma t}}{i\hbar}\int_0^tdt^{\prime}\left[\hat{H}_{E,I}(t^{\prime}),e^{\Gamma t}\hat{\rho}_I^{(0)}(t^{\prime})\right].
	\end{split}
	\label{solve5}
\end{equation} 
\subsection{The first-order correction of the density matrix in Bloch representation}
\subsubsection{Covariant derivative}
Before deriving the density matrix in detail, we introduce the Covariant derivative \cite{ref25,ref29,ref30,ref31}. 
In Bloch representation, the commutation between the position operator $\boldsymbol{r}$ and the operator $\hat{O}$ satisfied $\bra{\psi_m(\boldsymbol{k}^{\prime})}\hat{O}\ket{\psi_n(\boldsymbol{k})}=\bra{\psi_m(\boldsymbol{k})}\hat{O}\ket{\psi_n(\boldsymbol{k})}\delta(\boldsymbol{k}-\boldsymbol{k}^{\prime})$ is \cite{ref23}
\begin{equation}
	\begin{split}
		&\bra{\psi_m(\boldsymbol{k}^{\prime})}\left[\boldsymbol{r},\hat{O}\right]\ket{\psi_n(\boldsymbol{k})}\\
		&=\sum_{n_1}\int d^dk\left[\bra{\psi_m(\boldsymbol{k}^{\prime})}\boldsymbol{r}\ket{\psi_{n_1}(\boldsymbol{k}_1)}\bra{\psi_{n_1}(\boldsymbol{k}_1)}\hat{O}\ket{\psi_n(\boldsymbol{k})}-\bra{\psi_m(\boldsymbol{k}^{\prime})}\hat{O}\ket{\psi_{n_1}(\boldsymbol{k}_1)}\bra{\psi_{n_1}(\boldsymbol{k}_1)}\boldsymbol{r}\ket{\psi_n(\boldsymbol{k})}\right]\\
		&=\sum_{n_1}\int d^dk\left\{\left[i\partial_{\boldsymbol{k}^{\prime}}\delta(\boldsymbol{k}_1-\boldsymbol{k}^{\prime})\delta_{mn_1}+\boldsymbol{\mathcal{A}}_{mn_1}(\boldsymbol{k}^{\prime})\delta(\boldsymbol{k}_1-\boldsymbol{k}^{\prime})\right]\bra{\psi_{n_1}(\boldsymbol{k})}\hat{O}\ket{\psi_n(\boldsymbol{k})}\delta(\boldsymbol{k}-\boldsymbol{k}_1)\right.\\
		&\left.-\bra{\psi_m(\boldsymbol{k}^{\prime})}\hat{O}\ket{\psi_{n_1}(\boldsymbol{k}^{\prime})}\delta(\boldsymbol{k}_1-\boldsymbol{k}^{\prime})\left[i\partial_{\boldsymbol{k}_1}\delta(\boldsymbol{k}-\boldsymbol{k}_1)\delta_{n_1n}+\boldsymbol{\mathcal{A}}_{n_1n}(\boldsymbol{k}^{\prime})\delta(\boldsymbol{k}-\boldsymbol{k}_1)\right]\right\}\\
		&=\sum_{n_1}\int d^dk_1i\partial_{\boldsymbol{k}^{\prime}}\delta(\boldsymbol{k}_1-\boldsymbol{k}^{\prime})\delta_{mn_1}\bra{\psi_{n_1}(\boldsymbol{k})}\hat{O}\ket{\psi_n(\boldsymbol{k})}\delta(\boldsymbol{k}-\boldsymbol{k}_1)\\
		&+\sum_{n_1}\int d^dk_1\boldsymbol{\mathcal{A}}_{mn_1}(\boldsymbol{k}^{\prime})\delta(\boldsymbol{k}_1-\boldsymbol{k}^{\prime})\bra{\psi_{n_1}(\boldsymbol{k})}\hat{O}\ket{\psi_n(\boldsymbol{k})}\delta(\boldsymbol{k}-\boldsymbol{k}_1)\\
		&-\sum_{n_1}\int d^dk_1\bra{\psi_m(\boldsymbol{k}^{\prime})}\hat{O}\ket{\psi_{n_1}(\boldsymbol{k}^{\prime})}\delta(\boldsymbol{k}_1-\boldsymbol{k}^{\prime})i\partial_{\boldsymbol{k}_1}\delta(\boldsymbol{k}-\boldsymbol{k}_1)\delta_{n_1n}\\
		&-\sum_{n_1}\int d^dk_1\bra{\psi_m(\boldsymbol{k}^{\prime})}\hat{O}\ket{\psi_{n_1}(\boldsymbol{k}^{\prime})}\delta(\boldsymbol{k}_1-\boldsymbol{k}^{\prime})\boldsymbol{\mathcal{A}}_{n_1n}(\boldsymbol{k}^{\prime})\delta(\boldsymbol{k}-\boldsymbol{k}_1)\\
		&=\left[O_{mn}(\boldsymbol{k})-O_{mn}(\boldsymbol{k}^{\prime})\right]i\partial_{\boldsymbol{k}^{\prime}}\delta(\boldsymbol{k}-\boldsymbol{k}^{\prime})
		+\sum_{n_1}\left[\boldsymbol{\mathcal{A}}_{mn_1}(\boldsymbol{k}^{\prime})O_{n_1n}(\boldsymbol{k})
		-\boldsymbol{\mathcal{A}}_{n_1n}(\boldsymbol{k}^{\prime})O_{mn_1}(\boldsymbol{k}^{\prime})\right]\delta(\boldsymbol{k}-\boldsymbol{k}^{\prime})\\
		&=i\partial_{\boldsymbol{k}^{\prime}}\left[O_{mn}(\boldsymbol{k})-O_{mn}(\boldsymbol{k}^{\prime})\right]\delta(\boldsymbol{k}-\boldsymbol{k}^{\prime})
		+\delta(\boldsymbol{k}-\boldsymbol{k}^{\prime})i\partial_{\boldsymbol{k}^{\prime}}O_{mn}(\boldsymbol{k}^{\prime})
		+\sum_{n_1}\left[\boldsymbol{\mathcal{A}}_{mn_1}(\boldsymbol{k}^{\prime})O_{n_1n}(\boldsymbol{k})
		-\boldsymbol{\mathcal{A}}_{n_1n}(\boldsymbol{k}^{\prime})O_{mn_1}(\boldsymbol{k}^{\prime})\right]\delta(\boldsymbol{k}-\boldsymbol{k}^{\prime})\\
		&=\delta(\boldsymbol{k}-\boldsymbol{k}^{\prime})\left\{i\partial_{\boldsymbol{k}}O_{mn}(\boldsymbol{k})
		+\sum_{n_1}\left[\boldsymbol{\mathcal{A}}_{mn_1}(\boldsymbol{k})O_{n_1n}(\boldsymbol{k})
		-\boldsymbol{\mathcal{A}}_{n_1n}(\boldsymbol{k})O_{mn_1}(\boldsymbol{k})\right]\right\}.\\
		\label{[r,O]}
	\end{split}
\end{equation}
Here, $O_{mn}(\boldsymbol{k})=\bra{\psi_m(\boldsymbol{k})}\hat{O}\ket{\psi_n(\boldsymbol{k})}=\bra{u_m(\boldsymbol{k})}\hat{O}(\boldsymbol{k})\ket{u_n(\boldsymbol{k})}$, in which $\hat{O}(\boldsymbol{k})=e^{-i\boldsymbol{k}\cdot\boldsymbol{r}}\hat{O}e^{i\boldsymbol{k}\cdot\boldsymbol{r}}$.
$\boldsymbol{\mathcal{A}}_{mn}(\boldsymbol{k})=\bra{u_m(\boldsymbol{k})}i\partial_{\boldsymbol{k}}\ket{u_n(\boldsymbol{k})}$ is Berry connection of magnon, in which $\ket{u_n(\boldsymbol{k})}=e^{-i\boldsymbol{k}\cdot\boldsymbol{r}}\ket{\psi_n(\boldsymbol{k})}$.
$d$ is the dimension of the system.
So the covariant derivative can be expressed as \cite{ref25,ref25.1}
\begin{equation}
	\begin{split}
		\left[\boldsymbol{D}_{\boldsymbol{k}}O(\boldsymbol{k})\right]_{mn}
		&=\bra{u_m(\boldsymbol{k})}\left[\boldsymbol{r},\hat{O}(\boldsymbol{k})\right]\ket{u_n(\boldsymbol{k})}\\
		&=\bra{\psi_m(\boldsymbol{k})}\left[\boldsymbol{r},\hat{O}\right]\ket{\psi_n(\boldsymbol{k})}\\
		&=i\partial_{\boldsymbol{k}}O_{mn}(\boldsymbol{k})
		+\sum_{n_1}\left[\boldsymbol{\mathcal{A}}_{mn_1}(\boldsymbol{k})O_{n_1n}(\boldsymbol{k})
		-\boldsymbol{\mathcal{A}}_{n_1n}(\boldsymbol{k})O_{mn_1}(\boldsymbol{k})\right]\\
		\label{Dk}
	\end{split}
\end{equation}
which satisfy $\bra{\psi_m(\boldsymbol{k}^{\prime})}\left[\boldsymbol{r},\hat{O}\right]\ket{\psi_n(\boldsymbol{k})}=\left[\boldsymbol{D}_{\boldsymbol{k}}O(\boldsymbol{k})\right]_{mn}\delta(\boldsymbol{k}-\boldsymbol{k}^{\prime})$.
If $\boldsymbol{k}$ is regarded as discrete, Eq. \ref{[r,O]} satisfies \cite{ref23}
\begin{equation}
	%\begin{split}
		\bra{\psi_m(\boldsymbol{k}^{\prime})}\left[\boldsymbol{r},\hat{O}\right]\ket{\psi_n(\boldsymbol{k})}
		=\left[\boldsymbol{D}_{\boldsymbol{k}}O(\boldsymbol{k})\right]_{mn}\delta_{\boldsymbol{k}\boldsymbol{k}^{\prime}}.
	%\end{split}
\end{equation}
It means that $\bra{\psi_m(\boldsymbol{k})}\left[\boldsymbol{r},\hat{O}\right]\ket{\psi_n(\boldsymbol{k})}
=\left[\boldsymbol{D}_{\boldsymbol{k}}O(\boldsymbol{k})\right]_{mn}$, 
and the covariant derivative can be divided into the intraband part $\left[\boldsymbol{D}_{\boldsymbol{k}}O(\boldsymbol{k})\right]_{mn}^i$ and interband part $\left[\boldsymbol{D}_{\boldsymbol{k}}O(\boldsymbol{k})\right]_{mn}^e$
\begin{equation}
	\begin{cases}
	\left[\boldsymbol{D}_{\boldsymbol{k}}O(\boldsymbol{k})\right]_{mn}^i
	=i\partial_{\boldsymbol{k}}O_{mn}(\boldsymbol{k})\\
	\left[\boldsymbol{D}_{\boldsymbol{k}}O(\boldsymbol{k})\right]_{mn}^e
	=\sum_{n_1}\left[\boldsymbol{\mathcal{A}}_{mn_1}(\boldsymbol{k})O_{n_1n}(\boldsymbol{k})
	-\boldsymbol{\mathcal{A}}_{n_1n}(\boldsymbol{k})O_{mn_1}(\boldsymbol{k})\right].\\
	\end{cases}
	\label{Dkie}
\end{equation}
\subsubsection{\label{First-order correction of the density matrix}The first-order correction of the density matrix}

In Bloch representation, the zero-order density matrix is 
\begin{equation}
	%\begin{split}
	\bra{\psi_m(\boldsymbol{k}^{\prime})}\hat{\rho}^{(0)}\ket{\psi_n(\boldsymbol{k})}
	=f_n^B(\boldsymbol{k})\delta_{mn}\delta_{\boldsymbol{k}\boldsymbol{k}^{\prime}},
	%\end{split}
\end{equation}
in which $f_n^B(\boldsymbol{k})=1/\left[e^{\varepsilon_n(\boldsymbol{k})/k_BT}-1\right]$ is Bose distribution \cite{ref23,ref25}. Here, because the magnon number is not conserved, the chemical potential is zero.
The first-order correction of the density matrix is \cite{ref23}
\begin{equation}
	\begin{split}
		\bra{\psi_m(\boldsymbol{k}^{\prime})}\hat{\rho}^{(1)}(t)\ket{\psi_n(\boldsymbol{k})}
		&=\bra{\psi_m(\boldsymbol{k}^{\prime})}e^{-i\hat{H}_0t/\hbar}\hat{\rho}_I^{(1)}(t)e^{i\hat{H}_0t/\hbar}\ket{\psi_n(\boldsymbol{k})}\\
		&=e^{i\varepsilon_{n}(\boldsymbol{k}^{\prime})t/\hbar}e^{-i\varepsilon_{m}(\boldsymbol{k})t/\hbar}\bra{\psi_m(\boldsymbol{k}^{\prime})}\hat{\rho}_I^{(1)}(t)\ket{\psi_n(\boldsymbol{k})}\\
		&=\frac{e^{-\Gamma t}e^{i\varepsilon_{n}(\boldsymbol{k}^{\prime})t/\hbar}e^{-i\varepsilon_{m}(\boldsymbol{k})t/\hbar}}{i\hbar}\int_0^tdt^{\prime}\bra{\psi_m(\boldsymbol{k}^{\prime})}\left[\hat{H}_{IE}^{\prime}(t^{\prime}),e^{\Gamma t^{\prime}}\hat{\rho}_I^{(0)}\right]\ket{\psi_n(\boldsymbol{k})}\\
		&=\frac{e^{-\Gamma t}e^{i\varepsilon_{n}(\boldsymbol{k}^{\prime})t/\hbar}e^{-i\varepsilon_{m}(\boldsymbol{k})t/\hbar}}{i\hbar}\int_0^tdt^{\prime}\bra{\psi_m(\boldsymbol{k}^{\prime})}e^{i\hat{H}_0t^{\prime}/\hbar}\left[\hat{H}_{E}^{\prime}(t^{\prime}),e^{\Gamma t^{\prime}}\hat{\rho}^{(0)}\right]e^{-i\hat{H}_0t^{\prime}/\hbar}\ket{\psi_n(\boldsymbol{k})}\\
		&=\frac{e^{-\Gamma t}e^{i\varepsilon_{n}(\boldsymbol{k}^{\prime})t/\hbar}e^{-i\varepsilon_{m}(\boldsymbol{k})t/\hbar}}{i\hbar}\int_0^tdt^{\prime}e^{i\omega_m(\boldsymbol{k}^{\prime})t}e^{-i\omega_n(\boldsymbol{k})t}\bra{\psi_m(\boldsymbol{k}^{\prime})}\left[\hat{H}_{E}^{\prime}(t^{\prime}),e^{\Gamma t^{\prime}}\hat{\rho}^{(0)}\right]\ket{\psi_n(\boldsymbol{k})}\\
		&=\frac{g_J\mu_B}{i\hbar c_{lv}}e^{-\Gamma t}e^{i\varepsilon_{n}(\boldsymbol{k}^{\prime})t/\hbar}e^{-i\varepsilon_{m}(\boldsymbol{k})t/\hbar}\int_0^tdt^{\prime}e^{i\omega_m(\boldsymbol{k}^{\prime})t^{\prime}}e^{-i\omega_n(\boldsymbol{k})t^{\prime}}e^{\Gamma t^{\prime}}\tilde{\boldsymbol{E}}(t^{\prime})\cdot\bra{\psi_m(\boldsymbol{k}^{\prime})}\left[\boldsymbol{r},\hat{\rho}^{(0)}\right]\ket{\psi_n(\boldsymbol{k})}\\
		&=\frac{g_J\mu_B}{i\hbar c_{lv}}e^{-\Gamma t}e^{i\varepsilon_{nm}t/\hbar}\int_0^tdt^{\prime}e^{i\omega_m(\boldsymbol{k}^{\prime})t^{\prime}}e^{-i\omega_n(\boldsymbol{k})t^{\prime}}e^{\Gamma t^{\prime}}\tilde{\boldsymbol{E}}(t^{\prime})\cdot
		\left[\boldsymbol{D}_{\boldsymbol{k}}\rho^{(0)}(\boldsymbol{k})\right]_{mn}\delta_{\boldsymbol{k}\boldsymbol{k}^{\prime}}\\
		&=\frac{g_J\mu_B}{i\hbar c_{lv}}e^{-\Gamma t}e^{i\varepsilon_{nm}t/\hbar}\int_0^tdt^{\prime}e^{i\omega_m(\boldsymbol{k}^{\prime})t^{\prime}}e^{-i\omega_n(\boldsymbol{k})t^{\prime}}e^{\Gamma t^{\prime}}\tilde{\boldsymbol{E}}(t^{\prime})\\
		&\cdot\left\{i\partial_{\boldsymbol{k}}f_n(\boldsymbol{k})\delta_{mn}
		+\sum_{n_1}\left[\boldsymbol{\mathcal{A}}_{mn_1}(\boldsymbol{k})f_n(\boldsymbol{k})\delta_{n_1n}
		-\boldsymbol{\mathcal{A}}_{n_1n}(\boldsymbol{k})f_m(\boldsymbol{k})\delta_{mn_1}\right]\right\}\delta_{\boldsymbol{k}\boldsymbol{k}^{\prime}}\\
		&=\frac{g_J\mu_B}{i\hbar c_{lv}}e^{-\Gamma t}e^{i\varepsilon_{nm}t/\hbar}\int_0^tdt^{\prime}e^{i\omega_{mn}(\boldsymbol{k})t^{\prime}}e^{\Gamma t^{\prime}}\tilde{\boldsymbol{E}}(t^{\prime})\cdot
		\left[i\partial_{\boldsymbol{k}}f_n(\boldsymbol{k})\delta_{mn}
		+\boldsymbol{\mathcal{A}}_{mn}(\boldsymbol{k})f_{nm}(\boldsymbol{k})
		\right]\delta_{\boldsymbol{k}\boldsymbol{k}^{\prime}}\\
		&=\frac{g_J\mu_B}{\hbar c_{lv}}\sum_i\frac{e^{-i\omega_it}-e^{-i(-\omega_{nm}(\mathbf{k})-i\Gamma) t}}{\omega_i-\omega_{mn}+i\Gamma}\tilde{\boldsymbol{E}}(\omega_i)\cdot\left[i\partial_{\mathbf{k}}f_n(\boldsymbol{k})\delta_{mn}
		+\boldsymbol{\mathcal{A}}_{mn}(\mathbf{k}) f_{nm}(\boldsymbol{k})\right]\delta_{\boldsymbol{k}\boldsymbol{k}^{\prime}}.\\
		\label{rho1mn1}
	\end{split}
\end{equation}
Here, $\bra{\psi_m(\boldsymbol{k}^{\prime})}\hat{\rho}^{(1)}(t)\ket{\psi_n(\boldsymbol{k})}=\rho_{mn}^{(1)}(\boldsymbol{k},t)\delta_{\boldsymbol{k}\boldsymbol{k}^{\prime}}$, and
\begin{equation}
	\begin{split}
		\rho_{mn}^{(1)}(\boldsymbol{k},t)
		&=\bra{\psi_m(\boldsymbol{k})}\hat{\rho}^{(1)}(t)\ket{\psi_n(\boldsymbol{k})}\\
		&=\bra{u_m(\boldsymbol{k})}\hat{\rho}^{(1)}(\boldsymbol{k},t)\ket{u_n(\boldsymbol{k})}\\
		&=\frac{g_J\mu_B}{\hbar c_{lv}}\sum_i\frac{e^{-i\omega_it}-e^{-i(-\omega_{nm}(\boldsymbol{k})-i\Gamma) t}}{\omega_i-\omega_{mn}(\boldsymbol{k})+i\Gamma}\tilde{\boldsymbol{E}}(\omega_i)\cdot\left[i\partial_{\mathbf{k}}f_n(\boldsymbol{k})\delta_{mn}
		+\boldsymbol{\mathcal{A}}_{mn}(\mathbf{k}) f_{nm}(\boldsymbol{k})\right],\\
	\end{split}
\end{equation}
in which $\hat{\rho}^{(1)}(\boldsymbol{k},t)=e^{-i\boldsymbol{k}\cdot\boldsymbol{r}}\hat{\rho}^{(1)}(t)e^{i\boldsymbol{k}\cdot\boldsymbol{r}}$. $\rho_{mn}^{(1)}(\boldsymbol{k},t)$ can be divided into intraband part $\rho_{mn}^{(1)i}(\boldsymbol{k},t)$ and interband part $\rho_{mn}^{(1)e}(\boldsymbol{k},t)$
\begin{equation}
	\begin{cases}
		\rho_{mn}^{(1)i}(\boldsymbol{k},t)
		=\frac{g_J\mu_B}{\hbar c_{lv}}\sum_i\frac{e^{-i\omega_it}-e^{-i(-\omega_{nm}(\boldsymbol{k})-i\Gamma) t}}{\omega_i-\omega_{mn}(\boldsymbol{k})+i\Gamma}\tilde{\boldsymbol{E}}(\omega_i)\cdot i\partial_{\boldsymbol{k}}f_n(\boldsymbol{k})\delta_{mn}
		\\
		\rho_{mn}^{(1)e}(\boldsymbol{k},t)
		=\frac{g_J\mu_B}{\hbar c_{lv}}\sum_i\frac{e^{-i\omega_it}-e^{-i(-\omega_{nm}(\boldsymbol{k})-i\Gamma) t}}{\omega_i-\omega_{mn}+i\Gamma}\tilde{\boldsymbol{E}}(\omega_i)\cdot\boldsymbol{\mathcal{A}}_{mn}(\mathbf{k}) f_{nm}(\boldsymbol{k}).\\
	\end{cases}
	\label{rho1ie}
\end{equation}
Here, the effect of the perturbation term and the scatter term are considered in Eq.(\ref{rho1ie}). So we only consider the Hamiltonian of the magnetic system in Eq.(\ref{model-study}).

\section{The velocity and energy velocity of magnon
} \label{MVMEV}
\subsection{The magnon velocity in the Bloch representation
} \label{the magnon velocity in the Bloch representation}

The operator of velocity is $\hat{\boldsymbol{v}}=\frac{1}{i\hbar}\left[\boldsymbol{r},\hat{H}\right]=\frac{1}{i\hbar}\left[\boldsymbol{r},\hat{H}_0+\frac{g_J\mu_B}{c_{lv}}\tilde{\boldsymbol{E}}\cdot\boldsymbol{r}\right]=\frac{1}{i\hbar}\left[\boldsymbol{r},\hat{H}_0\right]$.
The matrix element of magnon velocity operator $\hat{\boldsymbol{v}}=\frac{1}{i\hbar}\left[\boldsymbol{r},\hat{H}\right]$ in Bloch representation formally conforms to Eq. \ref{[r,O]}.
So \cite{ref23}
{\small
\begin{equation}
	\begin{split}
		\bra{\psi_m(\boldsymbol{k}^{\prime})}\hat{\boldsymbol{v}}\ket{\psi_n(\boldsymbol{k})}
		&=\frac{1}{i\hbar}\bra{\psi_m(\boldsymbol{k}^{\prime})}\left[\boldsymbol{r},\hat{H}_0\right]\ket{\psi_n(\boldsymbol{k})}\\
		&=\frac{1}{i\hbar}\left[\boldsymbol{D}_{\boldsymbol{k}}H_0(\boldsymbol{k})\right]_{mn}\delta_{\boldsymbol{k}\boldsymbol{k}^{\prime}}\\
		&=\frac{1}{i\hbar}\left\{i\partial_{\boldsymbol{k}}\varepsilon_n(\boldsymbol{k})\delta_{mn}
		+\boldsymbol{\mathcal{A}}_{mn}(\boldsymbol{k})\varepsilon_{nm}(\boldsymbol{k})\right\}\delta_{\boldsymbol{k}\boldsymbol{k}^{\prime}}, \\
		\label{vmn}
	\end{split}
\end{equation}
}
in which $\varepsilon_{nm}(\boldsymbol{k})=\varepsilon_{n}(\boldsymbol{k})-\varepsilon_{m}(\boldsymbol{k})$. 
Then we can define $\boldsymbol{v}_{mn}(\boldsymbol{k})=\bra{\psi_m(\boldsymbol{k})}\hat{\boldsymbol{v}}\ket{\psi_n(\boldsymbol{k})}=\frac{1}{i\hbar}\left\{i\partial_{\boldsymbol{k}}\varepsilon_n(\boldsymbol{k})\delta_{mn}
+\boldsymbol{\mathcal{A}}_{mn}(\boldsymbol{k})\varepsilon_{nm}(\boldsymbol{k})\right\}$ which satisfied $\bra{\psi_m(\boldsymbol{k}^{\prime})}\hat{\boldsymbol{v}}\ket{\psi_n(\boldsymbol{k})}=\boldsymbol{v}_{mn}(\boldsymbol{k})\delta_{\boldsymbol{k}\boldsymbol{k}^{\prime}}$. 
Here, $\boldsymbol{v}_{mn}(\boldsymbol{k})$ can be divided into intraband part $\boldsymbol{v}_{mn}^i(\boldsymbol{k})=\frac{1}{\hbar}\partial_{\boldsymbol{k}}\varepsilon_n(\boldsymbol{k})\delta_{mn}$  and interband part $\boldsymbol{v}_{mn}^e(\boldsymbol{k})=\frac{1}{i\hbar}\boldsymbol{\mathcal{A}}_{mn}(\boldsymbol{k})\varepsilon_{nm}(\boldsymbol{k})$. 
\subsection{The magnon energy velocity in the Bloch representation
} \label{The magnon energy velocity in the Bloch representation}
The operator of energy velocity is $\hat{\boldsymbol{v}}^E=\frac{1}{2}\left[\hat{\boldsymbol{v}},\hat{H}_0\right]_+$ \cite{ref9.1}, in which $\left[\hat{A},\hat{B}\right]_+=\hat{A}\hat{B}+\hat{B}\hat{A}$. The matrix element of the energy velocity in Bloch representation is 
\begin{equation}
	\begin{split}
		\bra{\psi_m(\boldsymbol{k}^{\prime})}\hat{\boldsymbol{v}}^E\ket{\psi_n(\boldsymbol{k})}
		&=\bra{\psi_m(\boldsymbol{k}^{\prime})}\frac{1}{2}\left[\hat{\boldsymbol{v}},\hat{H}_0\right]_+\ket{\psi_n(\boldsymbol{k})}\\
		&=\frac{1}{2}\bra{\psi_m(\boldsymbol{k}^{\prime})}\hat{\boldsymbol{v}}\hat{H}_0+\hat{H}_0\hat{\boldsymbol{v}}\ket{\psi_n(\boldsymbol{k})}\\
		&=\frac{1}{2}\left[\varepsilon_m(\boldsymbol{k}^{\prime})+\varepsilon_n(\boldsymbol{k})\right]\bra{\psi_m(\boldsymbol{k}^{\prime})}\hat{\boldsymbol{v}}\ket{\psi_n(\boldsymbol{k})}\\
		&=\frac{1}{2i\hbar}\left[\varepsilon_m(\boldsymbol{k})+\varepsilon_n(\boldsymbol{k})\right]\left\{i\partial_{\boldsymbol{k}}\varepsilon_n(\boldsymbol{k})\delta_{mn}
		+\boldsymbol{\mathcal{A}}_{mn}(\boldsymbol{k})\varepsilon_{nm}(\boldsymbol{k})\right\}\delta_{\boldsymbol{k}\boldsymbol{k}^{\prime}}\\
		&=\frac{1}{2}\left[\varepsilon_m(\boldsymbol{k})+\varepsilon_n(\boldsymbol{k})\right]\boldsymbol{v}(\boldsymbol{k})\delta_{\boldsymbol{k}\boldsymbol{k}^{\prime}}.\\
		\label{vEmn}
	\end{split}
\end{equation}
Then, we introduce $\boldsymbol{v}_{mn}^E(\boldsymbol{k})=\bra{\psi_m(\boldsymbol{k})}\hat{\boldsymbol{v}}^E\ket{\psi_n(\boldsymbol{k})}$ which satisfies $\bra{\psi_m(\boldsymbol{k}^{\prime})}\hat{\boldsymbol{v}}^E\ket{\psi_n(\boldsymbol{k})}=\boldsymbol{v}_{mn}^E(\mathbf{k})\delta_{\boldsymbol{k}\boldsymbol{k}^{\prime}}$, 
and $\boldsymbol{v}^E(\boldsymbol{k})$ satisfies 
\begin{equation}
	%\begin{split}
		\boldsymbol{v}^E(\boldsymbol{k})
		=\frac{\epsilon_n(\boldsymbol{k})+\epsilon_m(\boldsymbol{k})}{2}\boldsymbol{v}(\boldsymbol{k}).
	%\end{split}
\end{equation}
Similar to in the subsection \ref{the magnon velocity in the Bloch representation} for magnon velocity, we can divide $\boldsymbol{v}_{mn}^E(\boldsymbol{k})$ into intraband part $\boldsymbol{v}_{mn}^{E,i}(\boldsymbol{k})=\frac{1}{2\hbar}\left[\varepsilon_m(\boldsymbol{k})+\varepsilon_n(\boldsymbol{k})\right]\partial_{\boldsymbol{k}}\varepsilon_n(\boldsymbol{k})\delta_{mn}$ and the interband part $\boldsymbol{v}_{mn}^{E,e}(\boldsymbol{k})=\frac{1}{2i\hbar}\left[\varepsilon_m(\boldsymbol{k})+\varepsilon_n(\boldsymbol{k})\right]\boldsymbol{\mathcal{A}}_{mn}(\boldsymbol{k})\varepsilon_{nm}(\boldsymbol{k})$.

\section{The expressions of magnon photocurrent and magnon energy photocurrent
} \label{The expressions of magnon photocurrent and magnon energy photocurrent}
\subsection{The expressions of magnon spin photocurrent}\label{The expressions of magnon spin photocurrent}
As Eq. \ref{jtr}, the MSPC is
\begin{equation}
	\begin{split}
	\boldsymbol{j}
	=\hbar tr\left[\hat{\rho}\hat{\boldsymbol{v}}\right]
	&=\hbar\sum_{n\boldsymbol{k}} \bra{\psi_n(\boldsymbol{k})}\hat{\rho}\hat{\boldsymbol{v}}\ket{\psi_n(\boldsymbol{k})}\\
	&=\hbar\sum_{nm\boldsymbol{k}\boldsymbol{k}^{\prime}} \bra{\psi_n(\boldsymbol{k})}\hat{\rho}\ket{\psi_m(\boldsymbol{k}^{\prime})}\bra{\psi_m(\boldsymbol{k}^{\prime})}\hat{\boldsymbol{v}}\ket{\psi_n(\boldsymbol{k})}\\
	&=\hbar\sum_{nm\boldsymbol{k}} \rho_{nm}(\boldsymbol{k},t)\boldsymbol{v}_{mn}(\boldsymbol{k}),\\
	\label{jtr2}
	\end{split}
\end{equation}
in which $\rho_{nm}(\boldsymbol{k},t)=\bra{\psi_n(\boldsymbol{k})}\hat{\rho}\ket{\psi_m(\boldsymbol{k})}$ is the matrix element of density matrix and $\boldsymbol{v}_{mn}(\boldsymbol{k})=\bra{\psi_m(\boldsymbol{k})}\hat{\boldsymbol{v}}\ket{\psi_n(\boldsymbol{k})}$ is the matrix element of velocity (for details, see Appendix \ref{First-order correction of the density matrix} and Appendix \ref{the magnon velocity in the Bloch representation}).
The density matrix can be expanded according to the order of the electric field $\rho_{nm}(\boldsymbol{k},t)=\rho_{nm}^{(0)}(\boldsymbol{k},t)+\rho_{nm}^{(1)}(\boldsymbol{k},t)$. The $\alpha$-order magnon current can be expressed as
\begin{equation}
	\begin{split}
		\boldsymbol{j}^{(\alpha)}
		&=\hbar\sum_{nm\boldsymbol{k}} \left\{\rho_{nm}^{(\alpha)}(\boldsymbol{k},t)\left[\boldsymbol{v}_{mn}^i(\boldsymbol{k})+\boldsymbol{v}_{mn}^e(\boldsymbol{k})\right]
		\right\}.\\
	\end{split}
\end{equation}
\subsubsection{The zero-order magnon current} \label{j(0)}
When $\alpha=0$, the zero-order magnon current is 
\begin{equation}
	\begin{split}
		\boldsymbol{j}^{(0)}
		&=\hbar\sum_{nm\boldsymbol{k}} \rho_{nm}^{(0)}(\boldsymbol{k},t)\left[\boldsymbol{v}_{mn}^i(\boldsymbol{k})+\boldsymbol{v}_{mn}^e(\boldsymbol{k})\right]\\
		&=\sum_{n\boldsymbol{k}} f_n^B(\boldsymbol{k})\partial_{\boldsymbol{k}}\varepsilon_n(\boldsymbol{k}),\\
	\end{split}
\end{equation}
in which $f_n^B(\boldsymbol{k})=1/\left[e^{\varepsilon_n(\boldsymbol{k})/k_BT}-1\right]$ is Bose-Einstein distribution. When bands satisfy $\varepsilon_n(\boldsymbol{k})=\varepsilon_n(-\boldsymbol{k})$, Bose-Einstein distribution satisfy $f_n(\boldsymbol{k})=f_n(-\boldsymbol{k})$. So $\boldsymbol{j}^{(0)}=0$.
\subsubsection{The first-order magnon current}\label{j(1)}
Because the first-order density matrix can be divided into intraband part and interband part $\rho_{mn}^{(1)}(\boldsymbol{k},t)=\rho_{mn}^{(1)i}(\boldsymbol{k},t)+\rho_{mn}^{(1)e}(\boldsymbol{k},t)$, the first-order magnon current is
\begin{equation}
	\begin{split}
		\boldsymbol{j}^{(1)}
		&=\hbar\sum_{nm}\int[dk] \left\{\left[\rho_{nm}^{(1)i}(\boldsymbol{k},t)+\rho_{nm}^{(1)e}(\boldsymbol{k},t)\right]\left[\boldsymbol{v}_{mn}^i(\boldsymbol{k})+\boldsymbol{v}_{mn}^e(\boldsymbol{k})\right]
		\right\}.\\
	\end{split}
\end{equation}
According to Eq. \ref{rho1ie}, Eq. \ref{vmn} and Eq. \ref{tildeEE}, the first-order magnon current $\boldsymbol{j}^{(1)}$ can be divided into oscillating part $\boldsymbol{j}_O^{(1)}$ and damping part $\boldsymbol{j}_D^{(1)}$ 
\begin{equation}
	%\begin{split}
		j_{O,a}^{(1)}
		=\sum_{ib}\left[\chi_{O,ab}^i(\omega_i)+\chi_{O,ab}^e(\omega_i)\right]e^{-i\omega_i t}E_b(\omega_i),
	%\end{split}
\end{equation}
\begin{equation}
	%\begin{split}
		j_{D,a}^{(1)}
		=\sum_{ib}\left[\chi_{D,ab}^i(\omega_i)+\chi_{D,ab}^e(\omega_i,t)\right]e^{-\Gamma t}E_b(\omega_i),
		\label{j1Da}
	%\end{split}
\end{equation}
in which subscripts $a$ and $b$ mean the direction in the Cartesian coordinate system.
The magnon spin photoconductivity can be expressed as
	\begin{equation}
		\begin{cases}
			\chi_{O,ab}^i(\omega)=
			\nu\sum_{n,c}\int[dk]\epsilon_{bcz}\frac{\omega}{\omega+i\Gamma}\frac{\partial f_n(\boldsymbol{k})}{\partial k_c}\frac{\partial \omega_n(\boldsymbol{k})}{\partial k_a}\\
			\chi_{O,ab}^e(\omega)=
			\nu\sum_{n\ne m,c}\int[dk]\epsilon_{bcz}\omega
			\frac{\mathcal{A}_{a,mn}(\boldsymbol{k})\mathcal{A}_{c,nm}(\boldsymbol{k})}{\omega-\omega_{nm}+i\Gamma}f_{nm}(\boldsymbol{k})\omega_{nm}(\boldsymbol{k}) \\
			\chi_{D,ab}^i(\omega)
			=-\chi_{O,ab}^i(\omega)\\
			\chi_{D,ab}^e(\omega,t)
			=-\chi_{O,ab}^e(\omega)e^{-i\omega_{nm}(\boldsymbol{k})t}.\\
		\end{cases}
	\end{equation}
Here, subscripts $O$ and $D$ mean the oscillating part and the damping part.
Superscript $i$ and $e$ mean the intraband term and interband term, respectively.
$\epsilon_{bcz}$ is the Levi-Civita symbol.
$\omega_n(\boldsymbol{k})=\varepsilon_n(\boldsymbol{k})/\hbar$ and $\omega_{nm}(\boldsymbol{k})=\omega_n(\boldsymbol{k})-\omega_m(\boldsymbol{k})$. 
$\nu=\frac{g_J\mu_B}{c_{lv}^2}$, 
$E_b(\omega)$ is the $b$ direction component of the electric field complex amplitude. 
Here, we take the continuous limit, so we replace the summation over $\boldsymbol{k}$ with the integral measure $\int[dk]=\frac{1}{(2\pi)^d}\int d^dk$, in which $d$ is the dimension of the system.

According to Eq. \ref{j1Da}, $\boldsymbol{j}_D^{(1)}$ decreases exponentially with time. When the relaxation time $\tau$ is short enough ($\Gamma>>\omega_{gap}$) or the time evolution is long, we can ignore the damping part $\boldsymbol{j}_D^{(1)}$. 
When $\Gamma\to0$ and $\omega<<\omega_{gap}$, $\omega\omega_{nm}/(\omega-\omega_{nm}+i\Gamma)\approx-\omega$, we can obtain
\begin{equation}
	\begin{split}
		\chi_{O,ab}^e(\omega)
		&\approx-\nu\sum_{n\ne m,c}\int[dk]\epsilon_{bcz}\omega
		\mathcal{A}_{a,mn}(\boldsymbol{k})\mathcal{A}_{c,nm}(\boldsymbol{k})f_{nm}^B(\boldsymbol{k})\\
		&=-\nu\sum_{c}\int[dk]\epsilon_{bcz}\omega
		\left[\sum_{m\ne n}\mathcal{A}_{a,mn}(\boldsymbol{k})\mathcal{A}_{c,nm}(\boldsymbol{k})f_{n}^B(\boldsymbol{k})
		-\sum_{m\ne n}\mathcal{A}_{a,mn}(\boldsymbol{k})\mathcal{A}_{c,nm}(\boldsymbol{k})f_{m}^B(\boldsymbol{k})\right]\\
		&=\nu\sum_{c}\int[dk]\epsilon_{bcz}\omega
		\sum_{m\ne n}\left[\mathcal{A}_{a,nm}(\boldsymbol{k})\mathcal{A}_{c,mn}(\boldsymbol{k})
		-\mathcal{A}_{c,nm}(\boldsymbol{k})
		\mathcal{A}_{a,mn}(\boldsymbol{k})\right]f_{n}^B(\boldsymbol{k})\\
		&=-i\nu\omega\sum_{c}\epsilon_{bcz}
		\sum_{n}\int[dk]\Omega_n^{ac}(\boldsymbol{k})f_{n}^B(\boldsymbol{k}),\\
	\end{split}
\end{equation}
in which
\begin{equation}
	\begin{split}
		\Omega_n^{ac}(\boldsymbol{k})
		&=i\sum_{m(\ne n)}\left[\mathcal{A}_{a,nm}(\boldsymbol{k})\mathcal{A}_{c,mn}(\boldsymbol{k})
		-\mathcal{A}_{c,nm}(\boldsymbol{k})
		\mathcal{A}_{a,mn}(\boldsymbol{k})\right]\\
		&=i\sum_{m(\ne n)}\left[\frac{\bra{u_n(\boldsymbol{k})}\left[\partial_{k_a}\hat{H}(\boldsymbol{k})\right]\ket{u_m(\boldsymbol{k})}\bra{u_m(\boldsymbol{k})}\left[\partial_{k_c}\hat{H}(\boldsymbol{k})\right]\ket{u_n(\boldsymbol{k})}}{\left(\varepsilon_n(\boldsymbol{k})-\varepsilon_m(\boldsymbol{k})\right)^2}
		-h.c.
		\right]\\
		\label{Berrycurvature}
	\end{split}
\end{equation}
is the Berry curvature. 

\subsection{The expression of the magnon energy photocurrent}\label{The expression of magnon energy photocurrent}
Similar to the Appendix \ref{The expressions of magnon spin photocurrent}, next we consider the expression of magnon energy photocurrent.
The magnon energy photocurrent is
\begin{equation}
	\begin{split}
		\boldsymbol{j}^E
		=tr\left[\hat{\rho}\hat{\boldsymbol{v}}^E\right]
		&=\sum_{n\boldsymbol{k}} \bra{\psi_n(\boldsymbol{k})}\hat{\rho}\hat{\boldsymbol{v}}^E\ket{\psi_n(\boldsymbol{k})}\\
		&=\sum_{nm\boldsymbol{k}\boldsymbol{k}^{\prime}} \bra{\psi_n(\boldsymbol{k})}\hat{\rho}\ket{\psi_m(\boldsymbol{k}^{\prime})}\bra{\psi_m(\boldsymbol{k}^{\prime})}\hat{\boldsymbol{v}}^E\ket{\psi_n(\boldsymbol{k})}\\
		&=\sum_{nm\boldsymbol{k}} \rho_{nm}(\boldsymbol{k},t)\boldsymbol{v}_{mn}^E(\boldsymbol{k}),\\
		\label{jQtr2}
	\end{split}
\end{equation}
in which $\rho_{nm}(\boldsymbol{k},t)=\bra{\psi_n(\boldsymbol{k})}\hat{\rho}\ket{\psi_m(\boldsymbol{k})}$ is the matrix element of density matrix and $\boldsymbol{v}_{mn}^E(\boldsymbol{k})=\bra{\psi_m(\boldsymbol{k})}\hat{\boldsymbol{v}}^E\ket{\psi_n(\boldsymbol{k})}$ is the matrix element of energy velocity (detail see Appendix \ref{First-order correction of the density matrix} and Appendix \ref{The magnon energy velocity in the Bloch representation}).
The $\alpha$-order magnon energy photocurrent can be expressed as
\begin{equation}
	\begin{split}
		\boldsymbol{j}^{E(\alpha)}
		&=\sum_{nm\boldsymbol{k}} \left\{\rho_{nm}^{(\alpha)}(\boldsymbol{k},t)\left[\boldsymbol{v}_{mn}^{E,i}(\boldsymbol{k})+\boldsymbol{v}_{mn}^{E,e}(\boldsymbol{k})\right]
		\right\}.\\
	\end{split}
\end{equation}
\subsubsection{\label{Zero-order magnon energy photocurrent}The zero-order magnon energy photocurrent}
When $\alpha=0$, the zero-order magnon energy photocurrent is 
\begin{equation}
	\begin{split}
		\boldsymbol{j}^{E(0)}
		&=\sum_{nm\boldsymbol{k}} \rho_{nm}^{(0)}(\boldsymbol{k},t)\left[\boldsymbol{v}_{mn}^{E,i}(\boldsymbol{k})+\boldsymbol{v}_{mn}^{E,e}(\boldsymbol{k})\right]\\
		&=\frac{1}{\hbar}\sum_{n\boldsymbol{k}} f_n(\boldsymbol{k})\varepsilon_n(\boldsymbol{k})\partial_{\boldsymbol{k}}\varepsilon_n(\boldsymbol{k}),
		\\
	\end{split}
\end{equation}
in which $f_n(\boldsymbol{k})=1/\left[e^{\varepsilon_n(\boldsymbol{k})/k_BT}-1\right]$ is Bose-Einstein distribution. 
\subsubsection{The first-order magnon energy photocurrent}
Because the first-order density matrix can be divided into intraband part and interband part $\rho_{mn}^{(1)}(\boldsymbol{k},t)=\rho_{mn}^{(1)i}(\boldsymbol{k},t)+\rho_{mn}^{(1)e}(\boldsymbol{k},t)$, the first-order magnon energy photocurrent is
\begin{equation}
	\begin{split}
		\boldsymbol{j}^{E(1)}
		&=\sum_{nm\boldsymbol{k}} \left\{\left[\rho_{nm}^{(1)i}(\boldsymbol{k},t)+\rho_{nm}^{(1)e}(\boldsymbol{k},t)\right]\left[\boldsymbol{v}_{mn}^{E,i}(\boldsymbol{k})+\boldsymbol{v}_{mn}^{E,e}(\boldsymbol{k})\right]
		\right\}.\\
	\end{split}
\end{equation}
According to Eq. \ref{rho1ie}, Eq. \ref{vEmn} and Eq. \ref{tildeEE}, the first-order magnon energy photocurrent $\boldsymbol{j}^{E(1)}$ can be divided into oscillating part $\boldsymbol{j}_O^{E(1)}$ and damping part $\boldsymbol{j}_D^{E(1)}$ 
\begin{equation}
	\begin{split}
		j_{O,a}^{E(1)}
		&=\sum_{ib}\left[\chi_{O,ab}^{E,i}(\omega_i)+\chi_{O,ab}^{E,e}(\omega_i)\right]e^{-i\omega_i t}E_b(\omega_i),\\
	\end{split}
\end{equation}
\begin{equation}
	%\begin{split}
	j_{D,a}^{E(1)}
	=\sum_{ib}\left[\chi_{D,ab}^{E,i}(\omega_i)+\chi_{D,ab}^{E,e}(\omega_i,t)\right]e^{-\Gamma t}E_b(\omega_i),
	\label{jE1Da}
	%\end{split}
\end{equation}
in which subscripts $a$ and $b$ mean the direction in the Cartesian coordinate system.
The magnon energy photoconductivity can be expressed as
\begin{equation}
	\begin{cases}
		\chi_{O,ab}^{E,i}(\omega)=
		\nu\sum_{n,c}\int[dk] \frac{\omega\epsilon_{bcz}}{\omega+i\Gamma} \omega_n(\boldsymbol{k})\frac{\partial f_n(\boldsymbol{k})}{\partial_{k_c}}\frac{\partial\omega_n(\boldsymbol{k})}{\partial_{k_a}}\\
		\chi_{O,ab}^{E,e}(\omega)=\frac{\nu}{2}\sum_{n\ne m,c}\int[dk] \left[\frac{\omega\epsilon_{bcz}}{\omega-\omega_{nm}+i\Gamma}\mathcal{A}_{a,mn}(\boldsymbol{k})\mathcal{A}_{c,nm}(\boldsymbol{k})f_{nm}(\boldsymbol{k})\left[\omega_n^2(\boldsymbol{k})-\omega_m^2(\boldsymbol{k})\right]\right]\\
		\chi_{D,ab}^{E,i}(\omega)
		=-\chi_{O,ab}^{E,i}(\omega)\\
		\chi_{D,ab}^{E,e}(\omega,t)
		=-\chi_{O,ab}^{E,e}(\omega)e^{-i\varepsilon_{nm}(\boldsymbol{k})t/\hbar}.\\
	\end{cases}
\end{equation}
Here, subscripts $O$ and $D$ mean the oscillating part and the damping part. 
Superscript $i$ and $e$ mean the intraband term and interband term, respectively.
$\epsilon_{bcz}$ is the Levi-Civita symbol.
$\epsilon_{bcz}$ is the Levi-Civita symbol.
$\omega_n(\boldsymbol{k})=\varepsilon_n(\boldsymbol{k})/\hbar$ and $\omega_{nm}(\boldsymbol{k})=\omega_n(\boldsymbol{k})-\omega_m(\boldsymbol{k})$. 
$\nu=\frac{g_J\mu_B}{c_{lv}^2}$, $E_b(\omega_i)$ is the $b$ direction component of the electric field complex amplitude. 
Similar to the Appendix \ref{The expressions of magnon spin photocurrent}, here we take the continuous limit.

Similar to a first-order magnon photocurrent, according to Eq. \ref{jE1Da}, $\boldsymbol{j}_D^{E(1)}$ decreases exponentially with time. When the relaxation time $\tau$ is short enough, we can ignore the damping part $\boldsymbol{j}_D^{E(1)}$.
When $\Gamma<<\omega_{gap}$ and $\omega_i<<\omega_{gap}$, $\omega_i\omega_{nm}/(\omega_i-\omega_{nm}+i\Gamma)\approx-\omega_i$, we can obtain
\begin{equation}
	\begin{split}
		\chi_{ab}^{E,e}(\omega)
		&=\frac{\nu}{2}\sum_{n\ne m,c}\int[dk] \left[\frac{\omega\epsilon_{bcz}}{\omega-\omega_{nm}+i\Gamma}\mathcal{A}_{a,mn}(\boldsymbol{k})\mathcal{A}_{c,nm}(\boldsymbol{k})f_{nm}(\boldsymbol{k})\left[\omega_n^2(\boldsymbol{k})-\omega_m^2(\boldsymbol{k})\right]\right]\\
		&=\frac{\nu}{2}\sum_{n\ne m,c}\int[dk] \left[\frac{\omega\omega_{nm}\epsilon_{bcz}}{\omega-\omega_{nm}+i\Gamma}\mathcal{A}_{a,mn}(\boldsymbol{k})\mathcal{A}_{c,nm}(\boldsymbol{k})f_{nm}(\boldsymbol{k})\left[\omega_n(\boldsymbol{k})+\omega_m(\boldsymbol{k})\right]\right]\\
		&\approx-\frac{\nu}{2}\omega\sum_{n\ne m,c}\int[dk] \left\{\epsilon_{bcz}\mathcal{A}_{a,mn}(\boldsymbol{k})\mathcal{A}_{c,nm}(\boldsymbol{k}) f_{nm}(\boldsymbol{k})\left[\omega_n(\boldsymbol{k})+\omega_m(\boldsymbol{k})\right]\right\}.\\
	\end{split}
\end{equation}
\section{\label{appchern}The Chern number of magnon in the two-dimensional collinear ferromagnetic Hexagonal lattice}
According to Eq. \ref{BlochD}, singular points in the Brillouin zone (BZ) satisfy  $|\boldsymbol{d}|=\pm d_z$. 
%in other words the structure factor is zero, 
The singular points are related to the topological properties of the system. 
In the Hexagonal lattice, the singular points are point $K$ and point $K^{\prime}$. 
Around point $K$, $\boldsymbol{d}(\boldsymbol{k})$ can be expressed as the function of $\boldsymbol{q}=\boldsymbol{k}-\boldsymbol{k}_K$, in which $\boldsymbol{k}_K$ is the point of $K$.
So  
\begin{equation}
	\begin{cases}
		d_0=\vartriangle\\
		d_x(\boldsymbol{q})\approx\frac{3}{2}JSaq_x\\
		d_y(\boldsymbol{q})\approx-\frac{3}{2}JSaq_y\\
		d_z(\boldsymbol{q})\approx-3\sqrt{3}SD.\\
	\end{cases}
\end{equation}
In a similar way, around point $K^{\prime}$, $\boldsymbol{d}(\boldsymbol{k})$ can be expressed as  
\begin{equation}
	\begin{cases}
		d_0=\vartriangle\\
		d_x(\boldsymbol{q}^{\prime})\approx\frac{3}{2}JSaq_x^{\prime}\\
		d_y(\boldsymbol{q}^{\prime})\approx\frac{3}{2}JSaq_y^{\prime}\\
		d_z(\boldsymbol{q}^{\prime})\approx3\sqrt{3}SD.\\
	\end{cases}
\end{equation}
Here $\boldsymbol{q}^{\prime}=\boldsymbol{k}-\boldsymbol{k}_{K^{\prime}}$, in which $\boldsymbol{k}_{K^{\prime}}$ is the point of $K^{\prime}$.

When $D>0$ meV, point $K$ satisfies $|\boldsymbol{d}|=-d_z$, the singular point $K^{\prime}$ satisfies $|\boldsymbol{d}|=d_z$.
For band $\varepsilon_-(\boldsymbol{k})$, the responding Bloch state $\ket{u_-(\boldsymbol{k})}$ can not be defined well in the zone that contains point $K$. So we need to take a phase transition $e^{i\phi_-(\boldsymbol{k})}=\frac{\frac{d_z-|\boldsymbol{d}|}{-d_x+id_y}}{|\frac{d_z-|\boldsymbol{d}|}{-d_x+id_y}|}=-\frac{|q_x+iq_y|}{q_x+iq_y}=e^{-i(\theta+\pi)}$. So the Chern number of band $\varepsilon_-(\boldsymbol{k})$ is \cite{ref32}
\begin{equation}
	\begin{split}
		C_-
		&=\frac{1}{2\pi}\int\int_{BZ} dk_xdk_y\Omega_-(\boldsymbol{k})
		=1.\\
	\end{split}
\end{equation}
For band $\varepsilon_+(\boldsymbol{k})$, the Bloch state $\ket{u_+(\boldsymbol{k})}$ can not defined well in the zone that contain point $K^{\prime}$. We take a phase transition $e^{i\phi_+(\boldsymbol{k})}=\frac{\frac{d_z+|\boldsymbol{d}|}{-d_x+id_y}}{|\frac{d_z+|\boldsymbol{d}|}{-d_x+id_y}|}=-\frac{|d_x-id_y|}{d_x-id_y}=-\frac{|-q_x+iq_y|}{q_x-iq_y}=e^{i(\theta+\pi)}$. So the Chern number of band $\varepsilon_+(\boldsymbol{k})$ is 
\begin{equation}
	\begin{split}
		C_+
		&=\frac{1}{2\pi}\int\int_{BZ} dk_xdk_y\Omega_+(\boldsymbol{k})
		=-1.\\
	\end{split}
\end{equation}
In a similar way, when $D<0$ meV, $C_-=-1$ and $C_+=1$.

\end{widetext}


\begin{thebibliography}{99}
\bibitem{ref0} E. H. Hall, On a New Action of the Magnet on Electric Currents, American Journal of Mathematics 2 (3) (1879) 287–292.

\bibitem{ref0.1} E. H. Hall, XVIII. On the “Rotational Coefficient” in nickel and cobalt, The London, Edinburgh, and Dublin Philosophical Magazine and Journal of Science 12 (74), (1881) 157–172.

\bibitem{ref0.2} M.I. Dyakonov, V.I. Perel, Current-induced spin orientation of electrons in semiconductors, Physics Letters A 35 (6) (1971) 459-460.

\bibitem{ref0.3} K. v. Klitzing, G. Dorda, M. Pepper, New Method for High-Accuracy Determination of the Fine-Structure Constant Based on Quantized Hall Resistance, Phys. Rev. Lett. 45 (6) (1980) 494-497.

\bibitem{ref0.4} D. C. Tsui, H. L. Stormer, A. C. Gossard, Two-Dimensional Magnetotransport in the Extreme Quantum Limit, Phys. Rev. Lett. 48 (22) (1982) 1559-1562.

\bibitem{ref0.5} D. J. Thouless, M. Kohmoto, M. P. Nightingale, M. den Nijs, Quantized Hall Conductance in a Two-Dimensional Periodic Potential, Phys. Rev. Lett. 49 (6) (1982) 405-408.

\bibitem{ref0.6} Karube, K., Ōnuki, Y., Nakajima, T. \textit{et al.}. Giant Hall effect in a highly conductive frustrated magnet GdCu2. npj Quantum Mater. 10, 55 (2025).

\bibitem{ref0.7} Bhandari, H., Ning, Z., Chang, PH. \textit{et al.}. Three-dimensional nature of anomalous Hall conductivity in $YMn_6Sn_{6-x}Ga_x$, $x \approx 0.55$. npj Quantum Mater. 10, 99 (2025).

\bibitem{ref0.8} J. E. Hirsch, Spin Hall Effect, Phys. Rev. Lett. 83 (9) (1999) 1834-1837.

\bibitem{ref0.9} S. Murakami , N. Nagaosa, and S. Zhang, Dissipationless Quantum Spin Current at Room Temperature, Science 301 (5638) (2003) 1348-1351.

\bibitem{ref0.91} J. Sinova, D. Culcer, Q. Niu, N. A. Sinitsyn, T. Jungwirth, A. H. MacDonald, Universal Intrinsic Spin Hall Effect, Phys. Rev. Lett. 92 (12) (2004) 126603.

\bibitem{ref0.92} C. L. Kane, E. J. Mele, Quantum Spin Hall Effect in Graphene, Phys. Rev. Lett. 95 (22) (2005) 226801.

\bibitem{ref0.93} B. A. Bernevig, S. Zhang, Quantum Spin Hall Effect, Phys. Rev. Lett. 96 (10) (2006) 106802.

\bibitem{ref0.94} D. Xiao, Y. Yao, Z. Fang, Q. Niu, Berry-Phase Effect in Anomalous Thermoelectric Transport, Phys. Rev. Lett. 97 (2) (2006) 026603.

\bibitem{ref0.95} Y. Gao, D. Xiao, Orbital magnetic quadrupole moment and nonlinear anomalous thermoelectric transport, Phys. Rev. B 98 (6) (2018) 060402.

\bibitem{ref0.96} X. Q. Yu, Z. G. Zhu, J. S. You, T. Low, G. Su, Topological nonlinear anomalous Nernst effect in strained transition metal dichalcogenides, Phys. Rev. B 99 (20) (2019) 201410. 

\bibitem{ref0.97} Y. Gao, S. A. Yang, Q. Niu, Field Induced Positional Shift of Bloch Electrons and Its Dynamical Implications, Phys. Rev. Lett. 112 (16) (2014) 166601.

\bibitem{ref0.98} H. Liu, J. Zhao, Y. X. Huang, W. Wu, X. L. Sheng, C. Xiao, S. A. Yang, Intrinsic Second-Order Anomalous Hall Effect and Its Application in Compensated Antiferromagnets, Phys. Rev. Lett. 127 (27) (2021) 277202.

\bibitem{ref0.99} M. C. Chang, Q. Niu, Berry phase, hyperorbits, and the Hofstadter spectrum: Semiclassical dynamics in magnetic Bloch bands, Phys. Rev. B 53 (11) (1996) 7010-7023.

\bibitem{ref0.991} G. Sundaram and Q. Niu, Wave-packet dynamics in slowly perturbed crystals: Gradient corrections and Berry-phase effects, Phys. Rev. B 59 (23) (1999) 14915-14925.

\bibitem{ref0.9911} D. Xiao, J. Shi, Q. Niu, Berry Phase Correction to Electron Density of States in Solids, Phys. Rev. Lett. 95 (13) (2005) 137204.

\bibitem{ref0.9912} D. Xiao, M. C. Chang, Q. Niu, Berry phase effects on electronic properties, Rev. Mod. Phys. 82 (3) (2010) 1959-2007.






\bibitem{ref0.992} C. Strohm, G. L. J. A. Rikken, P. Wyder, Phenomenological Evidence for the Phonon Hall Effect, Phys. Rev. Lett. 95 (15) (2005) 155901 .

\bibitem{ref0.993} L. Sheng, D. N. Sheng, C. S. Ting, Theory of the Phonon Hall Effect in Paramagnetic Dielectrics, Phys. Rev. Lett. 96 (15) (2006) 155901.

\bibitem{ref0.994} Y. Kagan, L. A. Maksimov, Anomalous Hall Effect for the Phonon Heat Conductivity in Paramagnetic Dielectrics, Phys. Rev. Lett. 100 (14) (2008) 145902.

\bibitem{ref0.995} J. S. Wang, L. Zhang, Phonon Hall thermal conductivity from the Green-Kubo formula, Phys. Rev. B 80 (1) (2009) 012301.

\bibitem{ref0.996} L. Zhang, J. S. Wang, B. Li, Phonon Hall effect in four-terminal nano-junctions, New J. Phys. 11 (11) 113038 (2009).

\bibitem{ref0.997} L. Zhang, J. Ren, J. S. Wang, B. Li, Topological Nature of the Phonon Hall Effect, Phys. Rev. Lett. 105 (22) (2010) 225901.

\bibitem{ref0.998} L. Zhang, J. Ren, J. S. Wang, B. Li, The phonon Hall effect: theory and application, J. Phys.: Condens. Matter 23 (30) (2011) 305402.

\bibitem{ref0.999} K. Sun, Z. Gao, J. S. Wang, Phonon Hall effect with first-principles calculations, Phys. Rev. B 103 (21) (2021) 214301.

\bibitem{ref0.9991} Z.-X. Li, Yunshan Cao and Peng Yan, Topological insulators and semimetals in classical magnetic systems, Phys. Rep. 915 (2021) 1-64.

\bibitem{ref0.9992} Anjan Barman \textit{et al.}, The 2021 Magnonics Roadmap, J. Phys.: Condens. Matter 33 (41) (2021) 413001.

\bibitem{ref0.9993} F. Zhuo, J. Kang, A. Manchon, Z. Cheng, Topological Phases in Magnonics, Adv. Physics Res. 4 (3) (2023) 2300054.

\bibitem{ref0.9994} Feng Mei \textit{et al.}, Topological magnon insulator and quantized pumps from strongly-interacting bosons in optical superlattices, 2019 New J. Phys. 21 095002.

\bibitem{ref0.9995} I Diniz and A T Costa, Microscopic origin of subthermal magnons and the spin Seebeck effect, 2016 New J. Phys. 18 052002.

\bibitem{ref1} A. V. Chumak, V. I. Vasyuchka, A. A. Serga,  B. Hillebrands, Magnon spintronics, Nature Phys 11 (6) (2015) 453–461.

\bibitem{ref1.1} L. Chotorlishvili, X. G. Wang, A. Dyrda, G.-H. Guo, V. K. Dugaev, J. Barnaś, J. Berakdar, Rectification of the spin Seebeck current in noncollinear antiferromagnets, Phys. Rev. B 106 (1) (2022) 014417 .

\bibitem{ref2} H. Katsura, N. Nagaosa, P. A. Lee, Theory of the Thermal Hall Effect in Quantum Magnets, Phys. Rev. Lett. 104 (6) (2010) 066403.

\bibitem{ref3} Y. Onose, T. Ideue, H. Katsura, Y. Shiomi, N. Nagaosa, Y. Tokura, Observation of the Magnon Hall Effect, Science 329 (5989) (2010) 297-299.

\bibitem{ref4} T. Ideue, Y. Onose, H. Katsura, Y. Shiomi, S. Ishiwata, N. Nagaosa, Y. Tokura, Effect of lattice geometry on magnon Hall effect in ferromagnetic insulators, Phys. Rev. B 85 (13) (2012) 134411.

\bibitem{ref5} R. Matsumoto and S. Murakami, Rotational motion of magnons and the thermal Hall effect, Phys. Rev. B 84 (18) (2011) 184406.

\bibitem{ref6} L. Zhang, J. Ren, J. S. Wang, B. Li, Topological magnon insulator in insulating ferromagnet, Phys. Rev. B 87 (14) (2013) 144101.

%\bibitem{ref7} A. Rückriegel, A. Brataas, and R. A. Duine, Phys. Rev. B 97, 081106(R) (2018).

\bibitem{ref8} A. Mook, J. Henk, and I. Mertig, Edge states in topological magnon insulators, Phys. Rev. B 90 (2) (2014) 024412.

\bibitem{ref9} A. Mook, J. Henk and I. Mertig, Magnon Hall effect and topology in kagome lattices: A theoretical investigation, Phys. Rev. B 89 (13) (2014) 134409.

\bibitem{ref9.01} Y. Wang, Z. G. Zhu, G. Su, Quantum theory of nonlinear thermal response, Phys. Rev. B 106 (3) (2022) 035148.

\bibitem{ref9.1} H. Varshney, R. Mukherjee, A. Kundu, A. Agarwal, Intrinsic nonlinear thermal Hall transport of magnons: A quantum kinetic theory approach, Phys. Rev. B 108 (16) (2023) 165412.

\bibitem{ref9.2} J. C. Li, Z. G. Zhu, Intrinsic second-order magnon thermal Hall effect, J. Phys.: Condens. Matter 36 (39) (2024) 395802.

\bibitem{ref9.3} R. Cheng, S. Okamoto. D. Xiao, Spin Nernst Effect of Magnons in Collinear Antiferromagnets, Phys. Rev. Lett. 117 (21) (2016) 217202.

\bibitem{ref9.4} V. A. Zyuzin, A. A. Kovalev, Magnon Spin Nernst Effect in Antiferromagnets, Phys. Rev. Lett. 117 (21) (2016) 217203.

\bibitem{ref10} H. Kondo, Y. Akagi, Nonlinear magnon spin Nernst effect in antiferromagnets and strain-tunable pure spin current, Phys. Rev. Research 4 (1) (2022) 013186.

\bibitem{ref11} R. Mukherjee, S. Verma, A. Kundu, Nonlinear magnon transport in bilayer van der Waals antiferromagnets, Phys. Rev. B 107 (24) (2023) 245426.

\bibitem{ref11.1} Weißenhofer, M., Mrudul, M.S., Mankovsky, S. \textit{et al.} Magnon orbital Nernst effect in altermagnets. npj Quantum Mater. (2026).



\bibitem{ref12} M. M. Nayga, S. Rachel, M. Vojta, Magnon Landau Levels and Emergent Supersymmetry in Strained Antiferromagnets, Phys. Rev. Lett. 123 (20) (2019) 207204.

\bibitem{ref13} I. Proskurin, A. S. Ovchinnikov, J. I. Kishine, R. L. Stamps, Excitation of magnon spin photocurrents in antiferromagnetic insulators, Phys. Rev. B 98 (13) (2018) 134422.

\bibitem{ref14} H. Ishizuka, M. Sato, Theory for shift current of bosons: Photogalvanic spin current in ferrimagnetic and antiferromagnetic insulators, Phys. Rev. B 100 (22) (2019) 224411.

\bibitem{ref14.1} H. Ishizuka, M. Sato, Large Photogalvanic Spin Current by Magnetic Resonance in Bilayer Cr Trihalides, Phys. Rev. Lett. 129 (10) (2022) 107201.

\bibitem{ref14.2} E. V. Boström, T. S. Parvini, J. W. McIver, A. Rubio
, S. V. Kusminskiy, M. A. Sentef, All-optical generation of antiferromagnetic magnon currents via the magnon circular photogalvanic effect, Phys. Rev. B 104 (10) (2021) L100404.

\bibitem{ref15} Y. Aharonov, A. Casher, Topological Quantum Effects for Neutral Particles, Phys. Rev. Lett. 53 (4) (1984) 319-321.

\bibitem{ref16} K. Nakata, J. Klinovaja, D. Loss, Magnonic quantum Hall effect and Wiedemann-Franz law, Phys. Rev. B 95 (12) (2017) 125429.

\bibitem{ref17} K. Nakata, S. K. Kim, J. Klinovaja, Magnonic topological insulators in antiferromagnets, D. Loss, Phys. Rev. B 96 (22) (2017) 224414.

\bibitem{ref18} Q. Ma, A. G. Grushin, K. S. Burch, Topology and geometry under the nonlinear electromagnetic spotlight, Nat. Mater. 20 (12) (2021) 1601–1614.

\bibitem{ref19} H. Xu, H. Wang, J. Zhou, J. Li, Pure spin photocurrent in non-centrosymmetric crystals: bulk spin photovoltaic effect, Nat Commun 12 (1) (2021) 4330.

\bibitem{ref20} S. Patankar, L. Wu, B. Lu, M. Rai, J. D. Tran, T. Morimoto, D. E. Parker, A. G. Grushin, N. L. Nair \textit{et al.}, Phys. Rev. B 98 (16) (2018) 165113.

\bibitem{ref21} Q. Ma, S. Y. Xu, H. Shen, D. MacNeill, V. Fatemi, T. R. Chang, A. M. M. Valdivia, S. Wu, Z. Du, C.-H. Hsu, S. Fang, Q. D. Gibson, K. Watanabe, T. Taniguchi, R. J. Cava, E. Kaxiras, H. Z. Lu, H. Lin, L. Fu, N. Gedik, P. Jarillo-Herrero, Observation of the nonlinear Hall effect under time-reversal-symmetric conditions, Nature 565 (7739) (2019) 337–342.

\bibitem{ref22} P. He, H. Isobe, D. Zhu, C. H. Hsu, L. Fu, H. Yang, Quantum frequency doubling in the topological insulator $Bi_2Se_3$, Nat Commun. 12 (1) (2021) 698.

\bibitem{ref22.1} O. Matsyshyn, I. Sodemann, Nonlinear Hall Acceleration and the Quantum Rectification Sum Rule, Phys. Rev. Lett. 123 (24) (2019) 246602.

%{\color{green}\bibitem{ref22.2} H. Xu, H. Wang, J. Zhou, J. Li, Pure spin photocurrent in non-centrosymmetric crystals: bulk spin photovoltaic effect, Nat Commun. 12 (1) (2021) 4330.}

\bibitem{ref22.3} D. J. Passos, G. B. Ventura, J. M. Viana Parente Lopes, J. M. B. Lopes dos Santos, N. M. R. Peres, Nonlinear optical responses of crystalline systems: Results from a velocity gauge analysis, Phys. Rev. B 97 (23) (2018) 235446.

\bibitem{ref23} Y. Liu, Z. G. Zhu, G. Su, Photogalvanic effect and second-harmonic generation from radio to infrared wavelengths in the $WTe_2$ monolayer, Phys. Rev. B 109 (8) (2024) 085117.

\bibitem{ref24} Y. Wang, Z. G. Zhu, G. Su, Magnon Landau-Zener tunneling and spin-current generation by electric field, Phys. Rev. B 111 (16) (2025) 165150.

\bibitem{ref25} Y. Wang, Z. G. Zhu, G. Su, Magnon spin photogalvanic effect induced by Aharonov-Casher phase, Phys. Rev. B 110 (5) (2024) 054434.

\bibitem{ref25.01} L. Cornelissen, J. Liu, R. Duine, J. B. Youssef, B. Van Wees, Long-distance transport of magnon spin information in a magnetic insulator at room temperature, Nat. Phys. 11, 1022 (2015).

\bibitem{ref25.1} Yu, J., Bernevig, B.A., Queiroz, R. \textit{et al.} Quantum geometry in quantum materials. npj Quantum Mater. 10, 101 (2025).

\bibitem{ref26} H. Chen, Q. Niu, A. H. MacDonald, Anomalous Hall Effect Arising from Noncollinear Antiferromagnetism, Phys. Rev. Lett. 112 (1) (2014) 017205.


\bibitem{ref28} S. Toth, B. Lake, Linear spin wave theory for single-Q incommensurate magnetic structures, J. Phys.: Condens. Matter 27 (16) (2015) 166002 .

\bibitem{ref28.1} H. Watanabe, Y. Yanase, Chiral Photocurrent in Parity-Violating Magnet and Enhanced Response in Topological Antiferromagnet, Phys. Rev. X 11 (1) (2021) 011001.

\bibitem{ref28.2} T. Holder, D. Kaplan, B. Yan, Consequences of time-reversal-symmetry breaking in the light-matter interaction: Berry curvature, quantum metric, and diabatic motion,  Phys. Rev. Research 2 (3) (2020) 033100.

\bibitem{ref29} J. L. Cheng, N. Vermeulen, J. E. Sipe, Third-order nonlinearity of graphene: Effects of phenomenological relaxation and finite temperature, Phys. Rev. B 91 (23) (2015) 235320.

\bibitem{ref30} G. B. Ventura, D. J. Passos, J. M. B. Lopes dos Santos, J. M. Viana Parente Lopes, N. M. R. Peres, Gauge covariances and nonlinear optical responses, Phys. Rev. B 96 (3) (2017) 035431.

\bibitem{ref31} D. E. Parker, T. Morimoto, J. Orenstein, J. E. Moore, Diagrammatic approach to nonlinear optical response with application to Weyl semimetals, Phys. Rev. B 99 (4) (2019) 045121.

\bibitem{ref32} N Goldman \textit{et al.}, Measuring topology in a laser-coupled honeycomb lattice: from Chern insulators to topological semi-metals ,2013 New J. Phys. 15 013025.










\end{thebibliography}
\end{document}